\documentclass[fleqn,usenatbib]{mnras}

\usepackage{newtxtext,newtxmath}

\usepackage[T1]{fontenc}
\usepackage{graphicx}	
\usepackage{xcolor}	
\usepackage{amsmath}	
\usepackage{caption}    
\usepackage{subcaption} 
\usepackage{multicol}  
\usepackage{booktabs}  
\usepackage{longtable}
\usepackage{footmisc}
\usepackage{hyperref}
\usepackage{upgreek}
\usepackage[flushleft]{threeparttable}

\DeclareRobustCommand{\VAN}[3]{#2}
\let\VANthebibliography\thebibliography
\def\thebibliography{\DeclareRobustCommand{\VAN}[3]{##3}\VANthebibliography}

\newcommand{\kpc}{{\rm kpc}}

\newcommand{\stream}{\rm s}
\newcommand{\cgm}{\rm cgm}

\usepackage{graphicx}	
\usepackage{amsmath}	
\usepackage{verbatim}	

\title[Stability and Ly$\alpha$ emission of Cold Stream]{
Stability and Ly$\alpha$ emission of Cold Stream in the Circumgalactic Medium: \\ impact of magnetic fields and thermal conduction}

\author[N. Ledos et al.]{
Nicolas Ledos,$^{1\thanks{E-mail: ledos@astro-osaka.jp}}$
Shinsuke Takasao,$^{1}$
Kentaro Nagamine$^{1,2,3}$
\\
$^{1}$Theoretical Astrophysics, Department of Earth $\&$ Space Science, Graduate School of Science, Osaka University, 1-1 Machikaneyama,\\ Toyonaka, Osaka 560-0043, Japan \\
$^{2}$Kavli IPMU (WPI), The University of Tokyo, 5-1-5 Kashiwanoha, Kashiwa, Chiba 277-8583, Japan \\
$^{3}$Department of Physics $\&$ Astronomy, University of Nevada, Las Vegas, 4505 S. Maryland Pkwy, Las Vegas, NV 89154-4002, USA\\
}

\date{Accepted XXX. Received YYY; in original form ZZZ}

\pubyear{2023}

\begin{document}
\label{firstpage}
\pagerange{\pageref{firstpage}--\pageref{lastpage}}
\maketitle

\begin{abstract}
Cold streams of gas with temperatures around $10^4$\,K play a crucial role in the gas accretion on to high-redshift galaxies.
The current resolution of cosmological simulations is insufficient to fully capture the stability and Ly$\alpha$ emission characteristics of cold stream accretion, underscoring the imperative need for conducting idealized high-resolution simulations.
We investigate the impact of magnetic fields at various angles and anisotropic thermal conduction (TC) on the dynamics of radiatively cooling streams through a comprehensive suite of two-dimensional high-resolution simulations. 
An initially small magnetic field ($\sim 10^{-3} \, \upmu \rm G$), oriented non-parallel to the stream, can grow significantly, providing stability against Kelvin-Helmholtz instabilities and reducing the Ly$\alpha$ emission by a factor of $<20$ compared to the hydrodynamics case.
With TC, the stream evolution can be categorised into three regimes: (1) the Diffusing Stream regime, where the stream diffuses into the surrounding hot circumgalactic medium; (2) the  Intermediate regime, where TC diffuses the mixing layer, resulting in enhanced stabilization and reduced emissions; (3) the  Condensing Stream regime, where the impact of magnetic field and TC on the stream's emission and evolution becomes negligible. 
Extrapolating our findings to the cosmological context suggests that cold streams with a radius of $\leq 1 \rm \, \kpc$ may fuel galaxies with cold, metal-enriched, magnetized gas ($B \sim 0.1\text{--}1 \, \upmu \rm G$) for a longer time, leading to a broad range of Ly$\alpha$ luminosity signatures of $\sim 10^{37}\text{--}10^{41}\, \rm \, erg \, s^{-1}$.
\end{abstract}




\section{Introduction}
In the framework of the $\Lambda$ cold dark matter (CDM) paradigm, the process of gas accretion on to high-redshift galaxies at $z>2$ is primarily driven by cold gas flowing along dark matter filaments of the cosmic web. This cold gas is considered to be one of the main components contributing to the overall gas accretion phenomenon in the high-redshift universe  \citep[e.g.][]{Fardal2001,Keres2005,Dekel2006,Dekel2009}. 
Such accretion, often referred to as cold streams or cold flows, plays a crucial role in fueling galaxies with gas that is readily available for collapse and subsequent star formation. 
\citet{Dekel2006} provided key insights into the conditions required for the survival of cold streams within massive haloes.
On top of being ubiquitous in cosmological simulations, the cold stream accretion scenario provides a key physical mechanism for explaining the observed cosmic star-formation history \citep[e.g.][]{Reddy2009,Cucciati2012,Gruppioni2013,Madau2014}, the low-redshift galaxy color bimodality \citep[e.g.][]{Strateva2001,Kauffmann2003,Blanton2003,bimodal_Baldry2004,bimodal_Bell2004,Nelson2018}, and the acquisition of galaxy angular momentum \citep[e.g.][]{Danovich2015}.

Recent observational data provide growing support for the cold stream accretion scenario \citep{Behroozi2019,Daddi2022a,Daddi2022b}. However, direct observations of cold streams remain relatively scarce. Observational support for cold accretion using absorption line spectroscopy of background quasars or galaxies primarily involves \ion{Mg}{II} and \ion{Fe}{II} lines \citep{Giavalisco2011,Rubin2012,Martin2012,Bouche2013,Bouche2016,Zabl2019}, as well as \ion{H}{I} gas \citep{Turner2017,Chen2020,Fu2021}.
The limited number of detections of cold accreting gas through absorption line systems can be attributed to their small covering factor compared to the surface area of the halo \citep{Faucher-Giguere2011}. 
On the other hand, emissions-line observations of Ly$\alpha$ emitters have revealed filamentary structures in or around the halos of high-redshift galaxies \citep[e.g.,][]{Cantalupo2014,Fumagalli2016,Borisova2016,Umehata2019}, as well as emission from cold gas consistent with cold stream emission\citep[e.g.,][]{Battaia2018,ChristopherMartin2019}. \citet{Daddi2021} utilized observations of Ly$\alpha$ emission to identify the presence of clear cold filamentary gas structures surrounding massive galaxies at a redshift of $z=2.9$.
\citet{Emonts2023}, on the other hand, detected cold filamentary gas structures using observations of neutral carbon (\ion{C}{i}) emission at $z=3.8$. In contrast, \citet{Zhang2023}, through Ly$\alpha$ and metal lines, detected emissions consistent with inspiraling streams around a galaxy at redshift $z=2.3$.
These observations provide direct evidence for the existence of cold streams near these galaxies, although they also highlight the challenge of relating the emissions to cold streams. Hence, comprehending the emission signature of cold streams is a crucial step in establishing their widespread occurrence beyond the realm of cosmological simulations.

Concurrently, there have been recent endeavors to investigate the influence of simulation resolution on the properties of gas within the halo, specifically the circumgalactic medium (CGM) \citep{Peeples2019,VandeVoort2019,Hummels2019,Suresh2019,Nelson2020,Bennett2020}. \citet{Bennett2020} find that increasing their mass resolution by a factor of $512$ near shocks increases the cold gas content in the CGM by $\sim 50\%$ and the inflow rate of cold gas by $\sim 25\%$ compared to typical case, which gives a much more multiphase and turbulent picture of the CGM than the usual one from the state-of-the-art cosmological simulations. However, they do not fully achieve convergence at their finest resolution.
\citet{Nelson2020} also shows that a resolution of $\Delta x < 100 \, \rm pc$ is needed to fully resolve the small-scale cold gas structure in the CGM of massive galactic haloes ($M_{\rm{h}}>10^{12}\, M_{\sun}$) at redshift $z=0.5$. In the case of cold stream accretion, the lack of resolution does not allow the development of Kelvin-Helmholtz Instabilities (KHI) at the interface between the cold dense gas of the stream and the hot diffuse CGM. These instabilities can shorten the lifetime of the stream as first studied by \citet{Mandelker2016}. 
Given that the typical resolution of CGM in most cosmological simulations near the virial radius is around $\sim 1\, \kpc$, it also suggests the need to study the cold stream evolution further using high-resolution simulations.

 To understand both the evolution and the emission signatures from cold streams, numerous works performed high-resolution simulations considering idealized stream geometry:  \citet{Mandelker2016} (linear analysis), \citet{Padnos2018} (2D hydrodynamic (HD) simulations), and \citet{Mandelker2019} (2D and 3D HD simulations) initiated such work with HD simulations and revealed that cold streams could be disrupted by both KHI surface modes and body modes (also called reflective modes). 
They concluded that surface modes have the highest growth rate and were the dominant mode that could alter the cold stream evolution. \citet{Aung2019} targeted the impact of self-gravity, which can cause the stream to fragment from gravitational instabilities. With 2D and 3D magnetohydrodynamic (MHD) simulations, \citet{bib_MHD_Berlok_2019b} studied the impact of the magnetic fields parallel to the stream, and found that it can help the stream to survive KHI if the field strength is strong enough ($B	\gtrsim 0.3\text{--}0.8\,  \upmu\rm G$). 
\citet{Vossberg2019} investigated via 2D HD simulations the appearance of over-dense stream regions from the growth of KHI. 
While these works all mainly investigated the cold stream evolution, \citet{Mandelker2020a} provided valuable insights into the emission signature of cold streams by incorporating radiative cooling into HD simulations, showing that the cooling emission scales with the cooling time as $\propto t_{\rm{cool}}^{-1/4}$. 
Further analytical considerations \citep{LyblobMandelker2020} demonstrate that the cold streams, with radius $R_{\stream}> 3 \rm \kpc$, exhibit Ly$\alpha$ luminosities exceeding $L_{\rm Ly\alpha} > 10^{42} \rm erg\, s^{-1}$ for halo masses $M_{\rm{h}}>10^{12} \rm M_{\sun}$ at $z\sim 2$.

The key emission mechanism comes from the mixing of the hot CGM gas ($T_{\cgm}\sim 10^6\, \rm K$) and the cold stream gas ($T_{\stream}\sim 10^4 \rm \, K$) which creates a gas mixture at an intermediate temperature ($T_{\rm{mix}} < 10^5 \rm \, K$) whose cooling rate becomes orders of magnitude higher \citep{Begelman1990}. \citet{Gronke2018,Gronke2020} investigated this mechanism from simulations of cold clouds embedded in a hot wind. The latter work found that both the cooling emission and condensation/mixing velocity of the cloud scaled with the cooling time in the mixing layer as $\propto t_{\rm{cool}}^{-1/4}$. From higher resolution shear layer simulations, \citet{Fielding2020} and \citet[][for strong cooling]{Tan2021} retrieve such scaling while \citet{Ji2019} (and \citet[][for weak cooling]{Tan2021}) found a scaling of $\propto t_{\rm{cool}}^{-1/2}$. Further simulations also investigated the impact of high Mach numbers \citep{Yang2023}. Hence, in the case of HD simulations with radiative cooling, one may predict the evolution of the cold stream in terms of mass flux and emission thanks to the estimated cooling time in the mixing layer.

The impact of additional physics, such as magnetic fields and thermal conduction, remains unanswered when combined with radiative cooling for studying cold streams. One can find some insights from simulations of cold clouds in the CGM. \citet[][2D HD simulations]{Armillotta2017} found that isotropic conduction can hinder the growth of KHI and increase the survival time of the cloud. \citet[][]{Hidalgo-Pineda2023} investigated with 3D simulations the impact of magnetic field and concluded that the magnetic field could help stabilize the cloud against KHI, allowing it to survive for a longer time scale. \citet{Bruggen2023} investigated both isotropic and anisotropic thermal conduction from 3D MHD simulations with different magnetic field angles. They found that both the magnetic field and thermal conduction can lower the KHI growth, i.e., the mixing of the cold cloud gas and the CGM, allowing the cloud to survive longer.
\citet{Sander2021} studied high-velocity clouds inside the CGM from 3D MHD simulations, including self-gravity, star formation, thermal conduction, and additional physics. They concluded that thermal conduction also helps stabilize the cloud but that it also diffuses the cold gas substructure, which has been detached from the cloud (in the cloud tail, for example).

To sum up, the mixing of the CGM gas and the streaming gas triggered by KHI is a crucial mechanism that can explain the stream emission signature and its evolution. In particular, the emission from the mixing layer is dominated by Ly$\alpha$ which might be linked to observed Ly$\alpha$ emitters. However, from simulations of cold clouds in the hot CGM, it appears that magnetic fields with various angles and thermal conduction can each affect the growth of the KHI. Reducing the KHI growth rate can increase the survival time of cold streams but may also decrease its emission. We hence intend to address this issue by performing a large suite of 2D MHD simulations ($\sim$ 120 simulations), including radiative cooling and anisotropic thermal conduction. Focusing on 2D simulations allows us to cover a wide range of parameters with different stream velocities, CGM/stream densities, and magnetic field angles, on HD, MHD, and MHD with thermal conduction (MHD+TC) simulations.

We start by describing the idealized cold stream model and the relevant time-scales in Section~\ref{sec2}. The numerical setup and the initial conditions of the numerical experiments are then described in Section~\ref{sec3}. Section~\ref{sec4} presents our results and analysis on the impact of magnetic fields and thermal conduction on the evolution of and emission from cold streams. Finally, we discuss the extrapolation of our results in a cosmological context and the caveats of our work in Section~\ref{sec5} before concluding in Section~\ref{sec6}.

\section{Cold stream model and relevant time-scales}\label{sec2}
This section discusses the cold stream model and the relevant time-scales for our simulations. We describe the model of the cold stream and the chosen parameters (Sec.~\ref{sec2:cold_stream}), the radiative cooling-heating model (Sec.~\ref{sec2:cooling}), the evolution of the mixing layer (Sec.~\ref{sec2:mixing_layer}), and the thermal conduction (Sec.~\ref{sec2:TC}). The section ends with a definition of the stream evolution regimes based on the important time-scales  (Sec.~\ref{sec2:timescale_comp}).

\subsection{Typical Parameters of Cold Streams}\label{sec2:cold_stream}
Our choice of parameters for the stream model is similar to previous numerical studies of idealised cold streams \citep[see, e.g.,][]{bib_MHD_Berlok_2019b,Mandelker2020a}.
Observations of cold inflow in the CGM typically target massive haloes with halo masses of $M_{\rm{h}} \gtrsim 10^{11}\, \rm M_{\sun} $, covering redshifts from $z\sim 0.4$ \citep{Martin2012,ChristopherMartin2019} to $z\sim 3.8$ \citep{Emonts2023}. The inferred \ion{H}{I} column densities of cold streams span over a wide range of $N_{\rm HI} \sim 10^{17} - 10^{21} \rm cm^{-2}$ with metallicities $Z \sim 10^{-3.8} \, \rm{Z_{\sun}} \text{--}1 \rm \, Z_{\sun}$. 

Building upon a cosmological simulation, \cite{Dekel2013} developed a simplified model for star-forming galaxies within massive haloes ($M_{\rm{h}}>10^{11}\rm M_{\sun}$) at $z>1$ with the following virial radius and velocity: 
\begin{equation} \label{eq:Rv}
R_{\rm{v}} \sim 100 \left(\frac{M_{\rm{v}}}{10^{12} \rm{M_{\sun}}} \right)^{1/3}\left(\frac{3}{1+z} \right) \, \rm kpc\, , 
\end{equation}
\begin{equation} \label{eq:Vr}
V_{\rm{v}} \sim 200 \left(\frac{M_{\rm{v}}}{10^{12} \rm{M_{\sun}}} \right)^{1/3}\left(\frac{3}{1+z} \right)^{1/2} \, \rm km\, s^{-1}\, , 
\end{equation}
where $M_{\rm{v}}$ is the virial mass of the halo and is taken as $10^{12} \, \rm M_{\sun}$.
For such halo mass at $z\sim 2$, the cosmological simulation \citep{Goerdt2010} and analytical extension of the model of \citet{Dekel2013} \citep[see][]{Padnos2018,Mandelker2020a,LyblobMandelker2020} give a stream number density $n_{\stream}\sim 10^{-3}\text{--}10^{-1} \, \rm cm^{-3}$, a density ratio of the stream density over the CGM density $\delta\sim 30\text{--}300$, {color{red} a stream radius $R_{\stream}\sim 3\text{--}50\, \kpc$}, and a Mach number based on the CGM sound speed $\mathcal{M}_{\cgm}\sim 0.75\text{--}2.25$. For the metallicities in the stream and in the CGM at the virial radius, cosmological simulations give tighter constraints than observations with $Z_{\stream}/Z_{\sun} \sim 10^{-2}\text{--}10^{-1}$ \citep{Goerdt2010}, $10^{-2}\text{--}10^{-1.5}$ \citep{Fumagalli2011,VanDeVoort2012,Roca-Fabrega2019} and $10^{-1.2}$ \citep{Strawn2021}, and $Z_{\cgm}/Z_{\sun}$ spanning from $10^{-2}$ \citep[][i.e., their lower value]{Roca-Fabrega2019} to $10^{-0.5}$ \citep[i.e., their average value]{Strawn2021}.

Consistently with previous work, we target three number densities $n_{\stream}=10^{-3},10^{-2},10^{-1} \, \rm cm^{-3}$, along with two density ratios $\delta \equiv \rho_{\stream} / \rho_{\cgm} = 30, 100$, one set of stream/CGM metallicities $\left(Z_{\stream},Z_{\cgm}\right) = \left(10^{-1.5},10^{-1}\right) \, \rm Z_{\sun}$, and three different stream Mach number $\mathcal{M}=\mathcal{M}_{\cgm}=0.5,1,2$. We also choose to fix the stream radius to $R_{\stream} = 1\, \kpc$.  While our chosen value $R_{\stream}$ is below the analytical estimate of \citet{LyblobMandelker2020}, using such a small value allows us to better study the effects of thermal conduction on stream evolution. Furthermore, cosmological simulations that hyper-refine streams in the CGM suggest that these may contain smaller-scale stream-like structures \citep{Bennett2020}. The impact of a larger radius is discussed in Sec.~\ref{sec2:timescale_comp} and in Sec.~\ref{sec6}.

Our understanding of the magnetic field in the large-scale structure of the Universe remains incomplete. The primordial magnetic field has been constrained to a lower limit of $\sim 10^{-10}\text{--}10^{-9}\, \rm \upmu\rm G$ 
 \citep{bib_Mobs_Neronov_2010,bib_Mobs_Dolag_2011}. 
Although the evolution of the primordial magnetic field has been theorized \citep{bib_Mth_Saveliev_2012}, it may be more reliable to focus on magnetic fields that have been studied in recent simulations and observations of the CGM.

Little is known about the properties of the magnetic field in the CGM of high-redshift galaxies. 
Most zoom-in simulations are focused on the magnetic field strength growth \citep{bib_MHD_Rieder_2017a,bib_MHD_Martin-Alvarez_2018} and morphology \citep{bib_MHD_Pakmor_2017,bib_MHD_Pakmor_2018,bib_MHD_Steinwandel_2019} in the galactic disc due to the small-scale dynamo.
Therefore, to better understand the composition and magnetic morphology of the CGM, we may rely on CGM-focused simulations. Simulations from the Auriga project \citep{bib_MHD_Pakmor_2020} and FIRE project \citep{Hopkins2020} have investigated the evolution of the magnetic field in the CGM, providing estimates of the magnetic field strength ranging from $10^{-3}\,  \upmu\rm G$ to $10^{-2}\,  \upmu\rm G$ at the virial radius and for redshift $z\sim 2$.

Observations of the near-centre CGM gas have put upper constraints on the magnetic field strength, but probing the magnetic field in high-redshift haloes is challenging. Using fast radio bursts, \citet{bib_Mobs_Prochaska_2019} have constrained the magnetic field to a range of $6\times 10^{-2}\text{--}2\, \upmu\rm G$ for electron density range of $10^{-5}\text{--}10^{-3}\,\rm{cm^{-3}}$ inside the hot halo ($T_{\cgm}\sim 10^6 \rm K$).
The study by \citet{bib_Mobs_Lan_2020} of over 1000 Faraday Rotation Measures in low-redshift galaxies ($z<1$) provides a lower constraint of $2\,  \upmu\rm G$ for the upper limit of the coherent magnetic field. 
Since the magnetic field in the halo grows with time, these upper limits motivate us to investigate the effects of a low magnetic field.

\cite{bib_MHD_Berlok_2019b} investigated the impact of a magnetic field aligned with the cold stream on the growth of KHI, using a magnetic field strength of approximately $1\,  \upmu \rm G$. 
This relatively high magnetic field strength ($\beta <100$) was chosen to explore a range of dynamically relevant magnetic field strengths for cold streams without radiative cooling, when the magnetic field is parallel to the stream. In our work, we investigate a magnetic field with various angles in which cases an additional amplification of the magnetic field strength can be expected due to the stretching of the magnetic field lines from the velocity difference at the interface between the stream and the CGM. Hence, we focus on a lower magnetic field strength of $B\sim 10^{-3}\,  \upmu\rm G$ defined by an initial ratio of thermal pressure over magnetic pressure $\beta = p / p_{\rm mag} = 10^5$ for our study.
This gives an Alfvenic time-scale,
\begin{equation} \label{eq:ta}
t_{\rm{a}} \sim R_{\stream}/v_{\rm a},
\end{equation}
with the Alfven speed $v_{\rm{a}} =B_0/\sqrt{\rho}$.
Defining the sound crossing time of the stream as $t_{\rm{sc}} = 2R_{\stream}/ c_{\stream}$, we have $t_{\rm{a}}\sim 0.5t_{sc}c_{\stream} / \left(2 \beta\right)^{0.5} \sim 100 \, t_{\rm{sc}} $, meaning that the Alfvenic time can be initially ignored.

\subsection{Radiative cooling-heating}\label{sec2:cooling}
\label{sec:rad_intro}

Tabulated cooling and heating rates are derived from the photoionization code CLOUDY \citep{Ferland2017}, which accounts for both atomic and metal cooling processes. 
The heating rates from \citet{Haardt2012} are employed for the UV background radiation from galaxies and quasars at redshift $z=2$.

\begin{figure*}
	\includegraphics[width=\columnwidth]{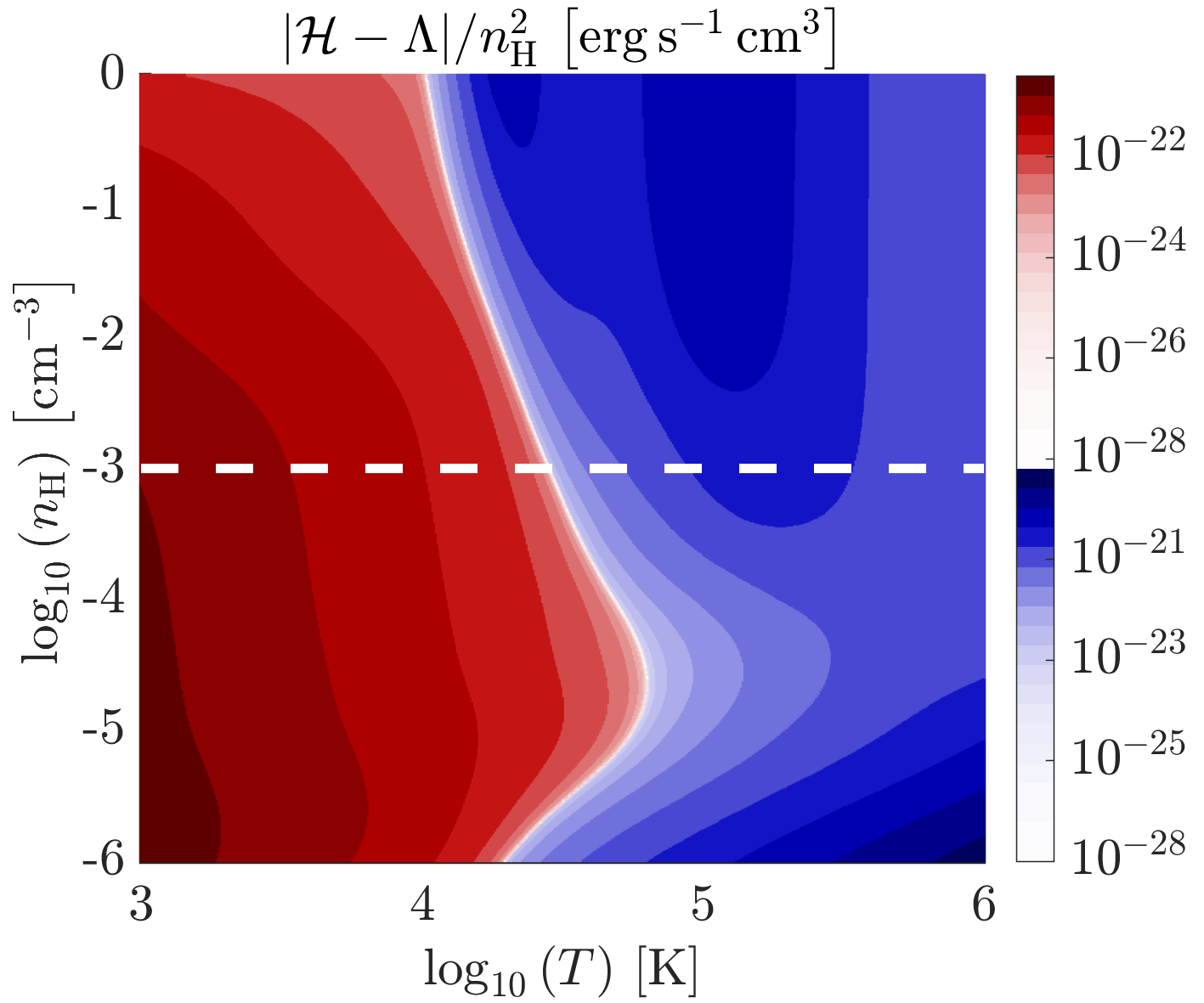}
	\includegraphics[width=\columnwidth]{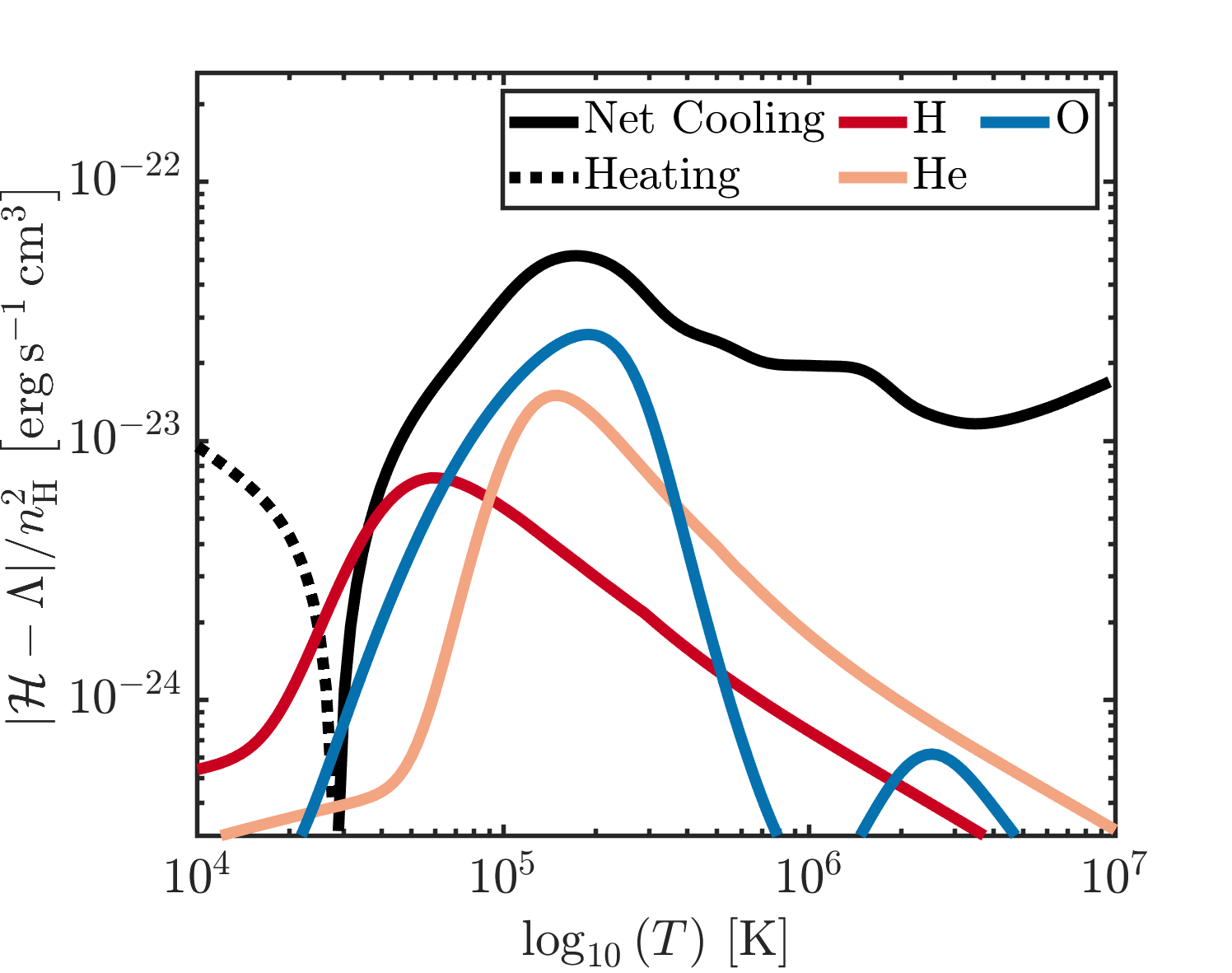}
    \caption{Net cooling-heating map from our CLOUDY model at a metallicity of $Z=0.1Z_{\sun}$. {\it Left panel:} the blue and red colormap refer to cooling and heating scaling, respectively. The white dashed line represents the heating-cooling curve at the fixed density of the right panel.
    {\it Right panel:} Net heating-cooling curve at a fixed number density $n_{\rm{H}} = 10^{-3} \, \rm cm^{-3}$. Plain lines and dashed line represents cooling and heating contributions, respectively. Colored lines represent the main species contributing to the cooling  under $10^6 \, \rm K$.}
    \label{fig:net_cool_heat_1}
\end{figure*}

Fig.~\ref{fig:net_cool_heat_1} presents the resulting cooling/heating map for the assumed CGM metallicity of $Z=0.1\, \rm Z_{\sun}$. 
The left panel presents logarithmic scales of the heating and cooling rates in red and blue color maps, respectively, where the white region denotes near-equilibrium states. The red region is dominated by heating, while the blue region is dominated by cooling.
The right panel shows the net cooling/heating curves of a typical gas in the mixing layer with number density $n_{\rm{mix}}=10^{-3} \, \rm cm^{-3}$.  The colored lines represent the total cooling from the main species in our temperature region. 

To ensure the thermal equilibrium of the cold stream in our initial conditions of the simulation, we determine its initial temperature from
\begin{equation} \label{eq:rad1}
\mathcal{H}_{n_{\rm{H}}}\left( T_{\stream}\right)-\Lambda_{n_{\rm{H}}}\left( T_{\stream}\right) = 0,
\end{equation}
where $\mathcal{H}$ is the heating rate and $\Lambda$ is the cooling rates, respectively. This gives stream temperatures of $T_{\stream}\sim (1.3,1.9,3.1)\times 10^4\, \rm K$ for $n_{\stream}=(10^{-1},10^{-2},10^{-3}) \, \rm cm^{-3}$, respectively.
 Assuming an isobaric cooling, the resulting cooling time for gas in the mixing layer at temperature $T_{\rm{mix}}$ is
\begin{equation} \label{eq:rad2}
t_{\rm{cool}} = \frac{ T_{\rm{mix}} k_{\rm{b}}}{\left(\gamma -1 \right)n_{\rm{mix}}\Lambda_{\rm net,mix}},
\end{equation}
where the net cooling-heating rate $\Lambda_{\rm net,mix}=\mathcal{H}_{n_{\rm{mix}}}\left( T_{\rm{mix}}\right)-\Lambda_{n_{\rm{mix}}}\left( T_{\rm{mix}}\right)$ is defined by the number density and temperature in the mixing layer between the stream and the CGM. The value of the mixing layer number density and temperature are defined in the subsequent section.

\subsection{Mixing layer}\label{sec2:mixing_layer}
We hereby summarise previous studies on the mixing layer relevant to our work. \citet{Begelman1990} and \citet{Slavin1993} first described the physical properties of the mixing of two gases from a shear interface in the case of the interstellar medium. In particular, considering the mass accretion rate $\dot{m}_{\rm c}$ and $\dot{m}_{\rm h}$ of the cold and the hot phase, they defined the temperature of the mixed phase as,
\begin{equation} \label{eq:T_mix}
T_{\rm{mix}} =\frac{\dot{m}_{\rm{c}}T_{\rm{c}} +  \dot{m}_{\rm{h}}T_{\rm{h}}}{\dot{m}_{\rm{c}} +  \dot{m}_{\rm{h}}}\sim \left(T_{\rm{c}}T_{\rm{h}}\right)^{\frac{1}{2}},
\end{equation}
assuming ideal mass accretion rate as $\dot{m}_{\rm{h-c}} \sim \rho_{\rm{h-c}} v_{\rm t,h-c}$ for both phases with $v_{\rm t,h-c}$ being the turbulent velocity for the hot and the cold phases, linked by equating their kinetic energies,  $\rho_{\rm{c}}v_{\rm{t,c}}^2 = \rho_{\rm{h}}v_{\rm{t,h}}^2$\footnote{The turbulent kinetic energy equality is not explicitly mentioned by \citet{Begelman1990} but is implicitly contained in his definition of $v_{\rm t,c}$.}.
This was later followed by detailed studies from simulations on interface geometry \citep[e.g.][]{Kwak2010,Ji2019,Fielding2020,Tan2021,Yang2023}, on the cold gas entrained in hot wind \citep{Gronke2018,Gronke2020}, and on cold streams \citep{Mandelker2020a}.

In the case where the radiative cooling is strong enough to condense gas from the hot phase, the mixing velocity or the inflow velocity of CGM gas into the mixing layer scales as \citep{Gronke2020,Mandelker2020a,Fielding2020,Tan2021}
\begin{equation} \label{eq:v_in}
v_{\rm{in}} \propto t_{\rm{cool}}^{-1/4}.
\end{equation}
The inflow of the hot gas in the mixing layer occurs at a steady rate. Once at the temperature $T_{\rm{mix}}$, the gas mixture cools efficiently, leading to a steady cooling emission which also scales as $L_{\rm{net}} \propto t_{\rm{cool}}^{-1/4}$.
From $v_{\rm{in}}$, one can also recover the mass evolution of the cold stream for sufficiently strong cooling due to the condensation of CGM gas,
\begin{equation} \label{eq:Mdot_th}
\dot{M}_{\stream} \sim \rho_{\cgm}v_{\rm{in}}S,
\end{equation}
where $S$ is the surface between the stream and the CGM.
We note that a different scaling is found by \citet{Ji2019} with $v_{\rm{in}} \propto t_{\rm{cool}}^{-1/2}$, similarly to the weak cooling case in \citet{Tan2021}.
While we discuss the scaling in our simulations, higher resolution simulations targeting specifically the mixing layer \citep{Fielding2020,Tan2021} might be needed to investigate the origin of the scaling properly in the presence of magnetic field and thermal conduction. Such work is beyond the scope of this paper.

The evolution of the radiative mixing layer is also important as it can stabilize the stream against the KHI. The initial growth of the mixing layer before its steady evolution due to cooling can be described by the shearing time \citep{Mandelker2019,Mandelker2020a},
\begin{equation} \label{eq:t_sh}
t_{\rm{sh}} = \frac{R_{\stream}}{\alpha v_{\stream,0}},
\end{equation}
with the dimensionless growth rate defined by the empirical fitting value $\alpha \sim 0.042+0.168\rm{exp}\left(-3\mathcal{M}_{\rm{tot}}^2\right)$ \citep{Dimotakis1991}, where the total Mach number is defined with the sound speed of both phases $\mathcal{M}_{\rm{tot}}= V_{\stream}/\left(c_{\stream}+c_{\cgm}\right)$. 

\subsection{Thermal conduction}\label{sec2:TC}

The thermal conduction time-scale for the stream is defined as
\begin{equation} \label{eq:tau_s}
\tau_{\stream} \equiv \frac{\rho_{\rm{mix}} R_{\stream} ^2}{\kappa_0 T_{\rm{mix}}^{2.5}}, 
\end{equation}
where we assume the Spitzer thermal conduction coefficient $\kappa = \kappa_0 T^{2.5}$ \citep{Spitzer1962}. Note that the coefficient is defined based on the mixing phase, as the temperature of the diffusion front is approximately the same.

This diffusion time, along with the cooling time, determines whether the stream will grow in mass or diffuse. However, we are also interested in the impact of thermal conduction on the mixing between the stream and the CGM due to the growth of the KHI. Therefore, we also define the diffusion time for a small perturbation of size $0.1R_{\rm \stream}$, as
\begin{equation} \label{eq:tau_p}
\tau_{\rm{p}} \equiv \frac{\rho_{\rm{mix}} \left(0.1R_{\stream}\right)^2}{\kappa_0 T_{\rm{mix}}^{2.5}},
\end{equation}
with the subscript $p$ standing for perturbation and where the diffusivity is also defined from the mixing phase.

\subsection{Time-scale comparison}\label{sec2:timescale_comp}
\begin{figure}
	\includegraphics[width=\columnwidth]{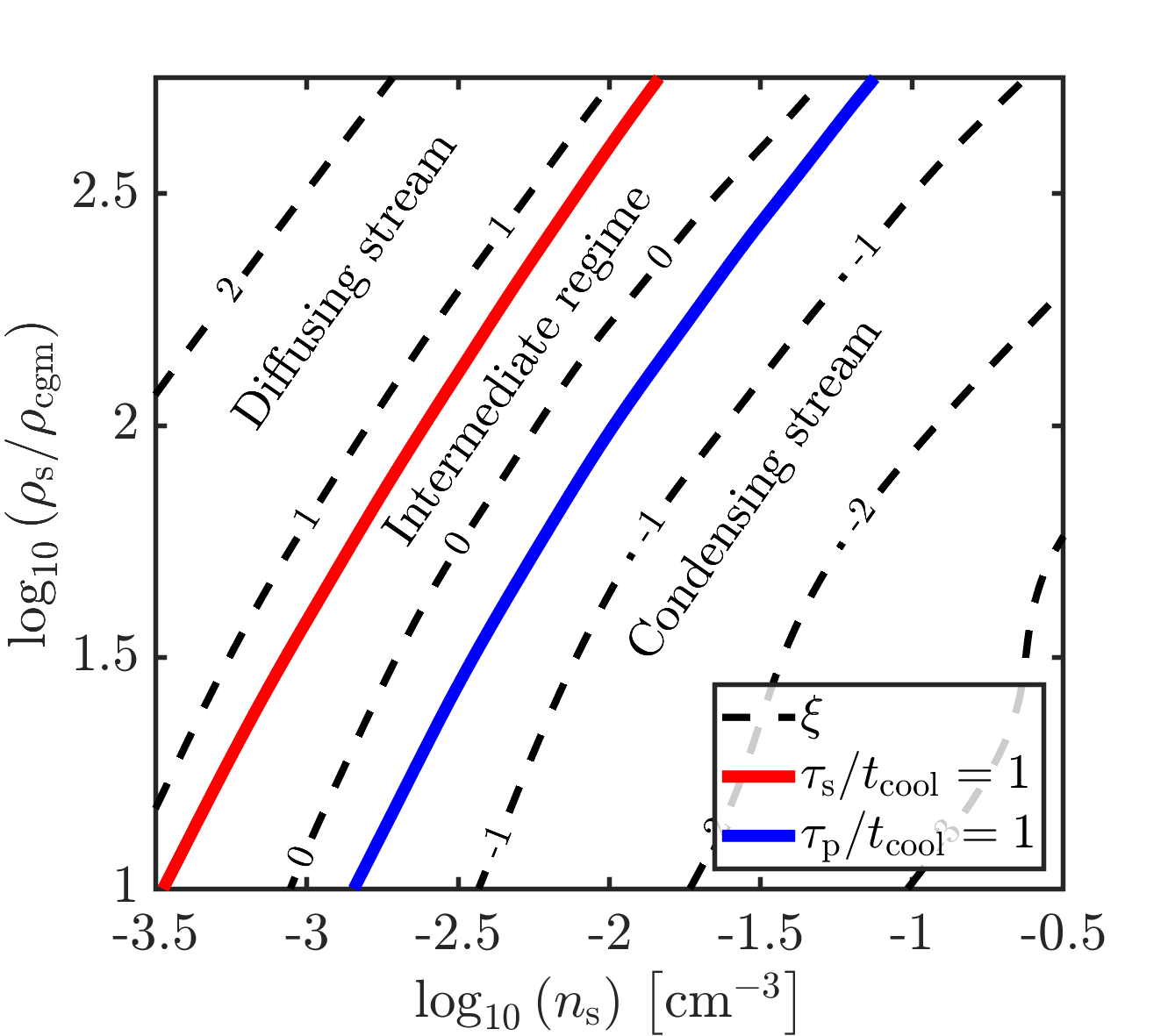}
    \caption{Contours of the ratio $\xi = t_{\rm{cool}} / t_{\rm{sh}}$ (dashed black lines) in log-scale as functions of the density ratio of the stream over CGM, and stream number density. The value $\log_{10}\left(\xi \right)$ is displayed on each contour line. The cold stream is assumed to be at the virial radius of a $M_{\rm{h}} = 10^{12} \rm{M_{\sun}}$ halo at $z=2$, with $R_{\stream} = 1\rm \, \kpc$, metallicity $Z_{\stream}/Z_{\sun}=10^{-1.5}$, Mach number $\mathcal{M}_{\cgm}=1$, and CGM metallicity $Z_{\cgm}/Z_{\sun}=10^{-1}$. The blue and red curves show the contours for which the diffusion times $\tau_{\stream}$ and $\tau_{\rm{p}}$ equal $t_{\rm cool}$. Those limits define three different regimes for the cold stream evolution in the presence of TC: {\it diffusing stream}, {\it intermediate}, and {\it condensing stream} regimes.}   
    \label{fig:time_comp}
\end{figure}

From the physical processes presented above, we can assume the evolution of the stream by considering the different time-scales $t_{\rm sh}$, $t_{\rm cool}$, $\tau_{\stream}$, and $\tau_{\rm{p}}$.
An important parameter to describe the stream evolution is the ratio of the cooling time over the shearing time,
\begin{equation} \label{eq:xi}
\xi =\frac{t_{\rm{cool}}}{t_{\rm{sh}}}.
\end{equation}
Fig.~\ref{fig:time_comp} plots contours of the ratio $\xi$ on the plane of stream density $n_{\stream}$ and density ratio $\delta = \rho_{\stream} / \rho_{\cgm}$.
In the hydrodynamic case, the stream evolution can be defined by $\xi$ as follows: 
\begin{itemize}
    \item {\it Disrupting stream} regime: $\xi>1$,
    \item {\it Condensing stream} regime: $\xi<1$.
\end{itemize}
In the presence of thermal conduction, the above categorization is modified by the ratios $\tau_{\stream}/t_{\rm cool}$ and $\tau_{\rm{p}}/t_{\rm cool}$, both shown in the plot. Those two ratios are an equivalent formulation to the Field length, which defines the limit at which a structure either diffuses or condenses. Hence, $\tau_{\stream}/t_{\rm cool}=1$ and $\tau_{\rm{p}}/t_{\rm cool}=1$ mark the limit for which the stream or a cold clump, respectively, either diffuse or survive.
Including thermal conduction with our cold stream parameters, we end up with three different regimes of interest: 
\begin{itemize}
    \item \textbf{{\it Diffusing stream} regime:} $\tau_{\rm{p}} < \tau_{\stream} < t_{\rm{sh}} < t_{\rm{cool}}$ \\
    In this case, the diffusion from thermal conduction (TC) is too rapid and overcomes all other processes. The stream diffuses faster than it can condense gas from itself or the CGM. This regime is analogous to a Field length larger than the stream radius. In practice for our model, this regime is obtained with the condition $\xi \gtrsim 8$.
    \item \textbf{{\it Intermediate} regime:}  $\tau_{\rm{p}} < t_{\rm{sh}} < t_{\rm{cool}} < \tau_{\stream} $ \\
    In this regime, the cooling time and the shearing time are about the same magnitude and are bigger than the diffusion time $\tau_{\rm p}$. Hence, the diffusion of perturbations or small clumps smaller than $0.1R_{\stream}$, happens faster than their growth, potentially shutting off the mixing and the subsequent condensation of CGM gas. In this regime, the mixing layer should diffuse and the stream remains at a constant mass while possibly fragmenting due to long-wavelength KHI modes. From our parameters, this regime can be roughly defined by $0.3<\xi <8$.
    \item \textbf{{\it Condensing stream} regime:} $t_{\rm{cool}} < t_{\rm{sh}} < \tau_{\rm{p}} < \tau_{\stream} $\\
    In this regime, radiative cooling occurs faster than both diffusion and KHI growth. The gas in the mixing layer cools efficiently, hence condensing on to the stream. We found that this regime is satisfied for $\xi \lesssim 0.3$.
\end{itemize}

The regime map of Fig.~\ref{fig:time_comp} is also dependent on the assumed stream radius $R_{\stream}$ and the metallicities of the CGM and stream. For a bigger or smaller radius, the {\it condensing stream} regime region in the $n_{\stream}-\delta$ plane would widen or shrink, respectively. Similarly, increasing or decreasing the metalicities would widen or shrink the {\it condensing stream} regime region, respectively.

\section{Numerical set-up}\label{sec3}

\subsection{Governing Equations and Computational Methods}\label{sec3:equations}
\label{sec:gov_eq} 
We solve the following normalized MHD equations in conservative form using the finite volume mesh code Athena++ \citep{Stone2020}, in which we implemented anisotropic thermal conduction and radiative cooling:
\begin{equation}
\left\{ \begin{array}{l}
\dfrac{\upartial\rho}{\upartial t}+\nabla\cdot\left(\rho\mathbfit{u}\right)=0\,,\\
\\
\dfrac{\upartial\left(\rho\mathbfit{u}\right)}{\upartial t}+\nabla\cdot\left[\rho\left(\mathbfit{u}\otimes\mathbfit{u}\right)-\left(\mathbfit{B}\otimes\mathbfit{B}\right) + p_{\rm{T}}\mathbb{I}\right] =0\,,\\
\\
\dfrac{\upartial\mathbfit{B}}{\upartial t}-\nabla\times\left(\mathbfit{u}\times\mathbfit{B} \right)=0 \, ,\\
\\
\dfrac{\upartial e}{\upartial t}+\nabla\cdot\left[\left(e+p_{\rm{T}}\right)\mathbfit{u}-\mathbfit{B}\left(\mathbfit{B}\cdot \mathbfit{u}\right) + \mathbfit{Q}  \right] = \rho^2\left(\mathcal{H} - \Lambda\right) \,,
\end{array}\right.
\label{eq:MHD_system_1}
\end{equation}
where $\mathbb{I}$ is the identity matrix. The total pressure $p_{\rm{T}}$ is defined as
\begin{equation} \label{eq:MHD_system_2}
p_{\rm{T}} = p + \frac{B^2}{2}.
\end{equation}
Notations $\rho$, $\mathbfit{u}$, $\mathbfit{B}$, $e$, $p$, stand for the mass density, velocity vector, magnetic vector, total energy density, and pressure, respectively. The total energy density is defined as
\begin{equation} \label{eq:MHD_system_3}
e = \frac{p}{\gamma - 1} + \frac{B^2}{2} + \rho\frac{u^2}{2}.
\end{equation}
The anisotropic thermal conduction term $\mathbfit{Q}$ is defined as
\begin{equation} \label{eq:MHD_system_4}
\mathbfit{Q} = -\kappa\mathbfit{b}(\mathbfit{b}\cdot \nabla T),
\end{equation}
with $\mathbfit{b} = \mathbfit{B} / B$ the magnetic field unit vector, and $\kappa = \kappa_0 T^{2.5}$ the Spitzer thermal conduction coefficient \citep{Spitzer1962}.
The radiative cooling terms $\mathcal{H}-\Lambda$ are interpolated from tables of our model described in Sec.~\ref{sec:rad_intro}.
We summarise the normalisation units in Table~\ref{tab:Norm_Unit}.

We also trace the mass fraction of gas in the initial stream and CGM gas using a passive scalar transport equation, defined as
\begin{equation} \label{eq:MHD_system_5}
\dfrac{\upartial\left(\rho \mu_{\stream}\right)}{\upartial t}+\nabla\cdot\left(\rho\mathbfit{u}\mu_{\stream}\right)=0,
\end{equation}
where $\mu_{\stream} = \rho_{\stream} / \rho$ is, in other words, the mass fraction of the gas at the initial stream metallicity. In practice, we use this scalar to compute the metallicity of the gas when computing the cooling rate.

\begin{table}
\caption{Normalisation Units.}
\label{tab:Norm_Unit}
\begin{tabular}{l|cc}
\hline
Quantity & Normalised unit & Values\\
\hline
Length & $L_0 = R_{\stream}$ & $1\, \kpc$ \\ 
Velocity & $c_0 = c_{\stream} = \left( \gamma T_0 k_{\rm{b}}/m \right)^{1/2}$ & $ 18\text{--}27\, \rm km\, s^{-1}$ \\ 
Time & $t_0 = L_0 / c_0 $ & $ 37\text{--}56\, \rm Myr$ \\ 
Temperature & $T_0 = T_{\stream}$ & $ \left(1.3\text{--}3.1\right)\times 10^4\, \rm K$ \\ 
Density & $\rho_{\stream,0} = n_{\stream}m_0$ & $ 10^{-27}\text{--}10^{-25}\, \rm g\, cm^{-3}$ \\ 
Pressure & $ P_0 = \rho_0 k_{\rm{b}} T_0 / m_0$ & $10^{-14}\text{--}10^{-13}\, \rm dyn\, cm^{-2}$ \\ 
Magnetic field & $B_0 = \left( 8\pi P_0/ \beta\right)^{1/2}$ & $ \left(1\text{--}6,8\right)\times 10^{-3}\,  \upmu\rm G$ \\ 
\hline
\end{tabular}
\end{table}

The conservative MHD formulation is solved using the HLLD Riemann solver from \cite{Miyoshi2005} using the spatial PLM reconstruction which is second-order accurate. The divergence-free constraint of the magnetic field is ensured with the Constrained-Transport-scheme introduced in \citet{Gardiner2005,Gardiner2008}. 
For time integration, the second-order Runge-Kutta scheme is used for simulations without thermal conduction. In the case of thermal conduction, the conduction time-step is proportional to $\Delta x ^2$, leading to high computational costs. To reduce the computational cost, the super-time-stepping (STS) Runge-Kutta-Legendre second-order solver from \cite{Meyer2014} is used to solve the thermal conduction equation. 
We apply the limiting scheme developed by \citet{Sharma2007} which avoids overestimation of the conduction flux when this one is anisotropic.

\begin{table*}
\caption{List of simulation parameters. From left to right, the physics ID stands for hydrodynamic (HD), magnetohydrodynamic (MHD), and MHD with anisotropic thermal conduction (MHD+TC); the density ratio $\delta$; the stream number density $n_{\stream}$; the magnetic field angle $\alpha$, such that $\alpha=0^\circ$ is a field parallel to the stream; the magnetic field initial strength, the stream Mach number with respect to the CGM sound speed, $\mathcal{M}=\mathcal{M}_{\cgm}$; the CGM sound speed; the ratio of the virial crossing time and $t_0$ for $\mathcal{M}=1$; the stream temperature; the ratio of the cooling time over the shearing time $\xi=t_{\rm{cool}} / t_{\rm sh}$; and the temperature at which the cooling is maximum.  For each row, simulations are performed for the three Mach numbers $\mathcal{M} =\left(0.5,1,2\right)$ or less if specified. The total number of simulations is about $120$.}
\label{tab:Sim_list}
\begin{tabular}{l|cccccccccc}
\hline
Physics ID & $\delta$ & $n_{\stream}$ & $\mathbf{B}$-field angle & $B_0$  & $\mathcal{M}$ & $c_{\cgm}$  &  $t_{\rm{vir}}/t_0\,\left(\mathcal{M}=1\right)$ &  $T_{\stream}$ & $\xi$ &$T_{\rm{cool,max}}$ \\
& $\rho_{\stream}/\rho_{\cgm}$ & $\left[\rm cm^{-3}\right]$ & $\alpha$ & $10^{-3} \,\upmu\rm{G}$ & [-] & $\rm \left[ km\, s^{-1}\right]$  & [-] & $\left[ 10^4\, \rm K \right]$ & [-] & $\left[ 10^4\, \rm K \right]$\\
\hline
HD&$30$&$10^{-3}$&-&-&($0.5,1,2$)&$146$&$18.26$&$3.11$&$2.9$&$4.46$ \\ 
HD&$30$&$10^{-2}$&-&-&($0.5,1,2$)&$116$&$18.26$&$1.96$&$0.6 \times 10^{-1}$&$2.99$ \\ 
HD&$30$&$10^{-1}$&-&-&($0.5,1,2$)&$ 96$&$18.26$&$1.33$&$2.5 \times 10^{-3}$&$2.27$ \\ 
HD&$100$&$10^{-3}$&-&-&($0.5,1,2$)&$267$&$10.00$&$3.11$&$2.1 \times 10^{1}$&$4.79$ \\ 
HD&$100$&$10^{-2}$&-&-&($0.5,1,2$)&$212$&$10.00$&$1.96$&$0.3$&$3.02$ \\ 
HD&$100$&$10^{-1}$&-&-&($0.5,1,2$)&$175$&$10.00$&$1.33$&$1.2 \times 10^{-2}$&$2.29$ \\ 
MHD&$30$&$10^{-3}$&$90 ^\circ$&$1.0 $&($0.5,1,2$)&$146$&$18.26$&$3.11$&$2.9$&$4.46$ \\ 
MHD&$30$&$10^{-3}$&$45 ^\circ$&$1.0 $&($0.5,1,2$)&$146$&$18.26$&$3.11$&$2.9$&$4.46$ \\ 
MHD&$30$&$10^{-3}$&$ 0 ^\circ$&$1.0 $&($0.5,1,2$)&$146$&$18.26$&$3.11$&$2.9$&$4.46$ \\ 
MHD&$30$&$10^{-2}$&$90 ^\circ$&$2.6 $&($0.5,1,2$)&$116$&$18.26$&$1.96$&$0.6 \times 10^{-1}$&$2.99$ \\ 
MHD&$30$&$10^{-2}$&$45 ^\circ$&$2.6 $&($0.5,1,2$)&$116$&$18.26$&$1.96$&$0.6 \times 10^{-1}$&$2.99$ \\ 
MHD&$30$&$10^{-2}$&$ 0 ^\circ$&$2.6 $&($0.5,1,2$)&$116$&$18.26$&$1.96$&$0.6 \times 10^{-1}$&$2.99$ \\ 
MHD&$30$&$10^{-1}$&$90 ^\circ$&$6.8 $&($0.5,1,2$)&$ 96$&$18.26$&$1.33$&$2.5 \times 10^{-3}$&$2.27$ \\ 
MHD&$30$&$10^{-1}$&$45 ^\circ$&$6.8 $&($0.5,1,2$)&$ 96$&$18.26$&$1.33$&$2.5 \times 10^{-3}$&$2.27$ \\ 
MHD&$30$&$10^{-1}$&$ 0 ^\circ$&$6.8 $&($0.5,1,2$)&$ 96$&$18.26$&$1.33$&$2.5 \times 10^{-3}$&$2.27$ \\ 
MHD&$100$&$10^{-3}$&$90 ^\circ$&$1.0 $&($0.5,1,2$)&$267$&$10.00$&$3.11$&$2.1 \times 10^{1}$&$4.79$ \\ 
MHD&$100$&$10^{-3}$&$45 ^\circ$&$1.0 $&($0.5,1,2$)&$267$&$10.00$&$3.11$&$2.1 \times 10^{1}$&$4.79$ \\ 
MHD&$100$&$10^{-3}$&$ 0 ^\circ$&$1.0 $&($0.5,1,2$)&$267$&$10.00$&$3.11$&$2.1 \times 10^{1}$&$4.79$ \\ 
MHD&$100$&$10^{-2}$&$90 ^\circ$&$2.6 $&($0.5,1,2$)&$212$&$10.00$&$1.96$&$0.3$&$3.02$ \\ 
MHD&$100$&$10^{-2}$&$45 ^\circ$&$2.6 $&($0.5,1,2$)&$212$&$10.00$&$1.96$&$0.3$&$3.02$ \\ 
MHD&$100$&$10^{-2}$&$ 0 ^\circ$&$2.6 $&($0.5,1,2$)&$212$&$10.00$&$1.96$&$0.3$&$3.02$ \\ 
MHD&$100$&$10^{-1}$&$90 ^\circ$&$6.8 $&($0.5,1,2$)&$175$&$10.00$&$1.33$&$1.2 \times 10^{-2}$&$2.29$ \\ 
MHD&$100$&$10^{-1}$&$45 ^\circ$&$6.8 $&($0.5,1,2$)&$175$&$10.00$&$1.33$&$1.2 \times 10^{-2}$&$2.29$ \\ 
MHD&$100$&$10^{-1}$&$ 0 ^\circ$&$6.8 $&($0.5,1,2$)&$175$&$10.00$&$1.33$&$1.2 \times 10^{-2}$&$2.29$ \\ 
MHD+TC&$30$&$10^{-3}$&$90 ^\circ$&$1.0 $&($0.5,1,2$)&$146$&$18.26$&$3.11$&$2.9$&$4.46$ \\ 
MHD+TC&$30$&$10^{-3}$&$45 ^\circ$&$1.0 $&($0.5,1,2$)&$146$&$18.26$&$3.11$&$2.9$&$4.46$ \\ 
MHD+TC&$30$&$10^{-3}$&$ 0 ^\circ$&$1.0 $&($0.5,1,2$)&$146$&$18.26$&$3.11$&$2.9$&$4.46$ \\ 
MHD+TC&$30$&$10^{-2}$&$90 ^\circ$&$2.6 $&($0.5,1,2$)&$116$&$18.26$&$1.96$&$0.6 \times 10^{-1}$&$2.99$ \\ 
MHD+TC&$30$&$10^{-2}$&$45 ^\circ$&$2.6 $&($0.5,1,2$)&$116$&$18.26$&$1.96$&$0.6 \times 10^{-1}$&$2.99$ \\ 
MHD+TC&$30$&$10^{-2}$&$ 0 ^\circ$&$2.6 $&($0.5,1,2$)&$116$&$18.26$&$1.96$&$0.6 \times 10^{-1}$&$2.99$ \\ 
MHD+TC&$30$&$10^{-1}$&$90 ^\circ$&$6.8 $&($0.5,1,2$)&$ 96$&$18.26$&$1.33$&$2.5 \times 10^{-3}$&$2.27$ \\ 
MHD+TC&$30$&$10^{-1}$&$45 ^\circ$&$6.8 $&($0.5,1,2$)&$ 96$&$18.26$&$1.33$&$2.5 \times 10^{-3}$&$2.27$ \\ 
MHD+TC&$30$&$10^{-1}$&$ 0 ^\circ$&$6.8 $&($0.5,1,2$)&$ 96$&$18.26$&$1.33$&$2.5 \times 10^{-3}$&$2.27$ \\ 
MHD+TC&$100$&$10^{-3}$&$90 ^\circ$&$1.0 $&($0.5,1,2$)&$267$&$10.00$&$3.11$&$2.1 \times 10^{1}$&$4.79$ \\ 
MHD+TC&$100$&$10^{-3}$&$45 ^\circ$&$1.0 $&($0.5,1$)&$267$&$10.00$&$3.11$&$1.6 \times 10^{1}$&$4.79$ \\ 
MHD+TC&$100$&$10^{-2}$&$90 ^\circ$&$2.6 $&($0.5,1,2$)&$212$&$10.00$&$1.96$&$0.3$&$3.02$ \\ 
MHD+TC&$100$&$10^{-2}$&$45 ^\circ$&$2.6 $&($0.5,1,2$)&$212$&$10.00$&$1.96$&$0.3$&$3.02$ \\ 
MHD+TC&$100$&$10^{-2}$&$ 0 ^\circ$&$2.6 $&($0.5,1,2$)&$212$&$10.00$&$1.96$&$0.3$&$3.02$ \\ 
MHD+TC&$100$&$10^{-1}$&$90 ^\circ$&$6.8 $&($0.5,1,2$)&$175$&$10.00$&$1.33$&$1.2 \times 10^{-2}$&$2.29$ \\ 
MHD+TC&$100$&$10^{-1}$&$45 ^\circ$&$6.8 $&($0.5,1,2$)&$175$&$10.00$&$1.33$&$1.2 \times 10^{-2}$&$2.29$ \\ 
MHD+TC&$100$&$10^{-1}$&$ 0 ^\circ$&$6.8 $&($0.5,1,2$)&$175$&$10.00$&$1.33$&$1.2 \times 10^{-2}$&$2.29$ \\ 
\hline
\end{tabular}
\end{table*}

\subsection{Initial and boundary conditions}\label{sec3:ICs}

\begin{figure}
    \includegraphics[width=1.0\columnwidth]{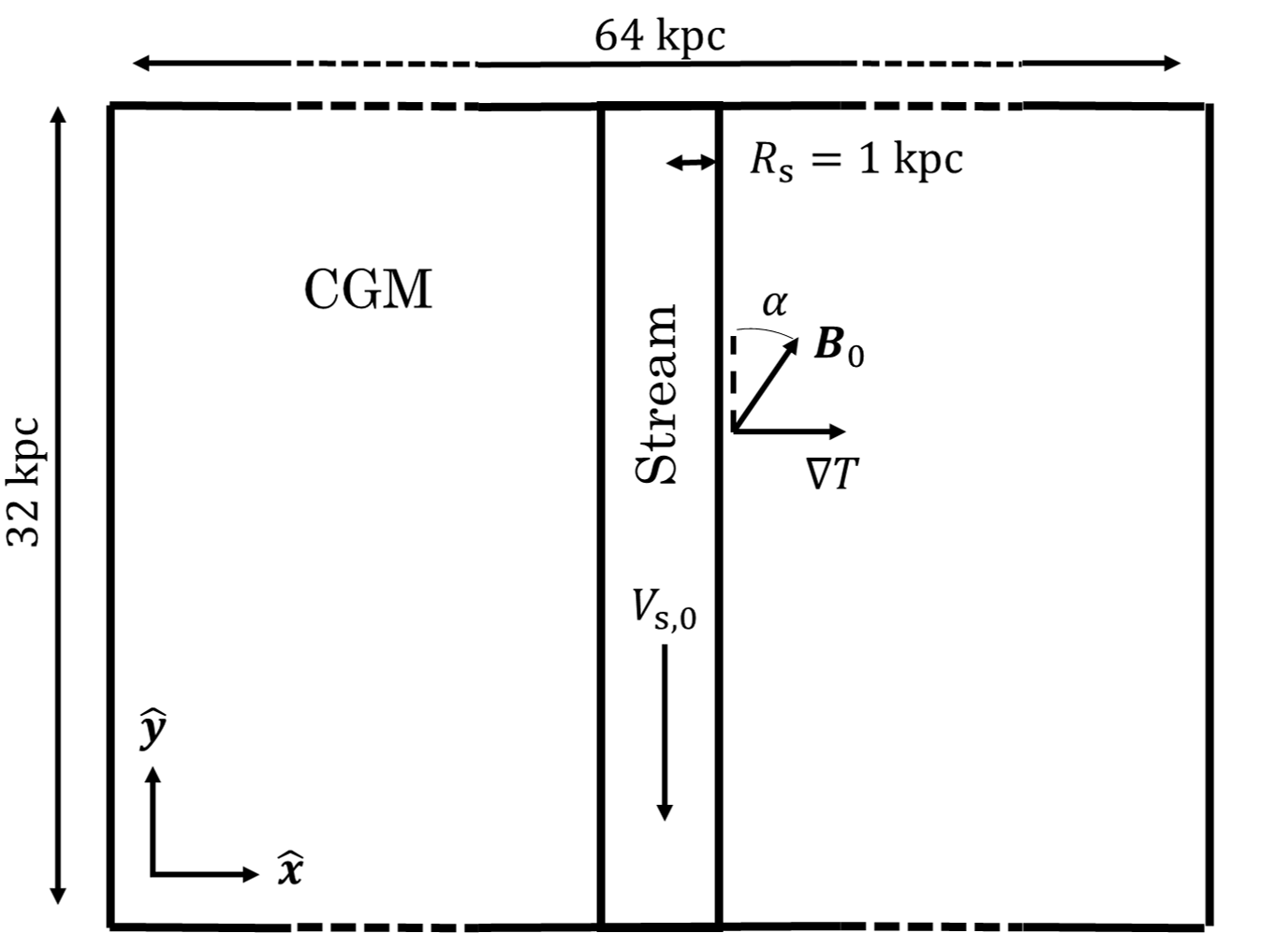}
    \caption{Illustration of the initial conditions of our simulations. The stream region is dense and cold, while the CGM is diffuse and hot. The angle $\alpha$ is the initial angle of the magnetic field with respect to the stream. The stream is moving downward inside the fixed CGM background. The direction of the initial temperature gradient is also shown.}\label{fig:ICs}
\end{figure}

Our initial conditions are similar to those used in the simulations from \citet{Mandelker2020a} and are summarised here. Fig.~\ref{fig:ICs} shows the initial setup.

The stream is initialized at the centre of a two-dimensional rectangular domain of size $64R_{\stream} \times 32 R_{\stream}$, the stream axis being the $y$-axis and with $R_{\stream}=1\, \kpc$. Boundary conditions are set as periodic in the stream parallel direction and as fixed CGM fluid values along the perpendicular directions. Compared to previous works \citep{Mandelker2019,bib_MHD_Berlok_2019b,Mandelker2020a}, we extended the domain transversal to the stream by a factor of two to avoid boundary effects on the stream due to thermal conduction and the fixed boundary conditions.
About the resolution, we use static mesh refinement with the highest resolution of $\Delta x = \Delta y = R_{\stream} / 64$ defined in the near-stream region ($<2R_{\stream}$). The grid size is then doubled up every $2R_{\stream}$ along the $x$-axis up to a maximum cell size of $R_{\stream}$.

The stream is defined in the density field by
\begin{equation} \label{eq:init_1}
\frac{\rho(x)}{\rho_{\stream,0}} = \frac{\rho_{\cgm}}{\rho_{\stream,0}} + \frac{1}{2}\left(1 - \frac{\rho_{\cgm}}{\rho_{\stream,0}}\right)  \left(1-\tanh\left(\frac{|x|-R_{\stream}}{h}\right)\right),
\end{equation}
where $h$ determines the smoothness of the transition\footnote{As discussed in \citet{Mandelker2020a}, this smoothness layer is not needed for HD simulations with cooling as the layer would shrink by condensation. However, in the presence of magnetic field and thermal conduction, we found that in the case of the Stream Diffusion regime, a step transition can lead to the divergence of the simulation. We hence kept the relatively sharp but non-zero smoothing layer from \citet{Mandelker2019}.} between the stream and the CGM gas, with $h=R_{\stream} / 32$.
In a similar way, the stream velocity is defined as
\begin{equation} \label{eq:init_3}
v(x) = - \frac{v_{s,0}}{2}\left(1.0 - \tanh\left(\frac{|x|-R_{\stream}}{h}\right)\right),
\end{equation}
where $v_{s,0} = \mathcal{M} c_{\cgm}$, 
and $\mathcal{M}$ is the Mach number of the stream based on the sound speed in the CGM. 
We considered values of $\mathcal{M} \in [0.5,1.0,2.0]$. 
To initialize the KHI, the transverse velocity field is initially seeded by perturbations,
\begin{equation} \label{eq:init_2}
u\left(x,y\right) = u_0\sum^{N_{\rm{k}}}_j \left[\cos{\left(k_j y\right)}\right]  \exp \left ( -\frac{(|x|-R_{\stream})^2}{8h^2} \right ),
\end{equation}
where $N_{\rm k}$ is the total number of perturbations defined by $k_j = 2\pi / \lambda_j$ with $\lambda_j \in [0.5R_{\stream},16R_{\stream}]$, 
and $u_0=0.01 c_{\stream}$ is the amplitude\footnote{To check the impact of the chosen value of $u_0$, we ran MHD simulations (with $\alpha=45^\circ$, $\mathcal{M}=1$, $n_{\stream}=0.01$) with $u_0$ $10$ times and $50$ times stronger, i.e., $u_0=0.1c_{\stream}$ and $u_0=0.5c_{\stream}$, respectively. The simulation with $u_0=0.1c_{\stream}$ does not show a significant difference compared to the fiducial one. The simulation with $u_0=0.5c_{\stream}$ exhibits stronger mixing between the stream and the CGM leading to higher cold stream mass growth.} of the perturbed modes.

The magnetic field is defined by an initial angle $\alpha$,
\begin{equation} \label{eq:init_4}
\mathbfit{B} = \left(B_0 \sin{\alpha}, B_0 \cos{\alpha} , 0\right)
\end{equation}
such as $\alpha=0^\circ$ corresponds to the case where the magnetic field is parallel to the stream axis $\hat{y}$ (anti-parallel to the flow) and perpendicular to the temperature gradient, and $\alpha=90^\circ$ corresponds to the case where the magnetic field is transverse to the stream (and aligned with the temperature gradient).
As the magnetic field is constant over the entire domain, hydrostatic equilibrium gives us a constant pressure $P_0$ both in the CGM and in the stream.

Table~\ref{tab:Sim_list} lists all simulations and their parameters. For each row of parameters, unless specified, three simulations with $\mathcal{M}=0.5,1$ and $2$ are run, leading to a total of about 120 simulations.
The resulting stream velocities span over $48\text{--}534\, \rm km s^{-1}$ in good agreement with observations.
Simulations are run for a total time of $t_{\rm{end}} = 22 t_0 \sim 0.8\text{--}1.2 \, \rm Gyr$.

\section{Results}\label{sec4}
We first describe the general evolution for HD, MHD and MHD+TC simulations, as well as the cooling emission signature and mass of the stream (Sec.~\ref{sec4:stream_evo}). Then, we focus on the impact of the magnetic field and thermal conduction (Sec.~\ref{sec4:Impact}, the magnetic field evolution (Sec.~\ref{sec4:mag}), and the turbulence in the mixing layer (Sec.~\ref{sec4:kin}).

\subsection{Stream evolution and cooling signature}\label{sec4:stream_evo}

\subsubsection{General evolution of HD and MHD cases without TC}\label{sec4:HH_MHD_evo}

\begin{figure*}
    \includegraphics[width=2.0\columnwidth]{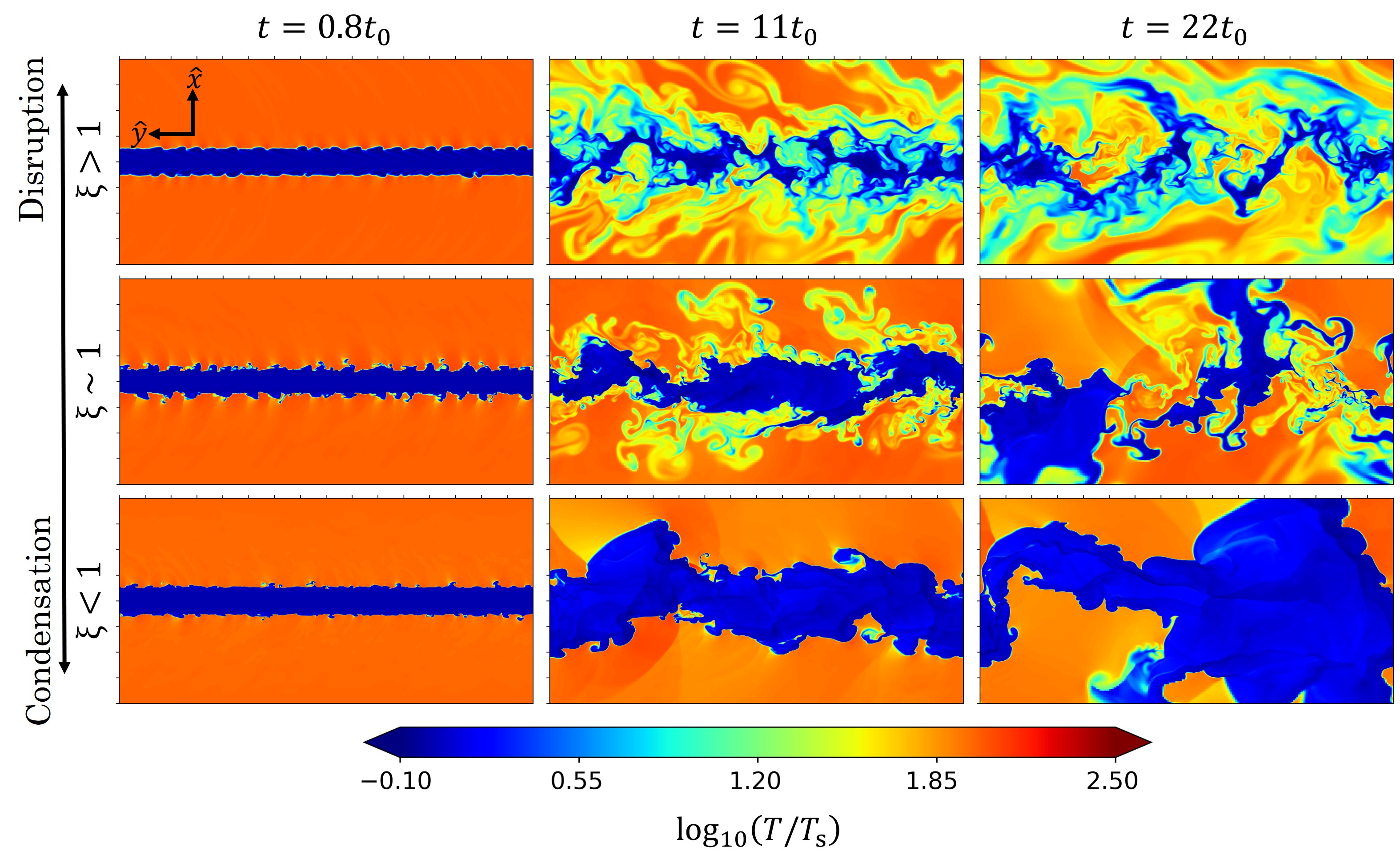}
    \caption{Maps of the gas temperature in the HD simulations at different times for the sonic case ($\mathcal{M}=1$), with a density contrast $\delta=100$, and for three different values of the $\xi$ ratio (cooling time over shearing time). Each panel shows the 32\,kpc full stream axis length in the horizontal direction and a zoomed region of 16\,kpc in the vertical direction. The contours displayed are rotated by $90^\circ$ compare to the illustration in Fig.~\ref{fig:ICs}.}
    \label{fig:Overall_1}
\end{figure*}

We present a brief showcase of the general evolution of the stream based on our HD and MHD simulations. 
In Fig.~\ref{fig:Overall_1}, we show temperature maps for HD simulations with different values of the ratio $\xi=t_{\rm{cool}}/t_{\rm{sh}}$, at the early, median, and final time, namely $t=0.8, 11, 22 t_0$, respectively. The displayed maps are rotated by $90^\circ$ compared to the illustration in Fig.~\ref{fig:ICs}. The fate of the stream is determined by the value of $\xi$. As described in Sec.~\ref{sec2:timescale_comp}, when $\xi<1$, ({\it condensing stream} regime), the CGM gas condenses onto the stream, resulting in the growth of the cold-stream mass.  For $\xi>1$ (the {\it disrupting stream} regime), an increasing amount of initially cold stream gas mixes into the CGM, leading to stream disruption. Such stream evolution is in good agreement with the findings presented by \citet{Mandelker2020a}. The simulation with $\xi\sim 1$ represents the limit where the cooling is strong enough to sustain the cold mass but not sufficient to cool the gas in the mixing layer efficiently, leading to a relatively thick mixing layer. At such a limit, the stream also fragments into cold and rather large clouds ($>R_{\stream}$) at a later time.

\begin{figure*}
	\includegraphics[width=2.0\columnwidth]{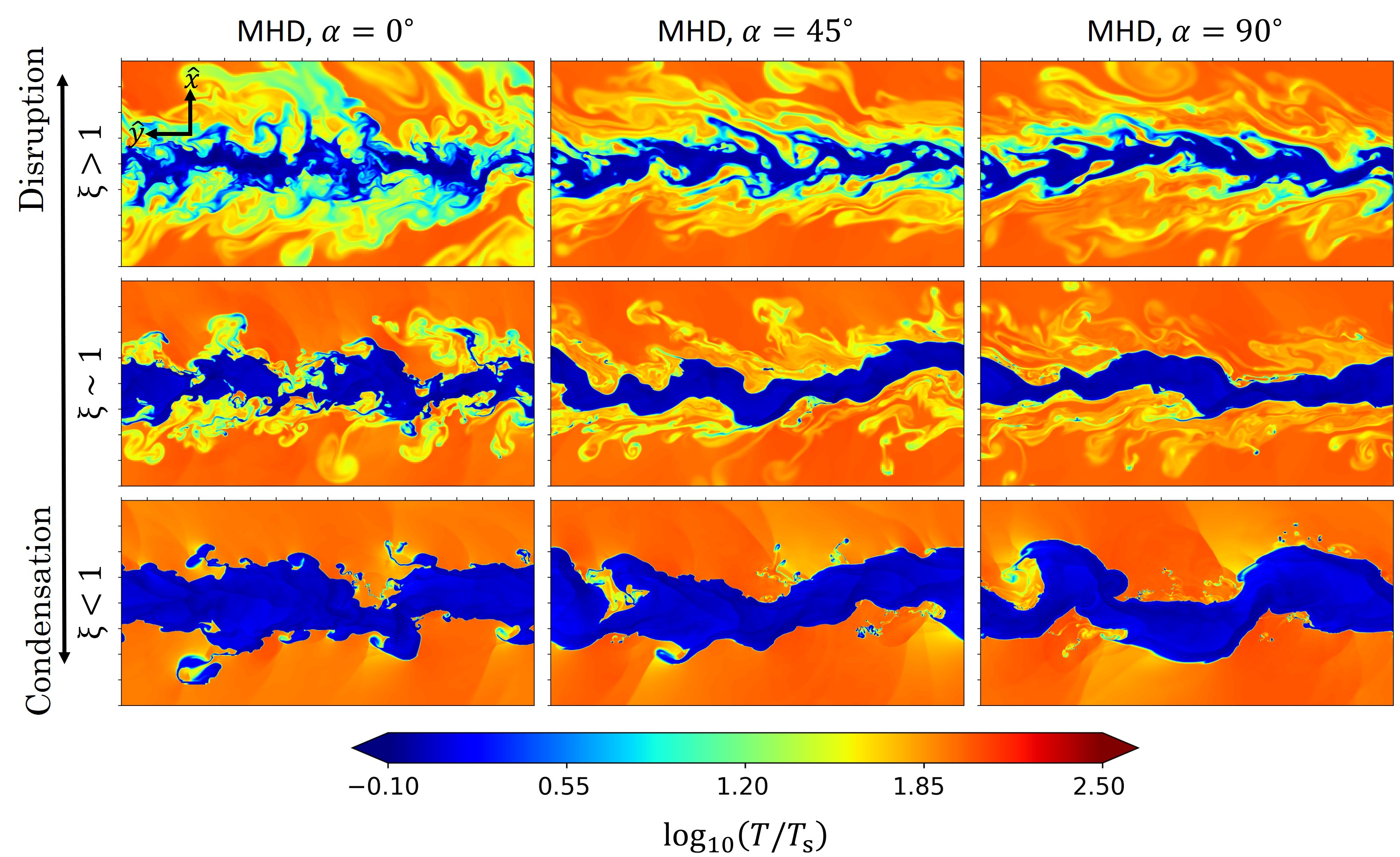}
    \caption{Maps of gas temperature in the MHD simulations with $\alpha=0^\circ , \, 45^\circ, \, 90^\circ$ for the sonic case ($\mathcal{M}=1$), with a density contrast $\delta=100$ and at a fixed time $t=11\, t_0$. The top, middle, and bottom rows correspond to the three different regimes of $\xi$. Each panel shows the 32\,kpc full stream axis length in the horizontal direction and a zoomed region of 16\,kpc in the vertical direction.}
    \label{fig:MHD_1}
\end{figure*}

Fig.~\ref{fig:MHD_1} illustrates the temperature maps of the MHD simulations at half-time $t =11t_0$ for $\mathcal{M}=1$ and $\delta=\rho_{\stream} / \rho_{\cgm} = 100$, considering different values of $\xi=t_{\rm{cool}}/t_{\rm{sh}}$ and various initial magnetic field angles. Comparing these with the middle column of Fig.~\ref{fig:Overall_1}, we see that the magnetic field significantly impacts the stream evolution only for an initial magnetic field not parallel to the stream ($\alpha \neq 0^\circ$) and when the stream is not in the {\it condensing stream} regime ($\xi \gtrsim1$).
The main visual difference is the decrease in the amount of gas in the mixing layer for $\alpha \neq 0^\circ$, particularly in the {\it disrupting stream} regime ($\xi>1$).
The magnetic field strength is initially insignificant ($\beta=10^5$), highlighting the need for a drastic increase of the field strength to impact the stream evolution, as one shall see in Sec.~\ref{sec4:mag}.

\subsubsection{General evolution of MHD cases with TC}\label{sec4:MHDTC_evo}

\begin{figure*}
	\includegraphics[width=2.0\columnwidth]{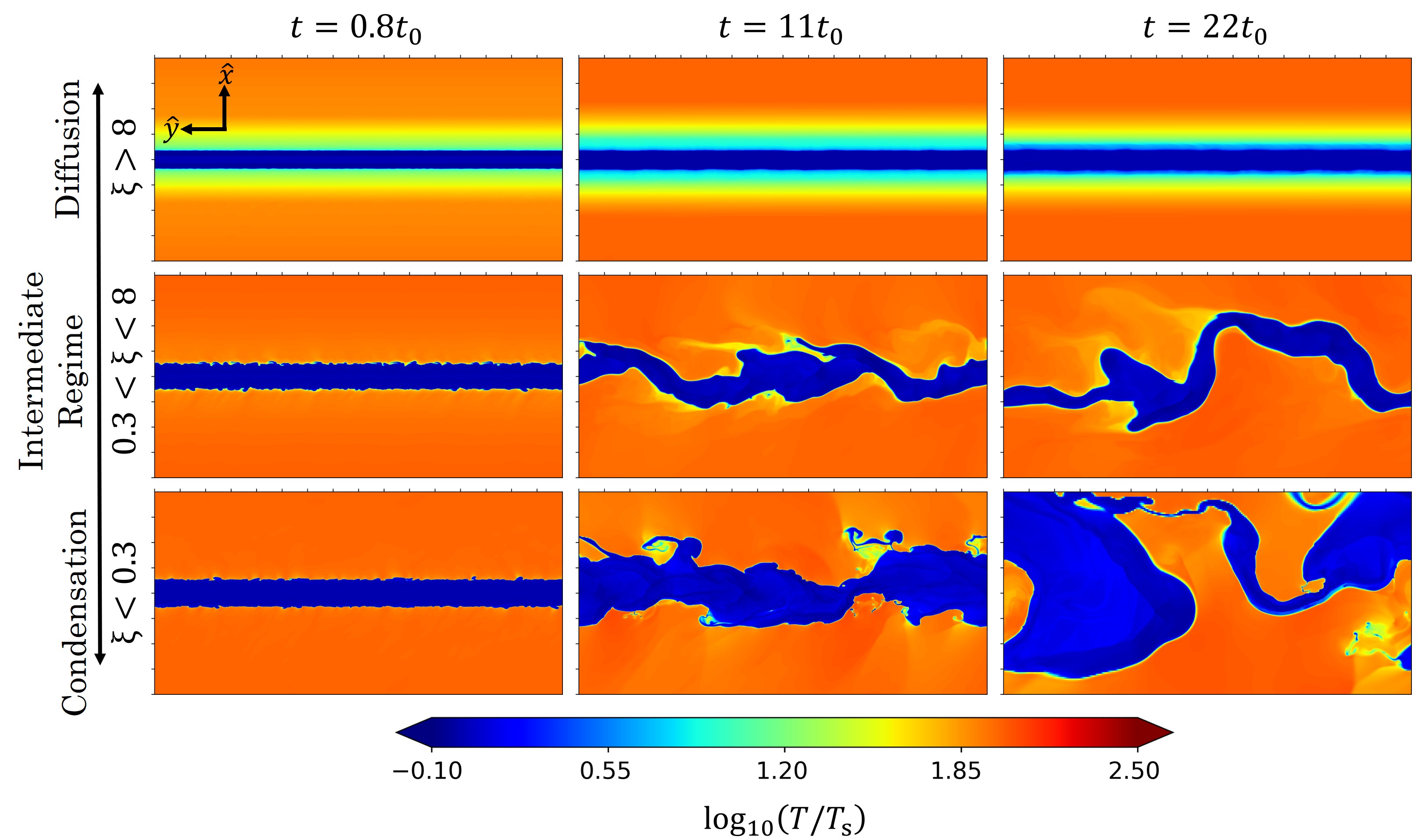}
    \caption{Maps of gas temperature in MHD+TC simulations at different times for the sonic case ($\mathcal{M}=1$) with an initial magnetic field angle $\alpha=45^\circ$, a density contrast $\delta=100$, and three different values of $\xi$ ratio (cooling time over shearing time) corresponding to the three different stream regimes in the presence of TC, i.e. {\it diffusing stream}, {\it intermediate}, {\it condensing stream} regimes (see Sec.~\ref{sec2:timescale_comp}). Each panel shows the 32\,kpc full stream axis length in the horizontal direction and a zoomed region of 16\,kpc in the vertical direction.}
    \label{fig:TC_1}
\end{figure*}
Fig.~\ref{fig:TC_1} presents the temperature maps for MHD+TC simulations with varying ratios of $\xi=t_{\rm{cool}}/t_{\rm{sh}}$, considering an initial magnetic field angle of $\alpha=45^\circ$, at early, median and final time, of $t=0.8, 11, 22 t_0$, respectively.
In the {\it diffusing stream} regime ($\xi >8$), as predicted, TC effectively hinders the growth of instabilities and diffuses the stream into the CGM. The rate of diffusion depends on the magnetic field angle and can be explicitly expressed as an efficiency parameter for the thermal conduction flux $\mathbf{Q}$. As the simulation progresses, the magnetic field lines bend at the interface between the stream and the CGM, diminishing the efficiency of the conduction. Consequently, at $t=22t_0$, the thermal conduction efficiency drops to a point where it no longer overcomes the cooling, even leading to a small cold gas mass increase at later time\footnote{Such increase is also visible in the time profile of the cold stream mass in Appendix~\ref{app:Res}}. As discussed later, the diffusion efficiency of the stream depends on $\alpha$ and on the stream velocity.
In the {\it intermediate} regime ($0.3<\xi <8$), TC diffuses the mixing layer and erases any small-scale perturbations. This is consistent with the fastest time-scale $\tau_{\rm{p}}$ in the {\it intermediate} regime which defines the diffusion of a small cold clump with a size of $0.1R_{\stream}$. Hence, the stream can be stabilized against small-scale surface modes, also meaning that the mixing layer and any small structures or perturbations diffuse in the CGM.
In the {\it condensing stream} regime ($\xi <0.3$), there are no substantial differences between HD, MHD and MHD+TC simulations. Notably, at the later stages of the MHD+TC simulations, the stream starts to fragment into large cold clumps. This evolution is also seen in the MHD simulations and originates from the magnetic field tension force which inhibits the smaller-scale perturbations, i.e., the short wavelength KHI modes \citep{bib_MHD_Berlok_2019b}. As a result, as time progresses, the longer wavelength KHI modes grow sufficiently to induce stream fragmentation.

\subsubsection{Cooling signature and stream mass}\label{sec4:MHDTC_impact}
To assess the cooling occurring in the mixing layer, we compute the net cooling emission of the gas in each cell below a temperature threshold $T_{\rm{u}} = \left(T_{\rm{mix}} + T_{\cgm}\right)/2$. This threshold targets only the emission in the mixing layer\footnote{We discuss and analyze these thresholds in Appendix~\ref{app:Mass_thr}.}. The gas in the stream, being in equilibrium between heating from the UV background and cooling, produces negligible net cooling, and therefore we do not define any lower bound. We found that a small variation of the threshold does not affect our results.
Note that we exclusively consider net cooling. Including cooling induced by photonionization from the UV background might result in an overestimation of the total cooling, as our simulations do not account for self-shielding.
The cooling rate is integrated over all gas under $T_{\rm{u}}$ and subsequently averaged over time, starting at the point where the mixing layer reaches the quasi-steady state: \footnote{in practice, a fixed time of $t=5t_0$ is used for all simulations. As depicted in the time profile plot in Appendix~\ref{app:Res} and Appendix~\ref{app:Mag_Mod}, the mixing layer in all simulations reaches a quasi-steady state (Equation \ref{eq:steadystate}), i.e., a roughly constant stream mass growth/loss, net cooling emissions, and magnetic field growth.}
\begin{equation}\label{eq:Lnet}
    L_{\rm{net}} = \left\langle \pi R_{\stream} \iint_{\rm{ML}} \rho^2 \left(\Lambda-\mathcal{H}\right)\, \rm{d} x\rm{d} y \right\rangle_{\rm{log_{10}}},
\end{equation}
where the factor $\pi R_{\stream}$ represents integrating around the stream axis to mimic a cylindrical stream. The stream radius is $R_{\stream}=1 \, \kpc$ in all our simulations. In practice, the integral over the mixing layer of an arbitrary variable $X$ on the domain ML is done such that $\iint_{\rm{ML}} X \mathrm{d}x\mathrm{d}y = \sum \Delta x\Delta y X\times \mathrm{bool}_i$, where the sum is done over all cells and where $\mathrm{bool}_i=1$ for a cell temperature $T_{\rm i}\in [T_{\rm{d}},T_{\rm{u}}]$ and $0$ if $T_{\rm i}\notin [T_{\rm{d}},T_{\rm{u}}]$. The brackets define the time log-average\footnote{In a few simulations with strong cooling ($\xi<10^{-2}$), the stream grows in size, and the mixing layer is shifted to regions with lower resolutions, leading to excessive cooling due to the lack of resolution \cite[see][]{Fielding2020}. The log-average allows us to smooth out the effect of those peaks, without changing results for all other simulations.},
\begin{equation}\label{eq:log_ave}
\left\langle X \right\rangle_{\rm{log_{10}}} = 10^{\sum \log_{10}\left(X_i\right) / N}
\end{equation}
 with $N$ the total number of averaged variables $X_i$.

\begin{figure}
	\includegraphics[width=\columnwidth]{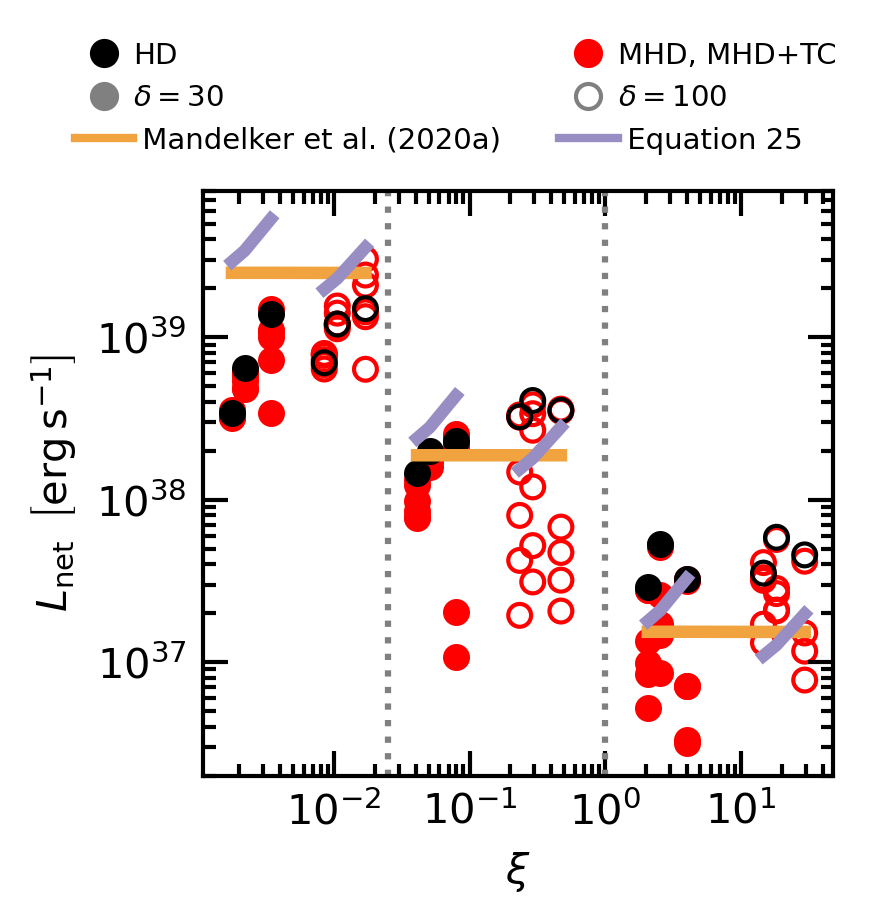}
    \caption{Average cooling emission in the mixing layer for all simulations. Red points are the MHD and MHD+TC simulations, and black points are the HD simulations. Analytical models using our fitted value of the condensation speed, $v_{\rm in}$, and models from \citet{Mandelker2020a} are shown by the purple and orange lines, respectively. {\it Detailed distribution of the points:} Filled and open circles represent simulations with $\delta=30$ and $100$, respectively. From left to right, the vertical grey dotted lines delimit simulations with $n_{\stream} = 10^{-1}, 10^{-2}, 10^{-3} \, \rm cm^{-3}$. Finally, for a subgroup with a given $\delta$ and $n_{\stream}$, points are differentiated from left to right by their Mach number $\mathcal{M}=0.5, 1, 2$.}
    \label{fig:CoolTot}
\end{figure}
The resulting net cooling emission is presented in Fig.~\ref{fig:CoolTot} as a function of $\xi$ for all simulations. The HD simulations generally exhibit the highest values of $L_{\rm net}$ compared to the MHD and MHD+TC simulations for almost all $\xi$ values.  The net cooling emissions in the HD simulations are in relative agreement but lower compared to those reported by \citet[][Fig.~$13$]{Mandelker2020a} from their three-dimensional simulations. The majority of their simulations correspond to our $n_{\stream}=10^{-2}\, \rm cm^{-3}$ cases, but with a larger radius of $R_{\stream}=3\, \kpc$. Our $L_{\rm{net}}$ values are relatively close to theirs after considering a factor of approximately $10$ multiplication, which accounts not only for equation~\ref{eq:Lnet} but also for the boost in condensation due to the larger $R_{\stream}$ resulting in lower a $\xi$. For $\xi \lesssim 0.05$, the HD simulations exhibit at most a factor of $5$ difference with MHD and MHD+TC simulations. For $\xi \gtrsim 0.05$, the discrepancies rise up to a factor of $\sim 20$, indicating a significant decrease of the stream emission signature when the magnetic fields and thermal conduction are considered. We shall see that this difference comes from both the magnetic fields and thermal conduction.
For comparison with analytical models, both the model from \citet[equation 31 and 37]{Mandelker2020a} and our fitted model are plotted. Similarly to recent simulations of individual planar mixing layers \citep{Fielding2020,Tan2021,Yang2023}, we model the cooling emission assuming that the mixing layer has reached a quasi-steady state in terms of its energetics. Physically, this quasi-steady state is the balance between the flux of kinetic energy and enthalpy and the net radiative cooling rate.
In the hydrodynamic case, the quasi-steady state can be described as,
\begin{equation}\label{eq:steadystate}
    \nabla\cdot\left[\left(e+p\right)\mathbfit{u}\right] = \dot{e}_{\rm cool} ,
\end{equation}
which, upon integration over the mixing layer, yields,
\begin{equation}\label{eq:Lcool}
     L_{\rm{cool}} =v_{\rm{in}}\left(\frac{\gamma}{\gamma-1}p+ \frac{\gamma}{2}p\mathcal{M}^2\right)S,
\end{equation}
with $S$ referring to the surface between the stream and the background, and $L_{\rm cool}$ the cooling emission in the mixing layer. To obtain $v_{\rm in}$, we first compute the stream mass growth rate $\dot{M}_{\stream}$ from the simulations (later shown in equation~\ref{eq:Mdot}), and then considering Equation~\ref{eq:Mdot_th}, we compute the simulation's $v_{\rm in,sim}$ taking the surface of a cylindrical stream with $S = 32R_{\stream}\times2\pi R_{\stream} \, \rm{kpc}^2$. The mixing velocity $v_{\rm in}$ is then directly fitted from the simulations values $v_{\rm in,sim}$ obtained using Equation \ref{eq:Mdot_th}. From the fit we obtain,
\begin{equation}\label{eq:vinth}
     v_{\rm{in}} \sim 0.4\xi^{-1/4}.
\end{equation}
Both models can roughly reproduce the expected emissions of the HD simulations, although they exhibit discrepancies. Our model shows a mean factor difference of  $\sim 1.5$ and a maximum factor difference of $\sim 7.3$. In contrast, the models presented by \citet[][equations 31 and 37]{Mandelker2020a} appear to exhibit a flat trend. This difference can be attributed to their utilisation of different definitions for the relevant timescales\footnote{\citet{Mandelker2020a} uses in their model the ratio of the minimum cooling time over the stream sound-crossing time. Such cooling time scale removes the dependency on $\delta$, while the sound-crossing time is independent of $\mathcal{M}$, leading to the flat trend.}. Furthermore, their model assumes that the cooling emission at late times is primarily a result of thermal energy loss by the hot CGM gas. In contrast, in our cases, we consider the energetic quasi-steady state within the mixing layer, linking the cooling emission to the mixing layer's gas enthalpy.

\begin{figure}
	\includegraphics[width=\columnwidth]{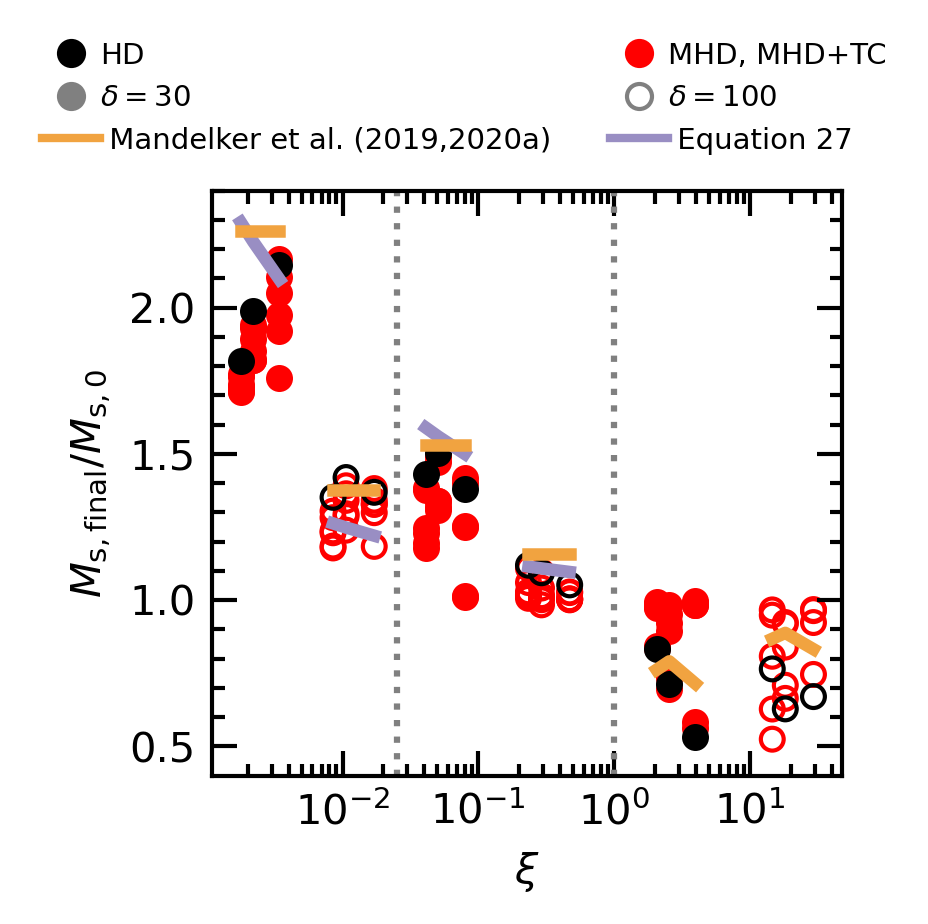}
    \caption{Final stream mass for all simulations. Red points are the MHD and MHD+TC simulations, and black points are the HD simulations. Analytical models using our fitting value of the condensation speed, $v_{\rm in}$, and models from \citet{Mandelker2019,Mandelker2020a} are shown by the purple and orange lines, respectively. The orange lines shows models from \citet{Mandelker2019} for $\xi>1$ and the one for \citet{Mandelker2020a} for $\xi<1$. The detailed distribution (grey dashed lines, filled and open circles) of the points is the same as in Fig.~\ref{fig:CoolTot}. }
    \label{fig:Mfinal_1}
\end{figure}
To investigate the stream mass evolution, we calculate the total cold mass in each simulation. The cold stream mass is defined by the gas below a temperature threshold $T_{\rm{d}} = \left(T_{\stream}T_{\rm{mix}}\right)^{1/2}$ \footnote{The impact of varying this threshold temperature on the cold mass is found only to affect our results by a few percent.}. The final mass $M_{\stream,\rm{final}}$ is obtained through integration over all domains. 
Fig.~\ref{fig:Mfinal_1} displays the final stream mass normalized by the initial value for all simulations. Once again, the HD values represent the maximum case of mass growth and loss among simulations with $\xi <1$ and $\xi >1$, respectively. Differences between HD and MHD/MHD+TC simulations are less pronounced than for the cooling emissions, showing a maximum difference of $\sim 0.4\, M_{\stream,0}$. Note that, for simulations with $\xi >1$, i.e., the {\it disrupting stream} regime for HD and MHD simulations, some simulations have a final stream mass very close to their initial value, indicating that MHD and MHD+TC in some cases can help the stream to stabilize against KHI.

We compare our numerical results to analytical models in Fig.~\ref{fig:Mfinal_1}. Thanks to our fitted value of $v_{\rm{in}}$ from equation~\ref{eq:vinth}, we recover the steady mass growth $\dot{M}_{\stream}$ from Equation \ref{eq:Mdot_th}. The theoretical final stream mass is then obtained as,
\begin{equation}
    \frac{M_{\stream,\rm{th}}\left(t\right)}{M_{\stream,0}} =  1 + \frac{\dot{M}_{\stream}}{M_{\stream,0}} \frac{t}{t_0},
\end{equation}
with $t=22t_0$ as a final simulation time. The predicted final mass from our model and the one from \citet{Mandelker2019}, for $\xi>1$, and \citet{Mandelker2020a}, for $\xi <1$, align relatively well with the results for the HD cases. For simulations $\xi>1$, the theoretical prediction does not hold as it assumes condensation of CGM gas on to the stream. Instead, we showcase the expected mass loss rate based on the deceleration of the stream $V_{\stream}\left(t\right)$ from \citet[equation 10 and 38]{Mandelker2019} for two dimensions. Quantitatively, the predicted mass loss rate also roughly agrees with the simulated values.

\subsection{Impacts of magnetic fields and thermal conduction}

\subsubsection{Impacts on the cooling emission and stream mass}\label{sec4:Impact}
\begin{figure*}
	\includegraphics[width=1.66\columnwidth]{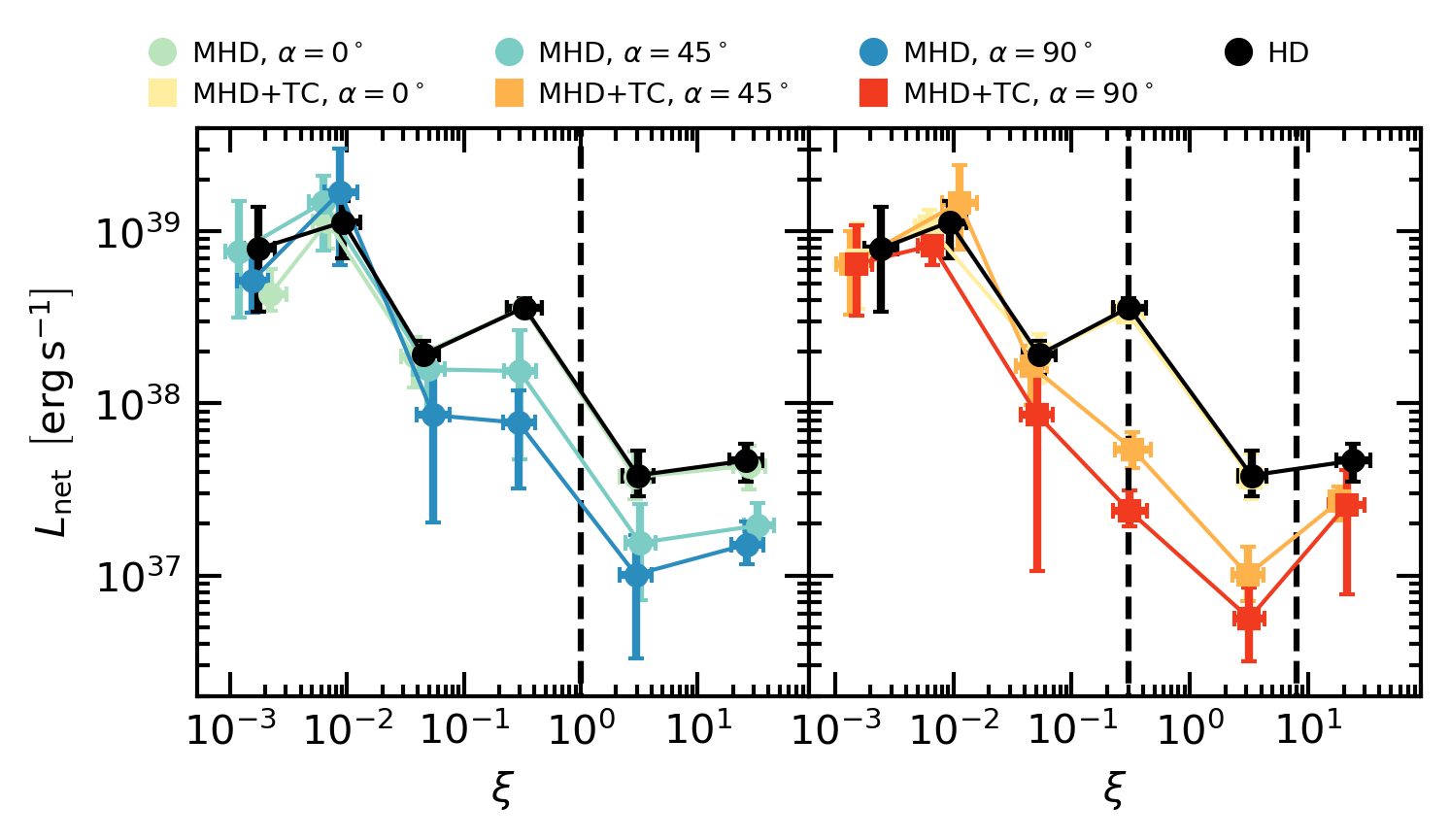}
    \caption{Net cooling emission in the mixing layer. {\it Left panel:} Comparison between HD and MHD simulations. The vertical black dashed line represents the boundary between the {\it condensing stream} regime ($\xi<1$) and the {\it disrupting stream} regime ($\xi>1$). {\it Right panel:} Comparison between HD and MHD+TC simulations. The dashed black lines indicate the limit between the {\it condensing stream} regime ($\xi<0.3$), the {\it intermediate} regime ($0.3<\xi<8$), and the {\it diffusing stream} regime ($\xi>8$). Each point shows an averaged value over the three Mach numbers $\mathcal{M}=0.5,1,2$, with the error bars representing the range of minimum and maximum values. A small random offset in the x coordinate is included for clarity.}
    \label{fig:Lnet_comp}
\end{figure*}
The net cooling emissions, $L_{\rm net}$, for our simulations are plotted in Fig.~\ref{fig:Lnet_comp}.
To reduce the number of points displayed for clarity, each point represents an averaged value over the three Mach number $\mathcal{M}=0.5,1,2$ for each set of simulations, i.e., each row of Table~\ref{tab:Sim_list}. The scatter represents the maximum and minimum values across the simulations with the three Mach numbers. A small random offset is added in the abscissa for clarity.

MHD starts to impact the cooling emission $L_{\rm{net}}$ at $\xi\sim 0.1$. Above this value, the emission gradually decreases for increasing $\alpha$, which can be directly linked to the reduction of the amount of gas in the mixing layer seen in Fig.~\ref{fig:MHD_1}. Across all $\xi$ values, the MHD simulations with $\alpha=0^\circ$ show identical $L_{\rm{net}}$ compared to the HD ones, while the decrease reaches a factor $10$ difference for MHD simulations with $\alpha=90^\circ$.

Similarly to the MHD simulations, MHD+TC also starts to impact the emission $L_{\rm{net}}$ at $\xi\sim 0.1$, causing a further decrease. In the {\it condensing stream} regime ($\xi<0.3$), thermal conduction does not significantly affect the emissions, except when $\alpha = 90^\circ$ and $\xi\sim 0.06$. In this case, the cooling emission is reduced by a factor of $20$ compared to the HD case, but, as indicated by the scatter, such a decrease only occurs for a specific Mach number ($\mathcal{M}=2$). The scatter can be attributed to the influence of the stream velocity on the magnetic field growth, as one shall see in the next section. 

In the {\it intermediate} Regime ($0.3<\xi<8$), thermal conduction further reduces the cooling emissions of MHD+TC by up to a factor of $\gtrsim 2$ compared to MHD case, for simulations with $\alpha \neq 0^\circ$. As can be seen in the temperature maps in Fig.~\ref{fig:TC_1}, and as expected from the definition of the {\it intermediate} regime in Sec.~\ref{sec2:timescale_comp}, thermal conduction diffuses the mixing layer leading to a smaller amount of gas in the temperature range $T_{\stream}\text{--}T_{\rm mix}$ that can efficiently cool and radiate. Notably, in this regime, thermal conduction also reduces the dependence of $L_{\rm net}$ on $\mathcal{M}$ compared to the MHD cases.

In the {\it diffusing stream} regime ($\xi \gtrsim 8$), the stream exhibits a slight enhancement of its emission for $\alpha \neq 0^\circ$  compared to the MHD+TC simulations at $\xi\sim 3$. As the stream is diffusing, a large amount of gas is heated above the radiative temperature equilibrium in the stream, $T_{\stream}\sim 10^4\, \rm K$, resulting in a higher amount of gas that can radiate. As shown in Appendix \ref{app:Lnet_nature}, in the {\it condensing stream} and {\it intermediate} regimes, the dominant cooling processes come from hydrogen.  However, in the case of the {\it diffusing stream} regime, the emission from hydrogen actually decreases and represents $\sim 4\text{--}10\%$ of the total cooling emission. The increase in the total cooling emission is due to Helium and numerous metals such as Oxygen and Neon.
\begin{figure*}
	\includegraphics[width=1.66\columnwidth]{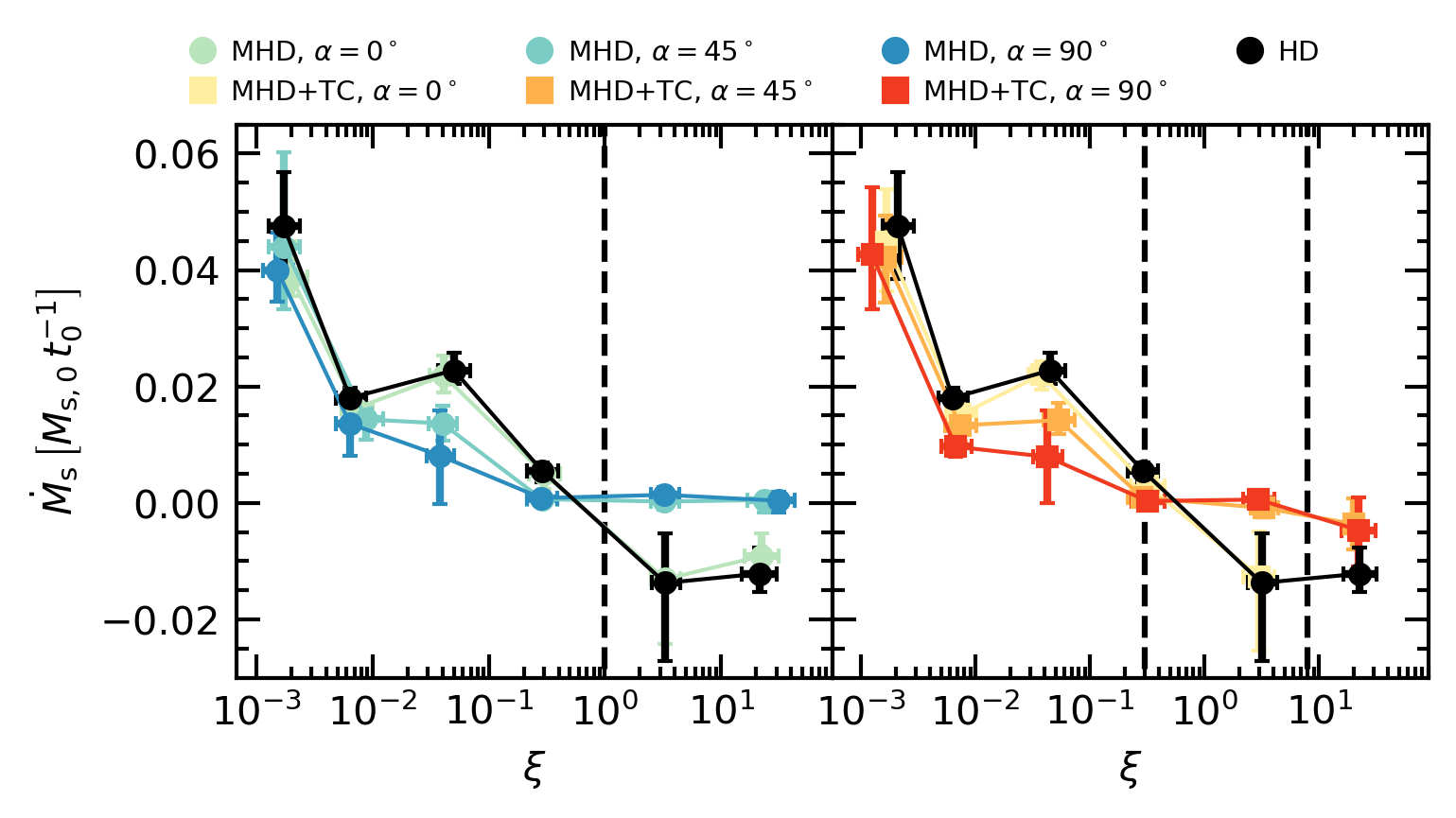}
    \caption{Stream mass rate in units of stream initial mass per $t_0=R_{\stream}/c_{\stream}$. {\it Left panel:} Comparison between HD and MHD simulations. The black dashed line represents the limit between the {\it condensing stream} regime ($\xi<1$) and the {\it disrupting stream} regime ($\xi>1$). {\it Right panel:} Comparison between HD and MHD+TC simulations. The dashed black lines indicate the limit between the {\it condensing stream} regime ($\xi<0.3$), the {\it intermediate} regime ($0.3<\xi<8$), and the {\it diffusing stream} regime ($\xi>8$). Each point shows an averaged value over the three Mach numbers $\mathcal{M}=0.5,1,2$, with the error bars representing the range of minimum and maximum values. A small random offset in the x coordinate is included for clarity.}
    \label{fig:Mdot_comp}
\end{figure*}

We also compute the mean stream mass growth/loss, to provide a better representation of the stream mass evolution,
\begin{equation} \label{eq:Mdot}
\dot{M}_{\stream} = \left\langle \dfrac{\mathrm{d}M_{\stream}}{\mathrm{d}t} \right\rangle,
\end{equation}
where the $\left\langle \right\rangle$ indicates the linear arithmetic averaging over time from the start of the quasi-steady state of the mixing layer to the end of the simulation, the same time frame over which we took the log-average of $L_{\rm net}$.

The stream mass growth/loss $\dot{M}_{\stream}$ is shown in Fig.~\ref{fig:Mdot_comp} for HD, MHD, and MHD+TC cases.
In the {\it condensing stream} regime ($\xi <1$), MHD simulations exhibit a small decrease of the stream mass growth of the order of $\lesssim 10\%$ compared to the HD simulations, except at $\xi\sim 0.06$ where MHD simulations gradually decrease the stream mass growth with increasing $\alpha$. There are almost no differences between HD and MHD simulations for $\alpha=0^\circ$.
In the {\it disrupting stream} regime ($\xi >1$), MHD simulations have a significant impact, as streams with $\alpha\neq 0^\circ$ do not experience mass loss compared to HD and MHD with a magnetic field parallel to the stream ($\alpha = 0^\circ$).

Thermal conduction affects the stream mass evolution only for $\xi>8$, i.e., for a stream in the {\it diffusing stream} regime. As expected, in such a case, the stream diffuses into the CGM. However, the effective conduction can be greatly reduced by the initial magnetic field angle and the bending of the field line over time, resulting in a stable stream with almost no mass loss as $\mathcal{M}$ and $\alpha$ increases.

\subsubsection{Magnetic field amplification}\label{sec4:mag}

\begin{figure*}
	\includegraphics[width=2.0\columnwidth]{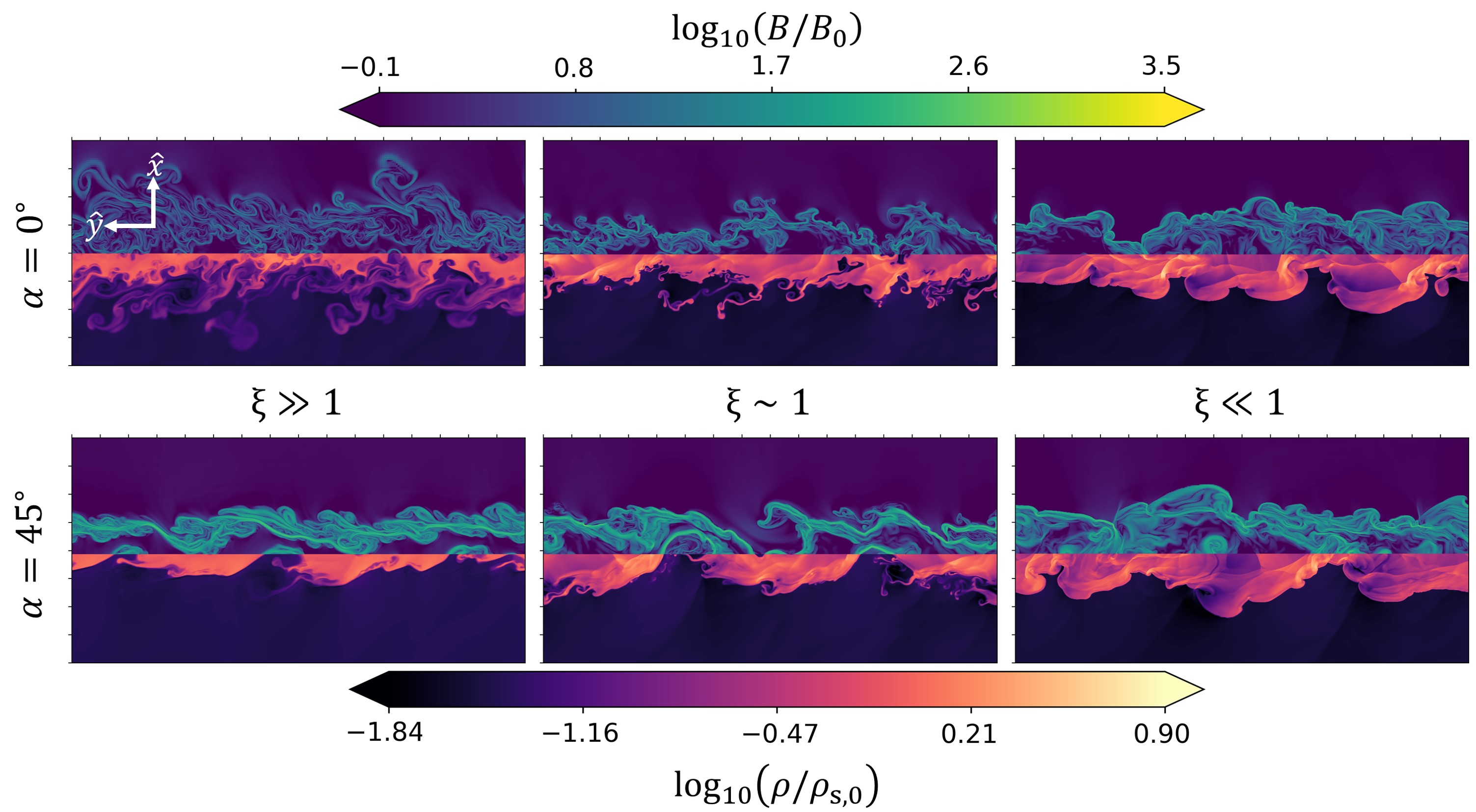}
    \caption{Comparison of magnetic field (top) and density (bottom) in MHD+TC simulations with $\alpha=0^\circ$ (field parallel to the stream) and $\alpha=45^\circ$ at the half time $t=11\, t_0$. From left to right, the contours show the simulations with decreasing $\xi$. Each panel shows the 32\,kpc full stream axis length in the horizontal direction and a zoomed region of 16\,kpc in the vertical direction.}
    \label{fig:BfieldContour}
\end{figure*}

We hereby investigate the amplification of the magnetic field.
Fig.~\ref{fig:BfieldContour} presents contours of the magnetic field strength and the density for angle $\alpha=0^\circ$ and $\alpha=45^\circ$, at different $\xi$ values. The contours are obtained from MHD+TC simulations with $\delta =30$. They do not exhibit a qualitative difference from the MHD and/or $\delta =100$ ones as long as $\xi<8$, i.e., as long as they are outside the {\it diffusing stream} regime. In all contours, the magnetic field increases at the interface of the stream and the CGM and within the stream.

For a field initially parallel to the stream ($\alpha=0^\circ$),  the magnetic field lines are stretched by the eddies that arise from the KHI. As the $\xi$ value decreases, the size of the mixing layer becomes smaller, leading to a slightly higher magnetic field amplification.

For $\alpha=45^\circ$, there is no significant difference in the magnetic field amplification across different $\xi$ values. The field is amplified by up to a factor of $\sim 500 B_0$, which is higher than the $\alpha=0^\circ$ simulations. This magnetic field increase can be explained by the presence of a component perpendicular to the stream, which is continuously stretched as the stream moves forward.

\begin{figure}
	\includegraphics[width=\columnwidth]{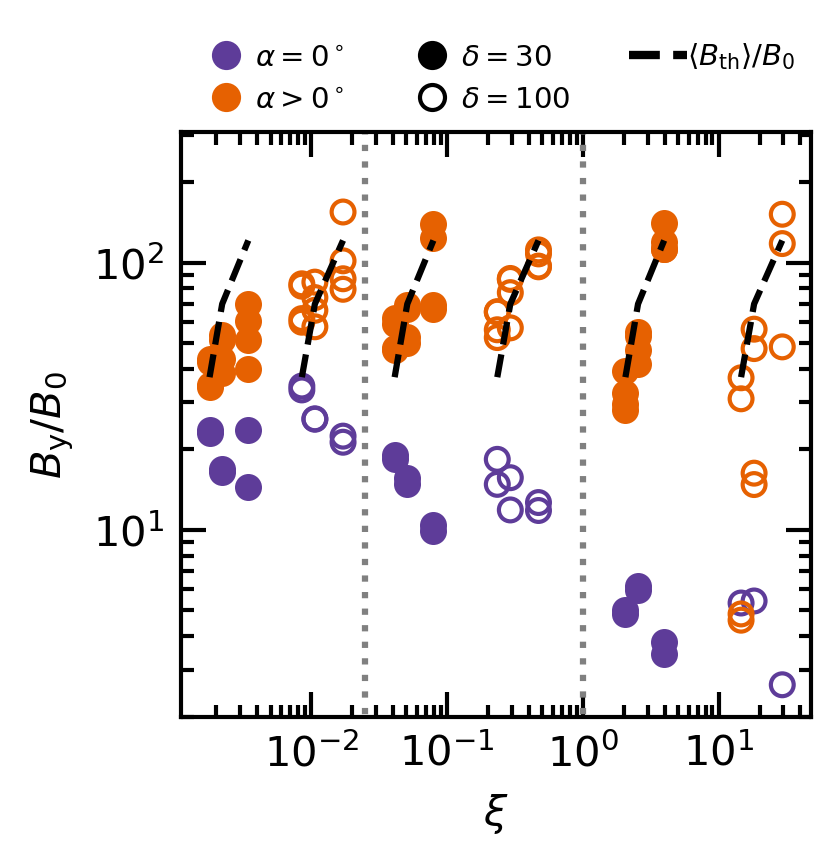}
    \caption{Mean magnetic field $B_{\rm y}$ normalized by $B_0$ in the stream and mixing layer region for all MHD and MHD+TC simulations as a function of $\xi$ parameter. The blue-navy points stand for simulations with initial magnetic field angle $\alpha=0^\circ$, and the orange points represent simulations with $\alpha\neq 0^\circ$ (i.e., $45^\circ$ or $90^\circ$). The black dashed lines represent the time-average of our model value $\left\langle B_{\rm{th}} \right\rangle$. The detailed distribution of the points (grey dashed line, filled and open circles) is the same as in Fig.~\ref{fig:CoolTot}. }
    \label{fig:Bfield}
\end{figure}

To quantify the growth of the magnetic field, the magnetic energy $E_{\rm m}$ is first averaged within the stream and the mixing layer, i.e., below the threshold temperature $T_{\rm u}$ defined in Sec.~\ref{sec4:MHDTC_impact}, 
\begin{equation} \label{eq:Bave}
B =  \left(2\left\langle E_{\rm{m}}\right\rangle\right)^{1/2} ,
\end{equation} 
where the brackets $\left\langle \right\rangle$ represents the linear-arithmetic time averaging as in Equation \ref{eq:Mdot}.
The results are plotted in Fig.~\ref{fig:Bfield} for the component $B_{\rm y}$. We focus only on this component because it is the one amplified by the velocity shear term which is the dominant amplification mechanism in our case\footnote{From the magnetic induction equation in Equation~\ref{eq:MHD_system_1}, taking our geometry and initial conditions, one can find that initially, $\partial_{\rm{t}} B_{\rm{y}} \sim B_{\rm{x,0}}\partial_{\rm{x}}v_{\rm{y,0}}$ is the dominant term when $\alpha \neq 0$. This is confirmed by Fig.~\ref{fig:Bfield} which shows that for $\alpha=0^\circ$, i.e. $B_{\rm{x,0}}=0$, the magnetic fields exhibit significantly lower amplification compared to $\alpha >0^\circ$ simulations.}. Considering $B_{\rm x}$ and plotting $B$ does not significantly change our results. The simulations with $\alpha=0^\circ$ and the one with $\alpha\neq 0^\circ$ exhibit two distinct trends, as qualitatively observed in Fig.~\ref{fig:BfieldContour}. Simulations with the magnetic field parallel to the stream reach a maximum of about $20\text{--}35B_0$ at $\xi\lesssim 2\times 10^{-2}$, followed by a decrease down to $B\sim3B_0$ for $\xi\sim 10$, following a roughly constant slope defined as $B\propto \xi^{-0.16}$. The scatter remains approximately constant along the slope and is primarily due to the variations in the Mach number $\mathcal{M}$ and the density ratio $\delta$ among the simulations.
In the case where the magnetic field has a component perpendicular to the stream ($\alpha> 0^\circ$), the mean magnetic field is amplified to $\sim 30\text{--}150B_0$ across $\xi$ values, with solely the scatter between the points increasing with $\xi$. Also, for a given $\delta$ and $n_{\stream}$, the magnetic field increases as $\xi$ increases (i.e., for increasing $\mathcal{M}$ with fixed $\delta$ and $n_{\stream}$). This trend is consistent with the fact that the growth of the magnetic field is mainly driven by the velocity shear at the interface of the stream and the CGM, for $\alpha\neq 0^\circ$.

Assuming that the velocity difference between the stream and the CGM is the main driver of the field line stretching, we model the field amplification by approximating a field line as a stretching flux tube \citep[see][for example]{Spruit2013}. The details of the models are derived and discussed in Appendix \ref{app:Mag_Mod}. As a result, the magnetic field can be expressed as,
\begin{equation}\label{eq:Bmodel1}
    \frac{B_{\rm{th}}\left(t\right)}{B_0} = \frac{1}{2}\left[ \left(\beta \frac{R_{\stream}}{R_{\stream} + v_{\stream,0}t}\right)^2 + 4\left(\beta+1\right)\right]^{1/2} - \frac{1}{2} \beta \frac{R_{\stream}}{R_{\stream} + v_{\stream,0}t} ,
\end{equation}
where $\beta =10^5$ represents the ratio of the thermal over the magnetic pressure, and the term $R_{\stream}/\left(R_{\stream}+v_{\stream,0}t\right)$ indicates the stretching of the field lines over time. For comparison, the average value of our model $\left\langle B_{\rm th} \right\rangle$ is plotted in Fig.~\ref{fig:Bfield} where the $\xi$ dependency of $B_{\rm{th}}$ is enforced in equation \ref{eq:Bmodel1} with $v_{\stream ,0} = \xi\times R_{\stream}/ \left(\alpha t_{\rm{cool}}\right)$ using equations \ref{eq:t_sh} and \ref{eq:xi}. Our model is consistent\footnote{As shown in Appendix \ref{app:Mag_Mod}, the model also captures well the time variation of the magnetic field..} with the simulations where $\alpha >0^\circ$.

It is worth noting that a few points for $\alpha>0^\circ$ and $\xi>10$ exhibit relatively small magnetic field growth. These points correspond to the MHD+TC simulation in the {\it diffusing stream} regime. As the stream diffuses in the CGM, the shear layer expands, which reduces the bending of the field lines due to the velocity difference, resulting in a smaller growth of $B$.

The approximately 100-fold increase in the magnetic field leads to a reduced value of average $\beta$ down to $\sim 10$ from its initial average value of $\sim 10^5$. This is because the magnetic field is not drastically amplified in the centre of the stream. The average thermal pressure remains constant in the stream centre. 
However, as observed in Fig.~\ref{fig:BfieldContour}, the field can be amplified up to $\gtrsim 200\text{--}500 B_0$ near the CGM and stream interface, giving $\beta \sim 1$ and a physical value $B>0.2\,  \upmu\rm G$. This magnified magnetic field can then explain the stream stabilization in the {\it disrupting stream} regime, as depicted in the mass rate plot of Fig.~\ref{fig:Mdot_comp}. These results are consistent with \citet{bib_MHD_Berlok_2019b}, who found that for $B \gtrsim 0.4\,  \upmu\rm G$ with an initial field parallel to the stream, the magnetic tension is strong enough to stabilize the stream against KHI.

Therefore, despite an initially low magnitude, the magnetic field can undergo significant amplification, thereby affecting both the stream emission and its evolution, as long as the magnetic field is not parallel to the stream. 

\subsubsection{Turbulent velocity in the mixing layer}\label{sec4:kin}

As mentioned previously, the mixing of CGM and stream gas is the key mechanism of the cold stream emission signature. In this section, we assess the turbulent velocity within the mixing layer using a mean-field approach. We describe below the procedure to compute the turbulent component of a given fluid value. Firstly, the values are averaged over the stream axis length, and for both sides of the stream axis to obtain a radial-dependent averaging,
\begin{equation} \label{eq:ReynoldsMean}
\overline{\rho}\left(r \right) = \frac{1}{L_{\rm{y}}}\int^{L_{\rm{y}}}_0 0.5\left(\rho(x,y) + \rho(-x,y)\right) \, \rm{d} y.
\end{equation} 
Next, the density-weighted averaged value is computed, as it is better suited for compressible flows \citep{Favre1969},
\begin{equation} \label{eq:FavreMean}
\widetilde{X}\left(r \right) = \frac{\overline{\rho X}}{\overline{\rho}}.
\end{equation}

The fluctuating field is recovered from the mean-field decomposition where $X'(x,y) = X(x,y) - \widetilde{X}(\left\vert x\right\vert)$.
The root-mean-square value of the fluctuating velocity field inside the mixing layer is derived in practice from the turbulent kinetic energy,
\begin{equation}\label{eq:VturbML}
    V'_{\rm{ML}} = \left(\frac{1}{r_{\rm{ML}}}\int \widetilde{u^{'2}} \rm{d} r \right)^{1/2},
\end{equation}
with $u^{'2}=u_{\rm{r}}^{'\, 2} +u_{\rm{y}}^{'2} $, and where the mixing layer is defined between the temperature thresholds $T_{\rm{d}}$ and $T_{\rm{u}}$ (see appendix~\ref{app:Mass_thr} for the threshold discussion). The mean mixing layer size is defined as $r_{\rm{ML}} = 1/L_{\rm{y}}\int \left( 0.5\int_{\rm{ML}} \rm{d}x \right)\rm{d}y$ , with the factor $0.5$ accounting for the average of the mixing layers at both sides of the stream and where the integration in performed similarly to the one in equation~\ref{eq:Lnet}. The integral in equation~\ref{eq:VturbML} is done in function of the radius $r$ because the density-weighted averaged values in Equation~\ref{eq:FavreMean} are a function of $r$. Here, $r_{\rm ML}$ represents the size of the mixing layer. It is important to note that our definition of the turbulent kinetic energy considers the mean radial velocity as part of the mean flow, so that the inflow velocity of the CGM gas $v_{\rm in}$ is accounted for in the bulk motion (cf. Sec.~\ref{sec2:mixing_layer}), rather than in the turbulent velocity term.

\begin{figure}
	\includegraphics[width=\columnwidth]{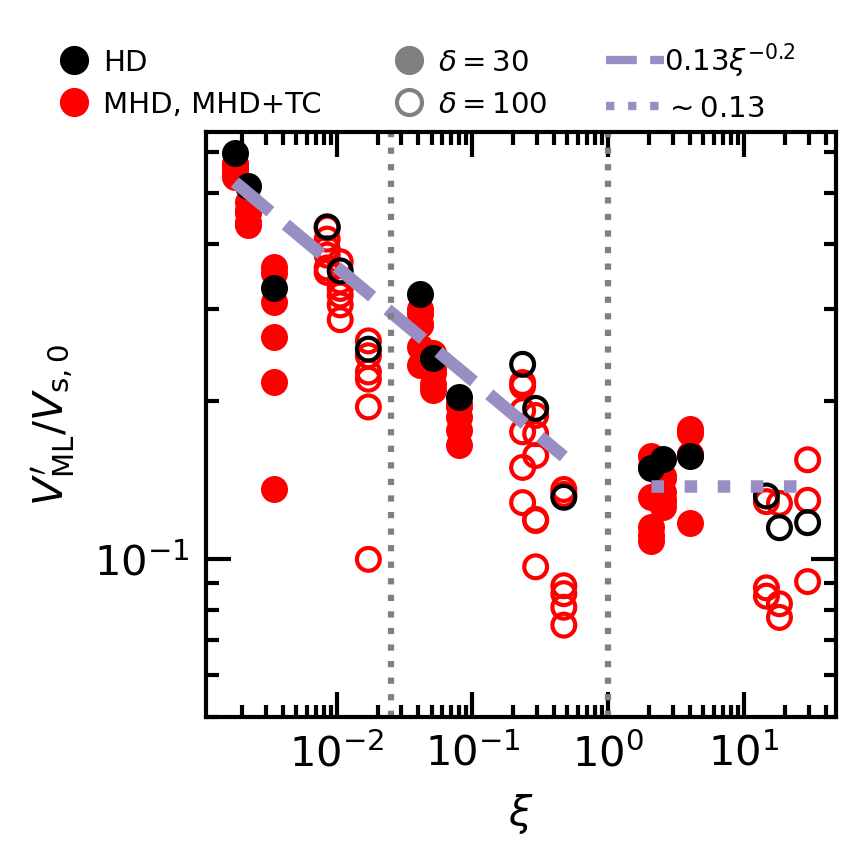}
    \caption{Turbulent velocity in the mixing layer. Red points are the MHD and MHD+TC simulations, and black points are the HD simulations. The thick dashed purple line shows a power-law fit for $\xi<1$, and the thick dotted purple line is a roughly constant fit for $\xi >1$. Both fit use only the HD simulations. The detailed distribution of the points (grey vertical dotted lines, filled and open circles) is the same as in Fig.~\ref{fig:CoolTot}. }
    \label{fig:Vturb_ML}
\end{figure}

The results are presented in Fig.~\ref{fig:Vturb_ML} for all simulations except the MHD+TC simulations in the {\it diffusing stream} regime\footnote{For the MHD+TC simulations in the {\it diffusing stream} regime, the thermal conduction diffuses all instabilities, resulting in a negligible turbulent velocity $V'_{\rm{ML}} < 10^{-2} v_{\stream,0}$. For clarity, we have omitted these simulations from the plot. \label{fn:Vturb}}.
The turbulent velocity exhibits a clear decreasing trend as $\xi$ increases. For $\xi<1$, the general decrease observed for the HD simulations can be fitted as $V'_{\rm{ML}} \propto v_{\stream,0}\xi^{-1/5}$, while for $\xi>1$ we found a roughly constant value around $0.11  v_{\stream,0}$ with a very weak dependency on $\xi$. This decreasing trend is consistent with the analytical model of \citet[equation $42$, Figure $12$]{Tan2021}, which was derived from previous simulations of cold--hot interface geometry. From their simulations, they also found the relation $V'_{\rm{ML}} \propto v_{\stream,0}\xi^{-1/5} \propto v_{\stream,0}^{4/5} t_{\rm{cool}}^{-1/5}$.

Furthermore, in line with the findings in previous sections, the HD simulations exhibit higher turbulent velocities than their MHD and MHD+TC counterparts. As observed previously, the presence of magnetic fields and thermal conduction hinders the growth of KHI for $\alpha\neq 0^\circ$, which directly correlates with the decrease of the magnitude of turbulence in the mixing layer.
In terms of the magnitude of turbulence, we found similar values to previous cold streams simulations from \citet[][without radiative cooling]{Mandelker2019} with $V'_{\rm{ML}} \sim 0.2 v_{\stream,0}$. However, our turbulent velocities are higher than in their simulations with radiative cooling \citep{Mandelker2020a} which exhibit $V'_{\rm{ML}} \sim 0.2 v_{\stream,0}\delta^{-1/2}\sim 0.02-0.03v_{\stream,0}$. This difference could be attributed to the different domains of integration. In their case, they consider both the stream and the mixing layer, while in our case, we strictly confine the domain inside the temperature thresholds of the mixing layer. When including the stream in the integral domain of Equation \ref{eq:VturbML}, we find $V'_{\rm{ML}} \sim 0.6 v_{\stream,0}\delta^{-1/2}$ which is smaller than our value within the mixing layer. Their simulations are in three dimensions while ours are in two dimensions; therefore, we do not expect a perfect match in the magnitude of the turbulence.

We found that the impact of MHD or MHD+TC on the turbulence magnitude correlates with the stream mass evolution. As the angle $\alpha$ tends to $90^\circ$, the strength of the turbulence decreases. We applied the fitting approach of the HD simulations in Fig.~\ref{fig:Vturb_ML} to the combined sample of MHD and MHD+TC simulations, with $\alpha\neq 0^\circ$. We exclude from the fits the simulations with $\alpha=0^\circ$ because their magnetic field does not increase enough to have a significant impact compared with the HD simulations. The resulting fit gives us,
\begin{equation}
V'_{\rm{ML}} \sim \left\{ \begin{array}{ll}
0.1\, v_{\stream,0} \, \xi^{-1/4} & \text{if} \, \, \xi <1,\\
0.1\, v_{\stream,0} & \text{if}\, \, 1<\xi <10.
\end{array}\right.
\label{eq:VturbMHDTC}
\end{equation}

The decrease in the turbulence magnitude is evident from the fit. Both HD, MHD and MHD+TC fits start at a similar turbulence magnitude at $\xi \sim 10^{-2}$ with $V'_{\rm{ML}}\sim 0.6 v_{\stream,0}$. However, the MHD and MHD+TC fit exhibits a steeper slope, resulting in a further decrease in the turbulent velocity down to $V'_{\rm{ML}}\sim 0.1 v_{\stream,0}$.

The more pronounced decrease in turbulent velocity illustrates the direct impact of MHD and TC as they dampen the KHI growth. In the case of a magnetic field, the amplified field creates a tension force that counteracts the KHI growth, hence, stabilising the stream. 
For thermal conduction, the diffusion of either the mixing layer or the stream can further reduce the KHI growth, thereby stabilising the stream against KHI and reducing the mixing of the stream and the CGM.

\section{Discussion}\label{sec5}

In this section, we extend our findings to the cosmological context of cold streams entering the halo of massive galaxies. We first discuss Ly$\alpha$ emission within the halo (Sec.~\ref{sec5:Rad}), followed by the properties of the cold streams penetrating the halo (Sec.~\ref{sec5:StreamProp}). Lastly, we address various limitations and caveats of our work (Sec.~\ref{sec5:Caveats}).

\subsection{Emission inside the halo}\label{sec5:Rad}

In Section~\ref{sec4:stream_evo}, we discussed the emission properties of a cold stream with a fixed radius of $R_{\stream} = 1 \, \mathrm{kpc}$ at the virial radius $R_{\rm{v}} = 100 \, \mathrm{kpc}$ of a massive galaxy residing in a $10^{12} \, \rm M_{\sun}$ halo.
As the stream survives and penetrates deeper into the halo, it becomes denser leading to an increase in its emission by a factor $10^2-10^3$  \citep{LyblobMandelker2020}.

In the {\it condensing stream} regime, characterised by high density, high metallicity, and/or large $R_{\stream}$, the impact of magnetic fields and thermal conduction on the stream's emission is negligible. 
However, in the {\it intermediate} or {\it diffusing stream} regimes, where the stream exhibits low number density, and/or low metallicity, and/or a small radius, the presence of magnetic fields and thermal conduction leads to a significant reduction in the stream emission by a factor of $10-20$.

Our analysis reveals that, on average, the hydrogen contribution to the stream's emission is $\sim 20\%$, with the highest cases reaching around 45\% (see Appendix \ref{app:Lnet_nature}). 
Given that the cooling emission is predominantly collisional, it is expected that the Ly$\alpha$ emission contributes to $\sim 50\%$ of the total hydrogen cooling rate \citep[][see figure 7]{Dijkstra2017}. 
Therefore, for MHD+TC simulations (considering only cases with $\alpha \neq 0^\circ$, which are more realistic than a purely parallel magnetic field), the total net emission from a cold stream near the galaxy ($0.1R_{\rm v}$ from the halo centre) is in the range of  $L_{\rm{net}}\left(0.1 R_{\rm{v}}\right) = \left(100\text{--}1000\right)\times L_{\rm{net}}\left(1 R_{\rm{v}}\right)$. The subsequent Ly$\alpha$ emission in the vicinity of the galaxy should be,
\begin{equation}
\begin{array}{ll}
L_{\rm{Ly\alpha}} & = a_{\rm{H}}a_{\rm{Ly\alpha}}L_{\rm{net}}\left(0.1R_{\rm{v}}\right) \\
 & \\
 & \sim \left\{ \begin{array}{ll}
10^{38}\text{--}5\times10^{41} \, \rm{erg\, s^{-1}} & \text{for} \, \, \xi <0.3,\\
4\times10^{37}\text{--}2.5\times10^{40} \, \rm{erg\, s^{-1}} & \text{for}\, \, \xi > 0.3.
\end{array}\right.
\end{array}
\end{equation}
with $a_{\rm{H}}=0.2\text{--}0.45$ the hydrogen contribution to the total net cooling emission, and $a_{\rm Ly\alpha} =0.5$ representing the Ly$\alpha$ contribution to the hydrogen cooling emission. The two cases $\xi<0.3$ and $\xi>0.3$ refer to the {\it condensing stream} regime, and the {\it intermediate} and {\it diffusing stream} regimes, respectively.

The stream's emission originates from the cooling layer, and as the stream radius increases, the volume occupied by the mixing layer surrounding the stream also increases. Consequently, the net emission $L_{\rm net}$  scales with $R_{\stream}^2$, i.e., the cross-sectional area of the stream. For $R_{\stream}= 10\,\rm kpc$, this implies that $ L_{\rm{Ly\alpha}} \sim 4\times 10^{39}\text{--}2.5\times 10^{42}\, \rm erg\, s^{-1}$ for a stream in the {\it intermediate} or {\it diffusing stream} regime, and $L_{\rm{Ly\alpha}} \sim 10^{40}\text{--}5\times 10^{43}\, \rm erg\, s^{-1}$ for a stream in the {\it condensing stream} regime, where in both case it is assumed that the stream remains in the regime it was with $R_{\stream}=1\,\kpc$.

Compared to the analytical model\footnote{Model derived from previous hydrodynamic simulations \citep[e.g.][]{Mandelker2020a} similar to our work.} of \citet{LyblobMandelker2020}, our cold stream emissions from MHD+TC are very similar for the dense and/or thick streams but is smaller by a factor of $\sim 100$ for diffuse and/or thin streams. This difference is due to both a lower $L_{\rm net}$ and a lower $a_{\rm H}$ \footnote{In their model, they attribute  $\sim 50\%$ of the emission to gas at a temperature of $T\sim \left(2\text{--}3\right) 10^4 \rm K$, which they associate with hydrogen emission. In our case, the hydrogen emission is directly computed from the cooling model,  leading to a lower percentage contribution.}.

It is worth noting that previous studies, such as \citet[e.g.][]{Goerdt2010,LyblobMandelker2020}, have often considered the feeding of galaxies by two or three prominent cold streams. 
Such a picture fits well with current cosmological simulations. 
However, more recent simulations with higher resolution in galactic haloes \citep{Hummels2019,Peeples2019,VandeVoort2019, Bennett2020, Nelson2020} have revealed the presence of a substantial number of small-scale ($<1\,\rm kpc$) cold structures within the CGM.
These smaller features become particularly evident when the resolution is increased near CGM shocks resulting from galactic feedback processes \citep[][see figure 5]{Bennett2020}. From \citet{Bennett2020}, the CGM features an almost a cold web-like structure which could increase the emission by a factor $>10$.
Consequently, the cold flow signature is not solely limited to large filamentary structures but encompasses the emission from a multitude of smaller, less dense streams. This implies that the overall cold flow emission can be significantly enhanced, by a factor of more than 10, due to the contribution of the potential  numerous thin cold streams within the CGM. Therefore, the emission properties associated with cold flows are more diverse and complex than previously considered, highlighting the importance of accounting for the full range of cold structures within the CGM.

\subsection{Properties of the cold stream inflow}\label{sec5:StreamProp}

We focus on the resulting properties of a stream entering a halo in terms of mass flux, metallicities, and magnetic field.
The mean\footnote{As our simulations have properties of a stream and a CGM at $R_{\rm v}$ over the all computational domain, we can average the mass flux over the y-direction to avoid dependency of the mass flux on the local cold mass rate.} cold mass inflow rate $j_{\stream}$ is computed as,
\begin{equation}
j_{\stream} =\frac{1}{L_{\rm{y}}} \int\left(\iint \rho v_{\rm{y}} \, \mathrm{d}A\right) \rm{d}y ,
\label{eq:js}
\end{equation}
where $L_{\rm y}= 32 \rm \, kpc$ is the stream length, $\mathrm{d}A=\mathrm{d}x\mathrm{d}z$ the cross-section perpendicular to the stream axis, and the integral is performed only for the gas defined as cold, with $T<T_{\rm d}$ (see Sec.~\ref{sec4:stream_evo}). The integration through $z$ is written for unit clarity but disappears in practice when normalizing $j_{\rm s}$ by its initial value. We then express the mass flow rate in units of the initial rate $j_{\stream,0}= \pi R_{\stream}^2 \rho_{\stream,0} v_{\stream,0} =0.01 \text{--} 1 \, \rm M_{\sun} yr^{-1}$ assuming a cylindrical stream of radius $R_{\stream}$.

\begin{figure*}
	\includegraphics[width=2.0\columnwidth]{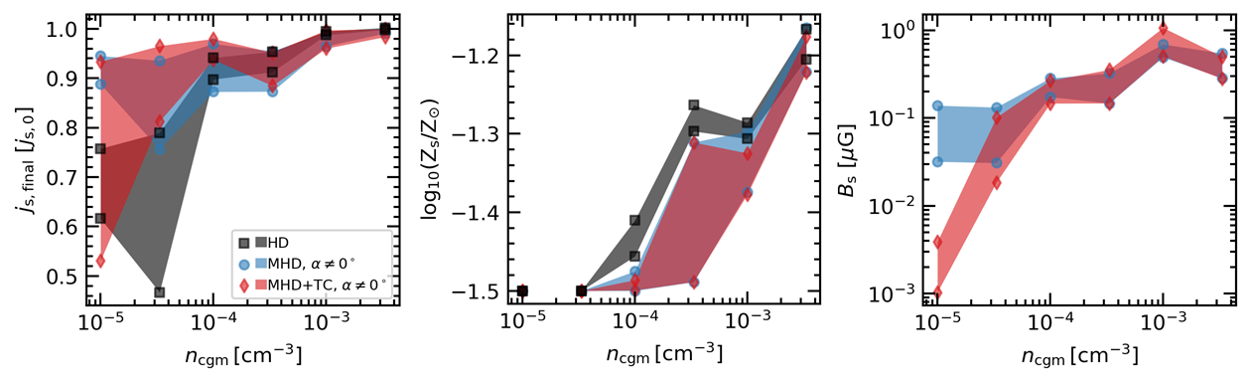}
    \caption{Properties of the stream entering the halo are depicted in the three panels, each using a common legend. 
    All quantities are plotted as a function of the CGM number density. 
    {\it Left panel}: Cold mass accretion rate at the final simulation time in units of initial cold accretion rate,  $j_{\stream,0}=m_{\rm{p}}\mu n_{\stream}\mathcal{M}c_{\cgm}\pi R_{\stream}^2 \sim 0.01\text{--}1 \, \rm M_{\sun} \, yr^{-1}$ for $R_{\stream}= 1 \, \rm kpc$ and $n_{\stream}=0.001\text{--}0.1 \, \rm cm^{-3}$.  
    {\it Middle panel}: Logarithm of the mean metallicity in the stream at the final time in Solar metallicity units.  The initial stream metallicity is $\log_{10}{\left( Z_{\stream,0}/Z_{\sun}\right)}=-1.5$, and the maximum value being the assumed CGM metallicity $\log_{10}{\left( Z_{\cgm,0}/Z_{\sun} \right)}=-1$. 
    {\it Right panel}: Mean magnetic field strength in the stream ($B_{\stream}$).   This value $B_{\stream}$ is reached in the early stage of the simulations at about $t\sim 0.1 t_{\rm end}$ and then remains roughly constant until the final time.}
    \label{fig:Maccret}
\end{figure*}
The metallicity in the stream is computed as a function of the stream mass as
\begin{equation}
Z_{\stream,final} \sim \left\{ \begin{array}{ll}
\frac{M_{\stream , \rm{final}} - M_{\stream,0}}{M_{\stream, \rm{final}}}Z_{\cgm} +  \frac{M_{\stream,0}}{M_{\stream, \rm{final}}}Z_{\stream} & \text{if} \, \, M_{\stream, \rm{final}} > M_{\stream,0},\\
Z_{\stream} & \text{if}\, \, M_{\stream, \rm{final}} < M_{\stream,0}.
\end{array}\right.
\label{eq:Zs}
\end{equation}
Here, we consider that if the stream grows, the additional cold mass is added to the stream with CGM metallicity, and if the stream loses mass, the remaining cold mass dwells at its initial stream metallicity.

The mean magnetic field is computed as in Equation~\ref{eq:Bave} and is presented in physical units at the final simulation time.

Fig.~\ref{fig:Maccret} illustrates the stream properties as a function of the CGM number density $n_{\cgm}= n_{\stream}\delta^{-1}$, which roughly scales as $\xi^{-0.5}$. Higher density leads to stronger cooling emission and smaller $\xi$ ratios, and vice versa. 
For the MHD and MHD+TC simulations, results are only shown for $\alpha \neq 0^\circ$, because there are no significant differences with the HD simulations for $\alpha=0^\circ$.
In the left panel, the impact of MHD on the cold mass accretion rate is evident, as it helps to retain $\sim 80\%$ of the initial mass flow in the low-density regime. This demonstrates that the presence of a magnetic field enhances the stability of the stream and sustains the cold mass accretion.

Thermal conduction only acts for very low density with $n_{\cgm}\sim 10^{-5}  \, \rm cm^{-3}$ where the stream diffuses, resulting in a reduced cold mass accretion rate of $\sim 0.5 j_{\stream ,0}$. However, as the conduction efficiency decreases with increasing velocity, the stream can maintain a cold mass accretion rate with $\sim 90\%$ of its initial value even at such low densities for $\mathcal{M}=2$. Therefore, compared to HD simulations, MHD+TC simulations show no significant difference in a high-density CGM with $\gtrsim 2\times 10^{-4} \, \rm cm^{-3}$, but TC can help the stream to survive below this number density threshold. 
These trends in our results agree with simulations of cold clouds embedded in galactic winds from MHD simulations by \citet{Hidalgo-Pineda2023} and MHD+TC simulations by \citet{Bruggen2023}, as well as resolution tests on cosmological simulations \citep{Hummels2019,Nelson2020,Bennett2020} 

From the metallicity plots, higher CGM density (stronger cooling) leads to higher metallicity in the stream, up to $\left( Z_{\stream,\rm{final}} - Z_{\stream}\right)/\left(Z_{\cgm}-Z_{\stream}\right) \sim 60\%$. By lowering the mixing of the gas, both MHD and MHD+TC simulations lower the metal enrichment of the stream. This metal pollution is counter-intuitive to the ideal picture of cold streams being pristine and can support some of the observed \citep{Bouche2013,Bouche2016} or assumed \citep{Giavalisco2011,Rubin2012,Martin2012,Zabl2019,Emonts2023} high metallicity of cold inflow.
From a cosmological simulation point of view, a higher metal enrichment of the cold inflow would also lead to higher star-formation efficiency. Combined with the cold stream's prolonged stability in the hot CGM, one may expect the star formation of massive galaxies to be sustained for a longer time.

In the case of high-density CGM, the magnetic field in the stream can undergo a significant enhancement, reaching $\sim 1\,  \upmu\rm G$. 
However, as the CGM density decreases, the magnetic field strength diminishes, reaching a value of $2 \times 10^{-2} \,  \upmu\rm G$ for MHD simulations. 
For MHD+TC simulations with $n_{\cgm} \sim 10^{-5} \, \rm cm^{-3}$, the magnetic field within the cold stream remains close to its initial value as the stream gradually diffuses.

From a cosmological simulation perspective, one may expect the magnetic field to increase again by compression once it reaches the ISM. Such magnetized cold inflow may take a longer time to collapse as they could be magnetically supported. 

The simplistic extrapolation of our findings raises an intriguing question about the potential oversights in cosmological simulations resulting from their resolution limitations. 
In addition to the prolonged sustenance of cold stream inflow, there is an important transformation in the properties of the cold material itself. It transitions from pristine cold streams to metal-enriched magnetized cold streams.

\subsection{Caveats: 2D vs. 3D and additional physics}\label{sec5:Caveats}

The first limitation of our work is that our simulations are conducted in a two-dimensional domain. 
In three dimensions, the KHI is expected to grow faster due to the appearance of additional instability modes. This faster growth results in a more efficient mixing between the stream and the surrounding medium, thereby altering the evolution of the stream and its observable properties \citep{Padnos2018,Mandelker2019}. 
Enhanced mixing would potentially lead to stronger emission and mass growth rates for the streams in the {\it condensing stream} regime ($\xi=t_{\rm{cool}}/t_{\rm{sh}} < 1$).

From MHD simulations with a magnetic field aligned with the stream, \citet{bib_MHD_Berlok_2019b} found that three-dimensional simulations exhibit an increased mixing. This is primarily driven by the growth of azimuthal KHI modes which are not inhibited by any magnetic tension force when the field is parallel with the stream. In future work, we will investigate the impact of magnetic fields not parallel to the stream using three-dimensional simulations. 

Additionally, a more realistic model should include additional physics. For example, our model ignores self-gravity, which may affect the stability of the stream \citep{Aung2019}. We also adopt a cooling-heating function (Fig.~\ref{fig:net_cool_heat_1}) assuming the gas is optically thin. However, the function should vary with the stream density because dense cold streams will self-shield against the UV background radiation. Since the gas temperature and density affect the importance of self-gravity in the stream \citep[see][]{Ostriker1964,Aung2019}, a detailed treatment of radiation heating will be important in evaluating the stability of the cold stream. We hypothesize that the self-shielding may not significantly affect the emission from the mixing layer because it is dominated by collisional cooling.

\section{Conclusions}\label{sec6}
Recent advancements in idealized high-resolution simulations have contributed to our understanding of cold streams and their emission signatures. To further enhance this knowledge, we conducted an extensive suite of two-dimensional simulations incorporating key physical processes, including radiative cooling, magnetic fields with varying angles, and anisotropic thermal conduction. The combination of these physics has not been explored comprehensively before. The simulations were performed using the Athena++ code and did not account for self-gravity or self-shielding.

\begin{figure*}
	\includegraphics[width=2.0\columnwidth]{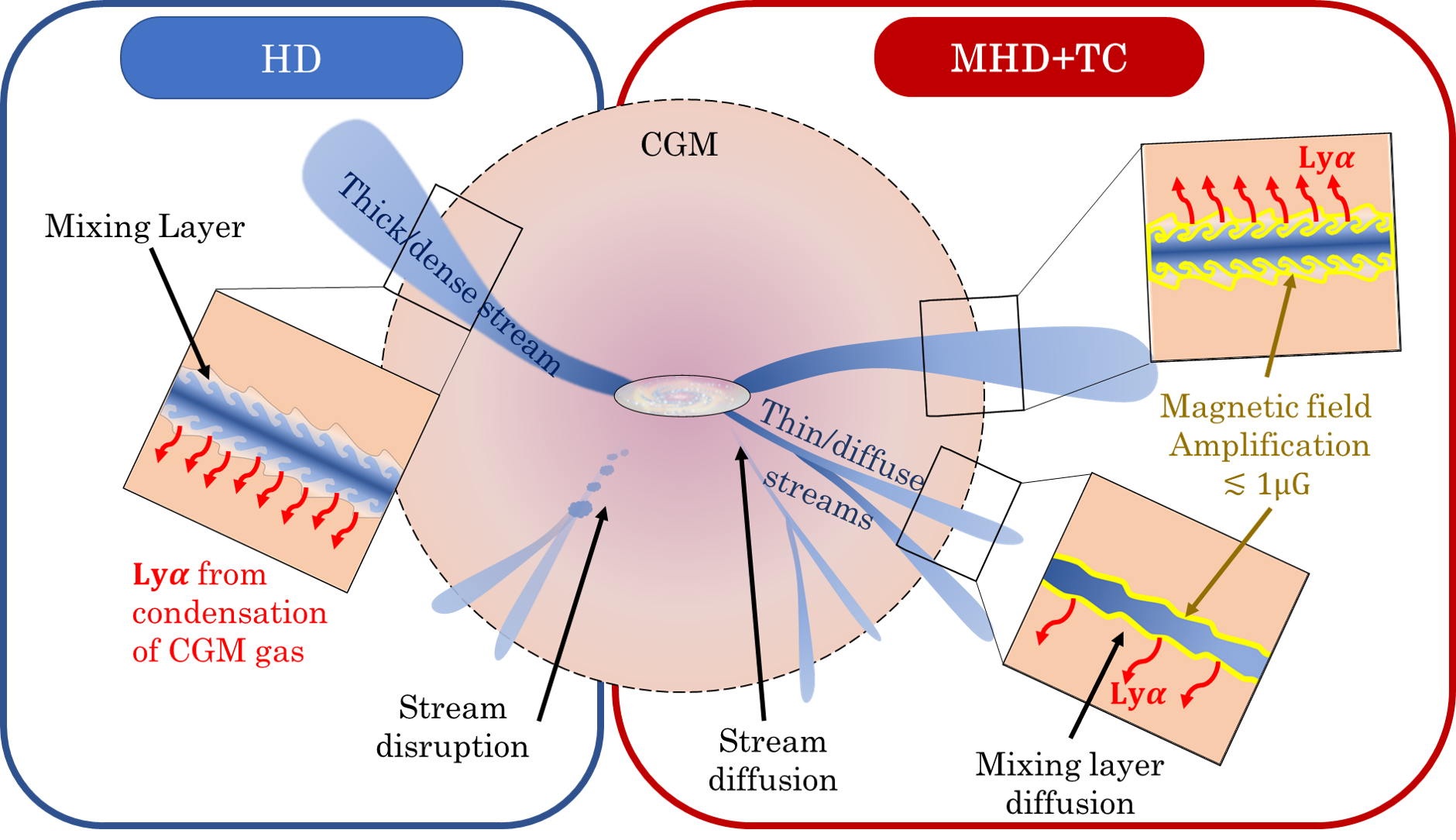}
    \caption{
    Schematic illustration comparing the fate of a cold stream in two scenarios: hydrodynamic (HD) and magnetohydrodynamic with thermal conduction (MHD+TC). In the HD case, thick or dense streams survive ({\it condensing stream} regime) while thin or diffuse streams disrupt within the CGM due to KHI ({\it disrupting stream} regime). 
    Ly$\alpha$ emissions originate from the mixing layer in both cases. In the MHD+TC case, thick dense streams persist, providing magnetized cold material to the galaxy ({\it condensing stream} regime).
    For thin diffuse streams, the combined effects of magnetic field amplification and thermal conduction stabilize the stream against Kelvin-Helmholtz instability (KHI), resulting in significantly reduced Ly$\alpha$ emissions ({\it intermediate} regime). 
    If the stream is too thin or diffuse, it may evaporate within the CGM, potentially impeding its reach towards the central galaxy ({\it diffusing stream} regime).}
    \label{fig:manga}
\end{figure*}

In our idealized simulations, we focused on a cold stream situated at the virial radius of a $10^{12}, \rm M_{\sun}$ halo at a redshift of $z=2$. 
We consider a stream of radius $R_{\stream} = 1\, \kpc$ and an initial magnetic field defined by the ratio of thermal pressure over magnetic pressure $\beta = 10^5$.
We summarise our findings in a schematic illustration in Fig.~\ref{fig:manga}.

\paragraph*{Cold streams regimes:} 
By including thermal conduction, the behaviour of the stream can be categorized into three regimes  (see Sec.~\ref{sec2:timescale_comp}), depending roughly on the ratio $\xi=t_{\rm{cool}}/t_{\rm{sh}}$, with $t_{\rm{cool}}$ the cooling time defined from the mixing layer (Equation~\ref{eq:rad2}), and $t_{\rm sh}$ the shearing time (Equation~\ref{eq:t_sh}).

(1) The {\it diffusing stream} regime ($\xi > 8$) corresponds to the thinnest and/or diffuse cold streams in Fig.~\ref{fig:manga}.
In this regime, thermal conduction dominates over other processes, impeding the growth of KHI and causing the stream to diffuse within the CGM. A faster stream can however significantly reduce the efficiency of the thermal conduction while still stabilizing it against KHI, allowing it to potentially reach the central galaxy.

(2) In the {\it intermediate} regime ($8>\xi>0.3$), radiative cooling can overcome thermal conduction within the stream, but not in the mixing layer, which continuously diffuses in the CGM. 

(3) The {\it condensing stream} regime ($\xi<0.3$) involves dense and/or thick cold streams as shown in Fig.~\ref{fig:manga}. In this scenario, cooling is highly efficient, leading to the condensation of CGM gas on to the stream. The key distinction from the hydrodynamic (HD) case in the {\it condensing stream} regime is that the stream becomes magnetized upon reaching the central galaxy.
    
\paragraph*{Emission signature:} 
In the {\it intermediate} and {\it diffusing stream} regimes, the emission signature experiences a significant decrease in the MHD+TC case compared to the HD case, by a factor of up to 20 (see Sec.~\ref{sec4:stream_evo}). This reduction in emission is attributed to two factors: the amplification of the magnetic field at the stream interface and the diffusion of the mixing layer, which is the source of the cooling emission.
In the {\it diffusing stream} regime, the emitting gas becomes hotter, increasing the cooling emissions from metals but a decrease in those from hydrogen.
In the {\it condensing stream} regime, the impact of thermal conduction and magnetic fields on the stream's emission is insignificant.
We observed that, outside the {\it diffusing stream} regime, approximately 20\% of the stream's cooling emission originates from hydrogen, which is lower than previous estimations found in the literature. 
This further diminishes the expected Ly$\alpha$ luminosity of cold streams.
    
\paragraph*{Cold stream evolution:} 
In the {\it intermediate} regime, the presence of magnetic fields and thermal conduction effectively suppresses the growth of the KHI (Sec.~\ref{sec4:mag} and \ref{sec4:kin}). As a result, the stream remains stable and does not experience mass loss, allowing it to survive for longer periods compared to the hydrodynamic case (Sec.~\ref{sec4:stream_evo}). 
In the {\it diffusing stream} regime, although the stream undergoes mass loss, the efficiency of thermal conduction decreases significantly as the stream velocity increases (see Sec.~\ref{sec4:MHDTC_evo}). 
As a result, streams with a Mach number $\mathcal{M}=2$ experience only minimal mass loss. 
It is worth noting that, similar to the emission signature, the presence of magnetic fields and thermal conduction has negligible effects on the evolution of the stream in the {\it condensing stream} regime or when the magnetic field is parallel to the stream ($\alpha = 0^\circ$).

\paragraph*{Cosmological implications:} 
By extrapolating our findings from idealized simulations to a cosmological context (Sec.~\ref{sec5}), we determined that the Ly$\alpha$ luminosity of cold streams within haloes falls within the range of $4\times 10^{37} \, \rm erg\, s^{-1}$ to $2.5\times 10^{40}\, \rm erg\, s^{-1}$, specifically for relatively small cold streams with a radius of $R_{\stream}=1\,\rm kpc$. Furthermore, we observed that the inflowing gas in these streams becomes enriched with metals and is magnetized, with the mean magnetic field strength in the stream reaching approximately $1\, \upmu \rm G$. These results provide insights into the properties and characteristics of cold streams in the cosmological context.

\section*{Acknowledgements}

Numerical computations were carried out on the Cray XC50 at the Center for Computational Astrophysics, National Astronomical Observatory of Japan, and the {\sc SQUID} at the Cybermedia Center, Osaka University as part of the HPCI system Research Project (hp200041, hp220044). 
This work is supported in part by the MEXT/JSPS KAKENHI grant numbers  19H05810, 20H00180, 22K21349 (K.N.). 
K.N. acknowledges the support from the Kavli IPMU, World Premier Research Center Initiative (WPI), where part of this work was conducted. 


\section*{Data Availability}

Data related to this publication and its figures are available on request from the corresponding author.


\bibliographystyle{mnras}
\bibliography{references} 

\begin{thebibliography}{}
\makeatletter
\relax
\def\mn@urlcharsother{\let\do\@makeother \do\$\do\&\do\#\do\^\do\_\do\%\do\~}
\def\mn@doi{\begingroup\mn@urlcharsother \@ifnextchar [ {\mn@doi@}
  {\mn@doi@[]}}
\def\mn@doi@[#1]#2{\def\@tempa{#1}\ifx\@tempa\@empty \href
  {http://dx.doi.org/#2} {doi:#2}\else \href {http://dx.doi.org/#2} {#1}\fi
  \endgroup}
\def\mn@eprint#1#2{\mn@eprint@#1:#2::\@nil}
\def\mn@eprint@arXiv#1{\href {http://arxiv.org/abs/#1} {{\tt arXiv:#1}}}
\def\mn@eprint@dblp#1{\href {http://dblp.uni-trier.de/rec/bibtex/#1.xml}
  {dblp:#1}}
\def\mn@eprint@#1:#2:#3:#4\@nil{\def\@tempa {#1}\def\@tempb {#2}\def\@tempc
  {#3}\ifx \@tempc \@empty \let \@tempc \@tempb \let \@tempb \@tempa \fi \ifx
  \@tempb \@empty \def\@tempb {arXiv}\fi \@ifundefined
  {mn@eprint@\@tempb}{\@tempb:\@tempc}{\expandafter \expandafter \csname
  mn@eprint@\@tempb\endcsname \expandafter{\@tempc}}}

\bibitem[\protect\citeauthoryear{Armillotta, Fraternali, Werk, Prochaska  \&
  Marinacci}{Armillotta et~al.}{2017}]{Armillotta2017}
Armillotta L.,  Fraternali F.,  Werk J.~K.,  Prochaska J.~X.,   Marinacci F.,
  2017, \mn@doi [MNRAS] {10.1093/mnras/stx1239}, 470, 114

\bibitem[\protect\citeauthoryear{Arrigoni-Battaia, Prochaska, Hennawi, Obreja,
  Buck, Cantalupo, Dutton  \& Macci{\`{o}}}{Arrigoni-Battaia
  et~al.}{2018}]{Battaia2018}
Arrigoni-Battaia F.,  Prochaska J.~X.,  Hennawi J.~F.,  Obreja A.,  Buck T.,
  Cantalupo S.,  Dutton A.~A.,   Macci{\`{o}} A.~V.,  2018, \mn@doi [MNRAS]
  {10.1093/mnras/stx2465}, 473, 3907

\bibitem[\protect\citeauthoryear{Aung, Mandelker, Nagai, Dekel  \&
  Birnboim}{Aung et~al.}{2019}]{Aung2019}
Aung H.,  Mandelker N.,  Nagai D.,  Dekel A.,   Birnboim Y.,  2019, \mn@doi
  [MNRAS] {10.1093/mnras/stz1964}, 490, 181

\bibitem[\protect\citeauthoryear{Baldry, Glazebrook, Brinkmann, Ivezi{\'{c}},
  Lupton, Nichol  \& Szalay}{Baldry et~al.}{2004}]{bimodal_Baldry2004}
Baldry I.~K.,  Glazebrook K.,  Brinkmann J.,  Ivezi{\'{c}} {\v{Z}}.,  Lupton
  R.~H.,  Nichol R.~C.,   Szalay A.~S.,  2004, \mn@doi [ApJ] {10.1086/380092},
  600, 681

\bibitem[\protect\citeauthoryear{Begelman}{Begelman}{1990}]{Begelman1990}
Begelman M.~C.,  1990, MNRAS, 000, 26

\bibitem[\protect\citeauthoryear{Behroozi, Wechsler, Hearin  \&
  Conroy}{Behroozi et~al.}{2019}]{Behroozi2019}
Behroozi P.,  Wechsler R.~H.,  Hearin A.~P.,   Conroy C.,  2019, \mn@doi
  [MNRAS] {10.1093/mnras/stz1182}, 488, 3143

\bibitem[\protect\citeauthoryear{Bell et~al.,}{Bell
  et~al.}{2004}]{bimodal_Bell2004}
Bell E.~F.,  et~al., 2004, \mn@doi [ApJ] {10.1086/420778}, 608, 752

\bibitem[\protect\citeauthoryear{Bennett \& Sijacki}{Bennett \&
  Sijacki}{2020}]{Bennett2020}
Bennett J.~S.,  Sijacki D.,  2020, \mn@doi [MNRAS] {10.1093/mnras/staa2835},
  499, 597

\bibitem[\protect\citeauthoryear{Berlok \& Pfrommer}{Berlok \&
  Pfrommer}{2019}]{bib_MHD_Berlok_2019b}
Berlok T.,  Pfrommer C.,  2019, \mn@doi [MNRAS] {10.1093/mnras/stz2347}, 489,
  3368

\bibitem[\protect\citeauthoryear{Blanton et~al.,}{Blanton
  et~al.}{2003}]{Blanton2003}
Blanton M.~R.,  et~al., 2003, \mn@doi [ApJ] {10.1086/375528}, 594, 186

\bibitem[\protect\citeauthoryear{Borisova et~al.,}{Borisova
  et~al.}{2016}]{Borisova2016}
Borisova E.,  et~al., 2016, \mn@doi [ApJ] {10.3847/0004-637x/831/1/39}, 831, 39

\bibitem[\protect\citeauthoryear{Bouch{\'{e}}, Murphy, Kacprzak, Contini  \&
  Martin}{Bouch{\'{e}} et~al.}{2013}]{Bouche2013}
Bouch{\'{e}} N.,  Murphy M.~T.,  Kacprzak G.~G.,  Contini T.,   Martin C.~L.,
  2013, Science, 341, 50

\bibitem[\protect\citeauthoryear{Bouch{\'{e}} et~al.,}{Bouch{\'{e}}
  et~al.}{2016}]{Bouche2016}
Bouch{\'{e}} N.,  et~al., 2016, \mn@doi [ApJ] {10.3847/0004-637x/820/2/121},
  820, 121

\bibitem[\protect\citeauthoryear{Br{\"{u}}ggen \& Scannapieco}{Br{\"{u}}ggen \&
  Scannapieco}{2023}]{Bruggen2023}
Br{\"{u}}ggen M.,  Scannapieco E.,  2023, {The Launching of Cold Clouds by
  Galaxy Outflows. V: The role of anisotropic thermal conduction}, preprint
  (\mn@eprint {arXiv} {2304.09881})

\bibitem[\protect\citeauthoryear{Cantalupo, Arrigoni-Battaia, Prochaska,
  Hennawi  \& Madau}{Cantalupo et~al.}{2014}]{Cantalupo2014}
Cantalupo S.,  Arrigoni-Battaia F.,  Prochaska J.~X.,  Hennawi J.~F.,   Madau
  P.,  2014, \mn@doi [Nature] {10.1038/nature12898}, 506, 63

\bibitem[\protect\citeauthoryear{Chen et~al.,}{Chen et~al.}{2020}]{Chen2020}
Chen Y.,  et~al., 2020, \mn@doi [MNRAS] {10.1093/mnras/staa2808}, 499, 1721

\bibitem[\protect\citeauthoryear{Cucciati et~al.,}{Cucciati
  et~al.}{2012}]{Cucciati2012}
Cucciati O.,  et~al., 2012, \mn@doi [{\aap}] {10.1051/0004-6361/201118010},
  539, 1

\bibitem[\protect\citeauthoryear{Daddi et~al.,}{Daddi et~al.}{2021}]{Daddi2021}
Daddi E.,  et~al., 2021, \mn@doi [\aap] {10.1051/0004-6361/202038700}, 649, A78

\bibitem[\protect\citeauthoryear{Daddi et~al.,}{Daddi
  et~al.}{2022a}]{Daddi2022b}
Daddi E.,  et~al., 2022a, \mn@doi [\aap] {10.1051/0004-6361/202243574}, 661, 1

\bibitem[\protect\citeauthoryear{Daddi et~al.,}{Daddi
  et~al.}{2022b}]{Daddi2022a}
Daddi E.,  et~al., 2022b, \mn@doi [\apjl.] {10.3847/2041-8213/ac531f}, 926, L21

\bibitem[\protect\citeauthoryear{Danovich, Dekel, Hahn, Ceverino  \&
  Primack}{Danovich et~al.}{2015}]{Danovich2015}
Danovich M.,  Dekel A.,  Hahn O.,  Ceverino D.,   Primack J.,  2015, \mn@doi
  [MNRAS] {10.1093/mnras/stv270}, 449, 2087

\bibitem[\protect\citeauthoryear{Dekel \& Birnboim}{Dekel \&
  Birnboim}{2006}]{Dekel2006}
Dekel A.,  Birnboim Y.,  2006, \mn@doi [MNRAS]
  {10.1111/j.1365-2966.2006.10145.x}, 368, 2

\bibitem[\protect\citeauthoryear{Dekel et~al.,}{Dekel et~al.}{2009}]{Dekel2009}
Dekel A.,  et~al., 2009, \mn@doi [Nature] {10.1038/nature07648}, 457, 451

\bibitem[\protect\citeauthoryear{Dekel, Zolotov, Tweed, Cacciato, Ceverino  \&
  Primack}{Dekel et~al.}{2013}]{Dekel2013}
Dekel A.,  Zolotov A.,  Tweed D.,  Cacciato M.,  Ceverino D.,   Primack J.~R.,
  2013, \mn@doi [MNRAS] {10.1093/mnras/stt1338}, 435, 999

\bibitem[\protect\citeauthoryear{Dijkstra}{Dijkstra}{2017}]{Dijkstra2017}
Dijkstra M.,  2017, {Saas-Fee Lecture Notes: Physics of Lyman Alpha Radiative
  Transfer}, preprint (\mn@eprint {arXiv} {1704.03416})

\bibitem[\protect\citeauthoryear{Dimotakis}{Dimotakis}{1991}]{Dimotakis1991}
Dimotakis P.~E.,  1991, Technical report, {Turbulent Free Shear Layer Mixing
  and Combustion}, \url
  {https://www.dimotakis.caltech.edu/pub/91/dimotakis.91d.html}.
AIAA, \url {https://www.dimotakis.caltech.edu/pub/91/dimotakis.91d.html}

\bibitem[\protect\citeauthoryear{Dolag, Kachelriess, Ostapchenko  \&
  Tom{\`{a}}s}{Dolag et~al.}{2011}]{bib_Mobs_Dolag_2011}
Dolag K.,  Kachelriess M.,  Ostapchenko S.,   Tom{\`{a}}s R.,  2011, \mn@doi
  [\apjl.] {10.1088/2041-8205/727/1/L4}, 727

\bibitem[\protect\citeauthoryear{Emonts et~al.,}{Emonts
  et~al.}{2023}]{Emonts2023}
Emonts B.~H.,  et~al., 2023, \mn@doi [Science] {10.1126/science.abh2150}, 379,
  1323

\bibitem[\protect\citeauthoryear{Fardal, Katz, Gardner, Hernquist, Weinberg  \&
  Dave}{Fardal et~al.}{2001}]{Fardal2001}
Fardal M.~A.,  Katz N.,  Gardner J.~P.,  Hernquist L.,  Weinberg D.~H.,   Dave
  R.,  2001, \mn@doi [ApJ] {10.1086/323519}, 562, 605

\bibitem[\protect\citeauthoryear{Faucher-Gigu\`{e}re \&
  Kere{\v{s}}}{Faucher-Gigu\`{e}re \& Kere{\v{s}}}{2011}]{Faucher-Giguere2011}
Faucher-Gigu\`{e}re C.~A.,  Kere{\v{s}} D.,  2011, \mn@doi [MNRAS]
  {10.1111/j.1745-3933.2011.01018.x}, 412, 118

\bibitem[\protect\citeauthoryear{Favre}{Favre}{1969}]{Favre1969}
Favre A.,  1969, Philadelphia: Society for Industrial and Applied Mathematics,
  pp 231--266

\bibitem[\protect\citeauthoryear{Ferland et~al.,}{Ferland
  et~al.}{2017}]{Ferland2017}
Ferland G.~J.,  et~al., 2017, \rmxaa, 53, 385

\bibitem[\protect\citeauthoryear{Fielding, Ostriker, Bryan  \& Jermyn}{Fielding
  et~al.}{2020}]{Fielding2020}
Fielding D.~B.,  Ostriker E.~C.,  Bryan G.~L.,   Jermyn A.~S.,  2020, \mn@doi
  [ApJ] {10.3847/2041-8213/ab8d2c}, 894, L24

\bibitem[\protect\citeauthoryear{Fu, Xue, Prochaska, Stockon, Ponnada, Lau,
  Cooray  \& Narayanan}{Fu et~al.}{2021}]{Fu2021}
Fu H.,  Xue R.,  Prochaska J.~X.,  Stockon A.,  Ponnada S.,  Lau M.~W.,  Cooray
  A.,   Narayanan D.,  2021, \mn@doi [ApJ] {10.3847/1538-4357/abdb32}, 908, 188

\bibitem[\protect\citeauthoryear{Fumagalli, Prochaska, Kasen, Dekel, Ceverino
  \& Primack}{Fumagalli et~al.}{2011}]{Fumagalli2011}
Fumagalli M.,  Prochaska J.~X.,  Kasen D.,  Dekel A.,  Ceverino D.,   Primack
  J.~R.,  2011, \mn@doi [MNRAS] {10.1111/j.1365-2966.2011.19599.x}, 418, 1796

\bibitem[\protect\citeauthoryear{Fumagalli, Cantalupo, Dekel, Morris, O'Meara,
  Prochaska  \& Theuns}{Fumagalli et~al.}{2016}]{Fumagalli2016}
Fumagalli M.,  Cantalupo S.,  Dekel A.,  Morris S.~L.,  O'Meara J.~M.,
  Prochaska J.~X.,   Theuns T.,  2016, \mn@doi [MNRAS] {10.1093/mnras/stw1782},
  462, 1978

\bibitem[\protect\citeauthoryear{Gardiner \& Stone}{Gardiner \&
  Stone}{2005}]{Gardiner2005}
Gardiner T.~A.,  Stone J.~M.,  2005, \mn@doi [J. Comput. Phys.]
  {10.1016/j.jcp.2004.11.016}, 205, 509

\bibitem[\protect\citeauthoryear{Gardiner \& Stone}{Gardiner \&
  Stone}{2008}]{Gardiner2008}
Gardiner T.~A.,  Stone J.~M.,  2008, \mn@doi [J. Comput. Phys.]
  {https://doi.org/10.1016/j.jcp.2007.12.017}, 227, 4123

\bibitem[\protect\citeauthoryear{Giavalisco et~al.,}{Giavalisco
  et~al.}{2011}]{Giavalisco2011}
Giavalisco M.,  et~al., 2011, \mn@doi [ApJ] {10.1088/0004-637X/743/1/95}, 743,
  95

\bibitem[\protect\citeauthoryear{Goerdt, Dekel, Sternberg, Ceverino, Teyssier
  \& Primack}{Goerdt et~al.}{2010}]{Goerdt2010}
Goerdt T.,  Dekel A.,  Sternberg A.,  Ceverino D.,  Teyssier R.,   Primack
  J.~R.,  2010, \mn@doi [MNRAS] {10.1111/j.1365-2966.2010.16941.x}, 407, 613

\bibitem[\protect\citeauthoryear{Gronke \& {Oh}}{Gronke \&
  {Oh}}{2018}]{Gronke2018}
Gronke M.,  {Oh} S.~P.,  2018, \mn@doi [MNRAS] {10.1093/mnrasl/sly131}, 480,
  L111

\bibitem[\protect\citeauthoryear{Gronke \& {Oh}}{Gronke \&
  {Oh}}{2020}]{Gronke2020}
Gronke M.,  {Oh} S.~P.,  2020, \mn@doi [MNRAS] {10.1093/mnras/stz3332}, 492,
  1970

\bibitem[\protect\citeauthoryear{Gruppioni et~al.,}{Gruppioni
  et~al.}{2013}]{Gruppioni2013}
Gruppioni C.,  et~al., 2013, \mn@doi [MNRAS] {10.1093/mnras/stt308}, 432, 23

\bibitem[\protect\citeauthoryear{Haardt \& Madau}{Haardt \&
  Madau}{2012}]{Haardt2012}
Haardt F.,  Madau P.,  2012, \mn@doi [ApJ] {10.1088/0004-637X/746/2/125}, 746

\bibitem[\protect\citeauthoryear{Hidalgo-Pineda, Farber  \&
  Gronke}{Hidalgo-Pineda et~al.}{2023}]{Hidalgo-Pineda2023}
Hidalgo-Pineda F.,  Farber R.~J.,   Gronke M.,  2023, {Better Together: The
  Complex Interplay Between Radiative Cooling and Magnetic Draping}, preprint
  (\mn@eprint {arXiv} {2304.09897})

\bibitem[\protect\citeauthoryear{Hopkins, Chan, Garrison-kimmel, Ji, Su,
  Hummels  \& Quataert}{Hopkins et~al.}{2020}]{Hopkins2020}
Hopkins P.~F.,  Chan T.~K.,  Garrison-kimmel S.,  Ji S.,  Su K.-y.,  Hummels
  C.~B.,   Quataert E.,  2020, \mn@doi [MNRAS] {10.1093/mnras/stz3321}, 492,
  3465

\bibitem[\protect\citeauthoryear{Hummels et~al.,}{Hummels
  et~al.}{2019}]{Hummels2019}
Hummels C.~B.,  et~al., 2019, \mn@doi [ApJ] {10.3847/1538-4357/ab378f}, 882,
  156

\bibitem[\protect\citeauthoryear{Ji, Oh  \& Masterson}{Ji
  et~al.}{2019}]{Ji2019}
Ji S.,  Oh S.~P.,   Masterson P.,  2019, \mn@doi [MNRAS]
  {10.1093/mnras/stz1248}, 487, 737

\bibitem[\protect\citeauthoryear{Kauffmann et~al.,}{Kauffmann
  et~al.}{2003}]{Kauffmann2003}
Kauffmann G.,  et~al., 2003, \mnras, 341, 33

\bibitem[\protect\citeauthoryear{Kere{\v{s}}, Katz, Weinberg  \&
  Dav{\'{e}}}{Kere{\v{s}} et~al.}{2005}]{Keres2005}
Kere{\v{s}} D.,  Katz N.,  Weinberg D.~H.,   Dav{\'{e}} R.,  2005, \mn@doi
  [MNRAS] {10.1111/j.1365-2966.2005.09451.x}, 363, 2

\bibitem[\protect\citeauthoryear{Kwak \& Shelton}{Kwak \&
  Shelton}{2010}]{Kwak2010}
Kwak K.,  Shelton R.~L.,  2010, \mn@doi [ApJ] {10.1088/0004-637X/719/1/523},
  719, 523

\bibitem[\protect\citeauthoryear{Lan \& Prochaska}{Lan \&
  Prochaska}{2020}]{bib_Mobs_Lan_2020}
Lan T.~W.,  Prochaska J.~X.,  2020, \mn@doi [MNRAS] {10.1093/mnras/staa1750},
  496, 3142

\bibitem[\protect\citeauthoryear{Madau \& Dickinson}{Madau \&
  Dickinson}{2014}]{Madau2014}
Madau P.,  Dickinson M.,  2014, \mn@doi [Annu. Rev. Astron. Astrophys.]
  {10.1146/annurev-astro-081811-125615}, 52, 415

\bibitem[\protect\citeauthoryear{Mandelker, Padnos, Dekel, Birnboim, Burkert,
  Krumholz  \& Steinberg}{Mandelker et~al.}{2016}]{Mandelker2016}
Mandelker N.,  Padnos D.,  Dekel A.,  Birnboim Y.,  Burkert A.,  Krumholz
  M.~R.,   Steinberg E.,  2016, \mn@doi [MNRAS] {10.1093/mnras/stw2267}, 463,
  3921

\bibitem[\protect\citeauthoryear{Mandelker, Nagai, Aung, Dekel, Padnos  \&
  Birnboim}{Mandelker et~al.}{2019}]{Mandelker2019}
Mandelker N.,  Nagai D.,  Aung H.,  Dekel A.,  Padnos D.,   Birnboim Y.,  2019,
  \mn@doi [MNRAS] {10.1093/mnras/stz012}, 484, 1100

\bibitem[\protect\citeauthoryear{Mandelker, Nagai, Aung, Dekel, Birnboim  \&
  {Van Den Bosch}}{Mandelker et~al.}{2020a}]{Mandelker2020a}
Mandelker N.,  Nagai D.,  Aung H.,  Dekel A.,  Birnboim Y.,   {Van Den Bosch}
  F.~C.,  2020a, \mn@doi [MNRAS] {10.1093/MNRAS/STAA812}, 494, 2641

\bibitem[\protect\citeauthoryear{Mandelker, {Van Den Bosch}, Nagai, Dekel,
  Birnboim  \& Aung}{Mandelker et~al.}{2020b}]{LyblobMandelker2020}
Mandelker N.,  {Van Den Bosch} F.~C.,  Nagai D.,  Dekel A.,  Birnboim Y.,
  Aung H.,  2020b, \mn@doi [MNRAS] {10.1093/mnras/staa2421}, 498, 2415

\bibitem[\protect\citeauthoryear{Martin-Alvarez, Devriendt, Slyz  \&
  Teyssier}{Martin-Alvarez et~al.}{2018}]{bib_MHD_Martin-Alvarez_2018}
Martin-Alvarez S.,  Devriendt J.,  Slyz A.,   Teyssier R.,  2018, \mn@doi
  [MNRAS] {10.1093/mnras/sty1623}, 479, 3343

\bibitem[\protect\citeauthoryear{Martin, Shapley, Coil, Kornei, Bundy, Weiner,
  Noeske  \& Schiminovich}{Martin et~al.}{2012}]{Martin2012}
Martin C.~L.,  Shapley A.~E.,  Coil A.~L.,  Kornei K.~A.,  Bundy K.,  Weiner
  B.~J.,  Noeske K.~G.,   Schiminovich D.,  2012, \mn@doi [ApJ]
  {10.1088/0004-637X/760/2/127}, 760, 127

\bibitem[\protect\citeauthoryear{Martin et~al.,}{Martin
  et~al.}{2019}]{ChristopherMartin2019}
Martin C.~D.,  et~al., 2019, Nat. Astron., 3, 822

\bibitem[\protect\citeauthoryear{Meyer, Balsara  \& Aslam}{Meyer
  et~al.}{2014}]{Meyer2014}
Meyer C.~D.,  Balsara D.~S.,   Aslam T.~D.,  2014, \mn@doi [J. Comput. Phys.]
  {https://doi.org/10.1016/j.jcp.2013.08.021}, 257, 594

\bibitem[\protect\citeauthoryear{Miyoshi \& Kusano}{Miyoshi \&
  Kusano}{2005}]{Miyoshi2005}
Miyoshi T.,  Kusano K.,  2005, \mn@doi [J. Comput. Phys.]
  {10.1016/j.jcp.2005.02.017}, 208, 315

\bibitem[\protect\citeauthoryear{Nelson et~al.,}{Nelson
  et~al.}{2018}]{Nelson2018}
Nelson D.,  et~al., 2018, \mn@doi [MNRAS] {10.1093/mnras/stx3040}, 475, 624

\bibitem[\protect\citeauthoryear{Nelson et~al.,}{Nelson
  et~al.}{2020}]{Nelson2020}
Nelson D.,  et~al., 2020, \mn@doi [MNRAS] {10.1093/mnras/staa2419}, 498, 2391

\bibitem[\protect\citeauthoryear{Neronov \& Vovk}{Neronov \&
  Vovk}{2010}]{bib_Mobs_Neronov_2010}
Neronov A.,  Vovk I.,  2010, Science, 328, 73

\bibitem[\protect\citeauthoryear{Ostriker}{Ostriker}{1964}]{Ostriker1964}
Ostriker J.~P.,  1964, ApJ, p.~1056

\bibitem[\protect\citeauthoryear{Padnos, Mandelker, Birnboim, Dekel, Krumholz
  \& Steinberg}{Padnos et~al.}{2018}]{Padnos2018}
Padnos D.,  Mandelker N.,  Birnboim Y.,  Dekel A.,  Krumholz M.~R.,   Steinberg
  E.,  2018, \mn@doi [MNRAS] {10.1093/mnras/sty789}, 477, 3293

\bibitem[\protect\citeauthoryear{Pakmor et~al.,}{Pakmor
  et~al.}{2017}]{bib_MHD_Pakmor_2017}
Pakmor R.,  et~al., 2017, \mn@doi [MNRAS] {10.1093/mnras/stx1074}, 469, 3185

\bibitem[\protect\citeauthoryear{Pakmor, Guillet, Pfrommer, G{\'{o}}mez, Grand,
  Marinacci, Simpson  \& Springel}{Pakmor et~al.}{2018}]{bib_MHD_Pakmor_2018}
Pakmor R.,  Guillet T.,  Pfrommer C.,  G{\'{o}}mez F.~A.,  Grand R.~J.,
  Marinacci F.,  Simpson C.~M.,   Springel V.,  2018, \mn@doi [MNRAS]
  {10.1093/mnras/sty2601}, 481, 4410

\bibitem[\protect\citeauthoryear{Pakmor et~al.,}{Pakmor
  et~al.}{2020}]{bib_MHD_Pakmor_2020}
Pakmor R.,  et~al., 2020, \mn@doi [MNRAS] {10.1093/mnras/staa2530}, 498, 3125

\bibitem[\protect\citeauthoryear{Peeples et~al.,}{Peeples
  et~al.}{2019}]{Peeples2019}
Peeples M.~S.,  et~al., 2019, \mn@doi [MNRAS] {10.3847/1538-4357/ab0654}, 873,
  129

\bibitem[\protect\citeauthoryear{Prochaska et~al.,}{Prochaska
  et~al.}{2019}]{bib_Mobs_Prochaska_2019}
Prochaska X.~J.,  et~al., 2019, Science, 73, 1

\bibitem[\protect\citeauthoryear{Reddy \& Steidel}{Reddy \&
  Steidel}{2009}]{Reddy2009}
Reddy N.~A.,  Steidel C.~C.,  2009, \mn@doi [ApJ]
  {10.1088/0004-637X/692/1/778}, 692, 778

\bibitem[\protect\citeauthoryear{Rieder \& Teyssier}{Rieder \&
  Teyssier}{2017}]{bib_MHD_Rieder_2017a}
Rieder M.,  Teyssier R.,  2017, \mn@doi [MNRAS] {10.1093/mnras/stx1670}, 471,
  2674

\bibitem[\protect\citeauthoryear{Roca-Fabrega et~al.,}{Roca-Fabrega
  et~al.}{2019}]{Roca-Fabrega2019}
Roca-Fabrega S.,  et~al., 2019, \mn@doi [Mon] {10.1093/mnras/stz063}, 484, 3625

\bibitem[\protect\citeauthoryear{Rubin, Prochaska, Koo  \& Phillips}{Rubin
  et~al.}{2012}]{Rubin2012}
Rubin K. H.~R.,  Prochaska J.~X.,  Koo D.~C.,   Phillips A.~C.,  2012, \mn@doi
  [\apjl.] {10.1088/2041-8205/747/2/L26}, 747, 26

\bibitem[\protect\citeauthoryear{Sander \& Hensler}{Sander \&
  Hensler}{2021}]{Sander2021}
Sander B.,  Hensler G.,  2021, \mn@doi [MNRAS] {10.1093/mnras/staa3952}, 501,
  5330

\bibitem[\protect\citeauthoryear{Saveliev, Jedamzik  \& Sigl}{Saveliev
  et~al.}{2012}]{bib_Mth_Saveliev_2012}
Saveliev A.,  Jedamzik K.,   Sigl G.,  2012, \mn@doi [Phys. Rev. D]
  {10.1103/PhysRevD.86.103010}, 86, 1

\bibitem[\protect\citeauthoryear{Sharma \& Hammett}{Sharma \&
  Hammett}{2007}]{Sharma2007}
Sharma P.,  Hammett G.~W.,  2007, \mn@doi [J. Comput. Phys.]
  {10.1016/j.jcp.2007.07.026}, 227, 123

\bibitem[\protect\citeauthoryear{Slavin, Shull  \& Begelman}{Slavin
  et~al.}{1993}]{Slavin1993}
Slavin J.~D.,  Shull M.~J.,   Begelman M.~C.,  1993, ApJ, 407, 83

\bibitem[\protect\citeauthoryear{Spitzer}{Spitzer}{1962}]{Spitzer1962}
Spitzer L.,  1962, Physics of Fully Ionized Gases.
Interscience, New York

\bibitem[\protect\citeauthoryear{Spruit}{Spruit}{2013}]{Spruit2013}
Spruit H.~C.,  2013, {Essential Magnetohydrodynamics for Astrophysics},
  preprint (\mn@eprint {} {1301.5572})

\bibitem[\protect\citeauthoryear{Steinwandel, Beck, Arth, Dolag, Moster  \&
  Nielaba}{Steinwandel et~al.}{2019}]{bib_MHD_Steinwandel_2019}
Steinwandel U.~P.,  Beck M.~C.,  Arth A.,  Dolag K.,  Moster B.~P.,   Nielaba
  P.,  2019, \mn@doi [MNRAS] {10.1093/mnras/sty3083}, 483, 1008

\bibitem[\protect\citeauthoryear{Stone, Tomida, White  \& Felker}{Stone
  et~al.}{2020}]{Stone2020}
Stone J.~M.,  Tomida K.,  White C.~J.,   Felker K.~G.,  2020, \mn@doi [ApJ]
  {10.3847/1538-4365/ab929b}, 249, 4

\bibitem[\protect\citeauthoryear{Strateva et~al.,}{Strateva
  et~al.}{2001}]{Strateva2001}
Strateva I.,  et~al., 2001, \mn@doi [\aj] {10.1086/323301}, 122, 1861

\bibitem[\protect\citeauthoryear{Strawn et~al.,}{Strawn
  et~al.}{2021}]{Strawn2021}
Strawn C.,  et~al., 2021, \mn@doi [MNRAS] {10.1093/mnras/staa3972}, 501, 4948

\bibitem[\protect\citeauthoryear{Suresh, Nelson, Genel, Rubin  \&
  Hernquist}{Suresh et~al.}{2019}]{Suresh2019}
Suresh J.,  Nelson D.,  Genel S.,  Rubin K.~H.,   Hernquist L.,  2019, \mn@doi
  [MNRAS] {10.1093/mnras/sty3402}, 483, 4040

\bibitem[\protect\citeauthoryear{Tan, {Oh}  \& Gronke}{Tan
  et~al.}{2021}]{Tan2021}
Tan B.,  {Oh} S.~P.,   Gronke M.,  2021, \mn@doi [MNRAS]
  {10.1093/mnras/stab053}, 502, 3179

\bibitem[\protect\citeauthoryear{Turner, Schaye, Crain, Rudie, Steidel, Strom
  \& Theuns}{Turner et~al.}{2017}]{Turner2017}
Turner M.~L.,  Schaye J.,  Crain R.~A.,  Rudie G.,  Steidel C.~C.,  Strom A.,
  Theuns T.,  2017, \mn@doi [MNRAS] {10.1093/mnras/stx1616}, 471, 690

\bibitem[\protect\citeauthoryear{Umehata et~al.,}{Umehata
  et~al.}{2019}]{Umehata2019}
Umehata H.,  et~al., 2019, \mn@doi [Science (80-. ).]
  {10.1126/science.aaw5949}, 366, 97

\bibitem[\protect\citeauthoryear{{Van De Voort} \& Schaye}{{Van De Voort} \&
  Schaye}{2012}]{VanDeVoort2012}
{Van De Voort} F.,  Schaye J.,  2012, \mn@doi [MNRAS]
  {10.1111/j.1365-2966.2012.20949.x}, 423, 2991

\bibitem[\protect\citeauthoryear{Vossberg, Cantalupo  \& Pezzulli}{Vossberg
  et~al.}{2019}]{Vossberg2019}
Vossberg A. C.~E.,  Cantalupo S.,   Pezzulli G.,  2019, \mn@doi [MNRAS]
  {10.1093/mnras/stz2276}, 489, 2130

\bibitem[\protect\citeauthoryear{Yang \& Ji}{Yang \& Ji}{2023}]{Yang2023}
Yang Y.,  Ji S.,  2023, \mn@doi [MNRAS] {10.1093/mnras/stad264}, 520, 2148

\bibitem[\protect\citeauthoryear{Zabl et~al.,}{Zabl et~al.}{2019}]{Zabl2019}
Zabl J.,  et~al., 2019, \mn@doi [MNRAS] {10.1093/mnras/stz392}, 485, 1961

\bibitem[\protect\citeauthoryear{Zhang et~al.,}{Zhang et~al.}{2023}]{Zhang2023}
Zhang S.,  et~al., 2023, \mn@doi [Science (80-. ).] {10.1126/science.abj9192},
  380, 494

\bibitem[\protect\citeauthoryear{van~de Voort, Springel, Mandelker, van~den
  Bosch  \& Pakmor}{van~de Voort et~al.}{2019}]{VandeVoort2019}
van~de Voort F.,  Springel V.,  Mandelker N.,  van~den Bosch F.~C.,   Pakmor
  R.,  2019, \mn@doi [MNRAS] {10.1093/mnrasl/sly190}, 482, L85

\makeatother
\end{thebibliography}


\newpage
\appendix

\section{Resolution test}\label{app:Res}
We conduct a convergence study of our simulations, focusing on a density ratio $\delta =100$, a Mach number $\mathcal{M}=1$, and a magnetic field angle $\alpha=45^\circ$. 
For both purely MHD and MHD with TC cases,
the number density is varied as $n_{\stream}={0.001,0.01,0.1} \rm cm^{-3}$, and the corresponding $\xi$ values are provided in Table~\ref{tab:Conv_tab}. 

We utilise grids with minimal resolution $\Delta_{\rm{min}}/R_{\stream}={1/32,1/16,1/8,1/4}$. 
Due to the use of static mesh refinement, each simulation employs one less level of refinement, resulting in a resolution in the stream region that is twice coarser.
Furthermore, as described in Sec.~\ref{sec3:ICs}, we apply smoothing to the interface using $h=R_{\stream}/32$, which determines the penetration depth of the perturbations in Equation~\ref{eq:init_2}. To ensure that the penetration depth remains above the minimum mesh size, the parameter $h$ is scaled by a factor of two between each grid enlargement.

\begin{figure*}
	\includegraphics[width=2.0\columnwidth]{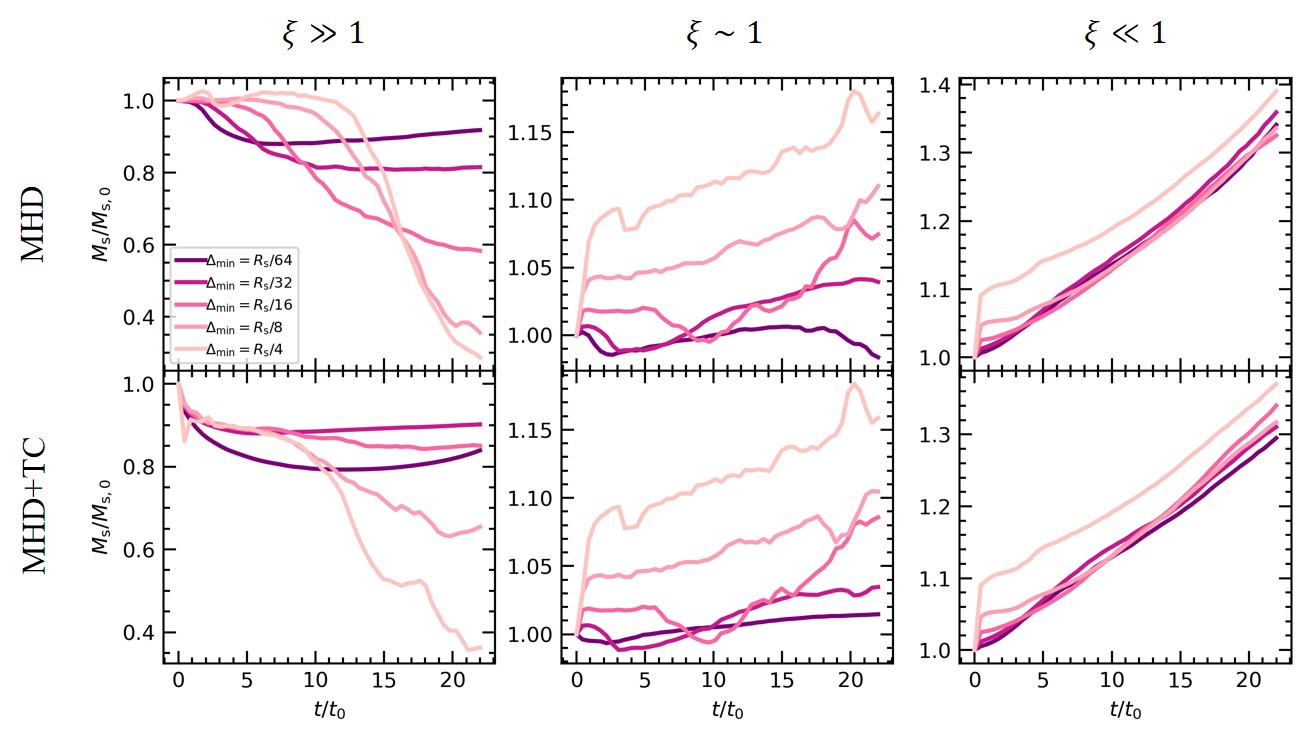}
    \caption{Convergence test results for both MHD (top panels) and MHD+TC (bottom rows) cases. Mass evolution of the stream in units of initial stream mass is shown as a function of time for different $\xi$ values.}
    \label{fig:A1_Mass_all}
\end{figure*}

Fig.~\ref{fig:A1_Mass_all} illustrates the temporal evolution of the stream mass.  Each panel depicts the fiducial (finest grid) and coarser grid cases using a dark to pale color scale, respectively. The convergence quality varies across the different regimes. 

In the {\it condensing stream} regime ($\xi \ll 1$), the results exhibit good agreement starting from $\Delta_{\rm{min}}=R_{\stream}/8$.  However, in the other regimes, the convergence is not yet fully achieved qualitatively by $\Delta_{\rm{min}}=R_{\stream}/32$.

Note that the scaling of the parameter $h$ also enlarges the size of the smoothing region. In the case of the {\it diffusing stream} regime, this enlargement leads to a modification in the initial temperature gradient, resulting in a slight impact on the stream diffusion. This explains the small break in convergence,  as from $\Delta_{\rm{min}}=R_{\stream}/4$ to $\Delta_{\rm{min}}=R_{\stream}/32$, the results start converging, but between $\Delta_{\rm{min}}=R_{\stream}/32$ and $\Delta_{\rm{min}}=R_{\stream}/64$, the difference between solutions suddenly increases. 
To further investigate this effect, we conducted simulations without scaling the parameter $h$.

To assess the quantitative convergence, we introduce the time-averaged relative error of a given quantity $X$ as follows: 
\begin{equation}
    \epsilon_{\rm{i}} =  \left\langle \frac{ X_{\rm{i}} - X_0 }{   X_0 } \right\rangle,
\end{equation}
where the subscript $\rm i$ represents the coarsening factor of the simulation, defined for $i=0,1,2,3,4$, such that  $\Delta_{\rm{min}}\left(i\right) / R_{\stream} =\left(1/64\right)\times 2^i$. The brackets indicate the arithmetic-linear time averaging over all simulation snapshots. 
This relative error is computed with respect to the fiducial value $X_0$ obtained from the finest resolution simulation.

\begin{table}
\centering
\caption{Simulation runs used for convergence test, along with the relative error of the stream mass expressed as a percentage of the fiducial value (finest grid case).  In the last row, the simulation with TC in the {\it diffusing stream} regime is shown, where the scaling of $h$ is removed.  Note that a good convergence is achieved when the scaling of $h$ is removed, as opposed to when it is kept.
}
\label{tab:Conv_tab}
\begin{tabular}{ll|cccc}
\hline
Runs & $\xi$ & $\epsilon_{1}$ & $\epsilon_{2}$ & $\epsilon_{3}$  & $\epsilon_{4}$ \\
 & & $\left[ \% \right]$ & $\left[ \% \right]$ & $ \left[ \% \right]$  & $\left[ \% \right]$ \\
\hline
MHD&$10^{-2.0}$&$0.9$&$0.6$&$0.8$&$5.3$ \\ 
MHD+TC&$10^{-2.0}$&$1.0$&$1.3$&$1.4$&$6.1$ \\ 
MHD&$10^{-0.5}$&$1.8$&$2.9$&$6.6$&$11.7$ \\ 
MHD+TC&$10^{-0.5}$&$1.1$&$2.3$&$5.6$&$10.8$ \\ 
MHD&$10^{1.3}$&$6.7$&$17.5$&$20.6$&$22.5$ \\ 
MHD+TC&$10^{1.3}$&$9.0$&$6.5$&$9.5$&$20.3$ \\ 
\hline
No scaling : & & & & &\\
MHD+TC &$10^{1.3}$&$0.5$&$0.6$&$3.4$&$3.4$ \\ 
\hline
\end{tabular}
\end{table}

The resulting relative error, expressed as a percentage of the fiducial value (finest grid case), is presented in Table~\ref{tab:Conv_tab} for the stream mass. 
In all cases and when considering the no-scaling case for the MHD+TC simulation in the {\it diffusing stream} regime ($\xi=10^{1.3}$), the mean error decreases with increasing resolution reaching a maximum of $\sim 7 \%$ by $\Delta_{\rm{min}}=R_{\stream}/32$.
The purely MHD simulation in the {\it disrupting stream} regime ($\xi \sim 10^{1.3}$) exhibits a relatively high error percentage at $\Delta_{\rm{min}}=R_{\stream}/32$. 
However, for all the other simulations, the errors are below a few percent, indicating a relatively good convergence.

The results obtained with and without scaling of the parameter $h$ (corresponding to the last two rows in the table) exhibit significant differences.  In this specific case, removing the scaling of $h$ results in the best convergence.  This finding confirms that the broadening of the smooth interface between the stream and the CGM is responsible for the discrepancy observed in Fig.~\ref{fig:A1_Mass_all}. 

In conclusion, our analysis indicates that all simulations, except for the MHD simulations in the {\it disrupting stream} regime ($\xi \gg 1$), demonstrate good convergence with relative errors below a few percent. 
For the MHD simulations in the {\it disrupting stream} regime, the fact that the simulations do not yet fully converge may change within $7\%$ percent the quantitative conclusion of our work for this particular case.

\section{Nature of the cooling emission}\label{app:Lnet_nature}
Complementary to the discussion in Sec.~\ref{sec5:Rad}, we hereby describe the contribution of Hydrogen to the total net cooling emission. From the cooling model described in Sec.~\ref{sec2:cooling}, we compute the ratio of the cooling emission from hydrogen over the total cooling emission for the mixing layer. This ratio is important for estimating the portion of the emission associated with Ly$\alpha$. Both cooling emissions are computed using equation \ref{eq:Lnet}.

\begin{figure}
	\includegraphics[width=1.0\columnwidth]{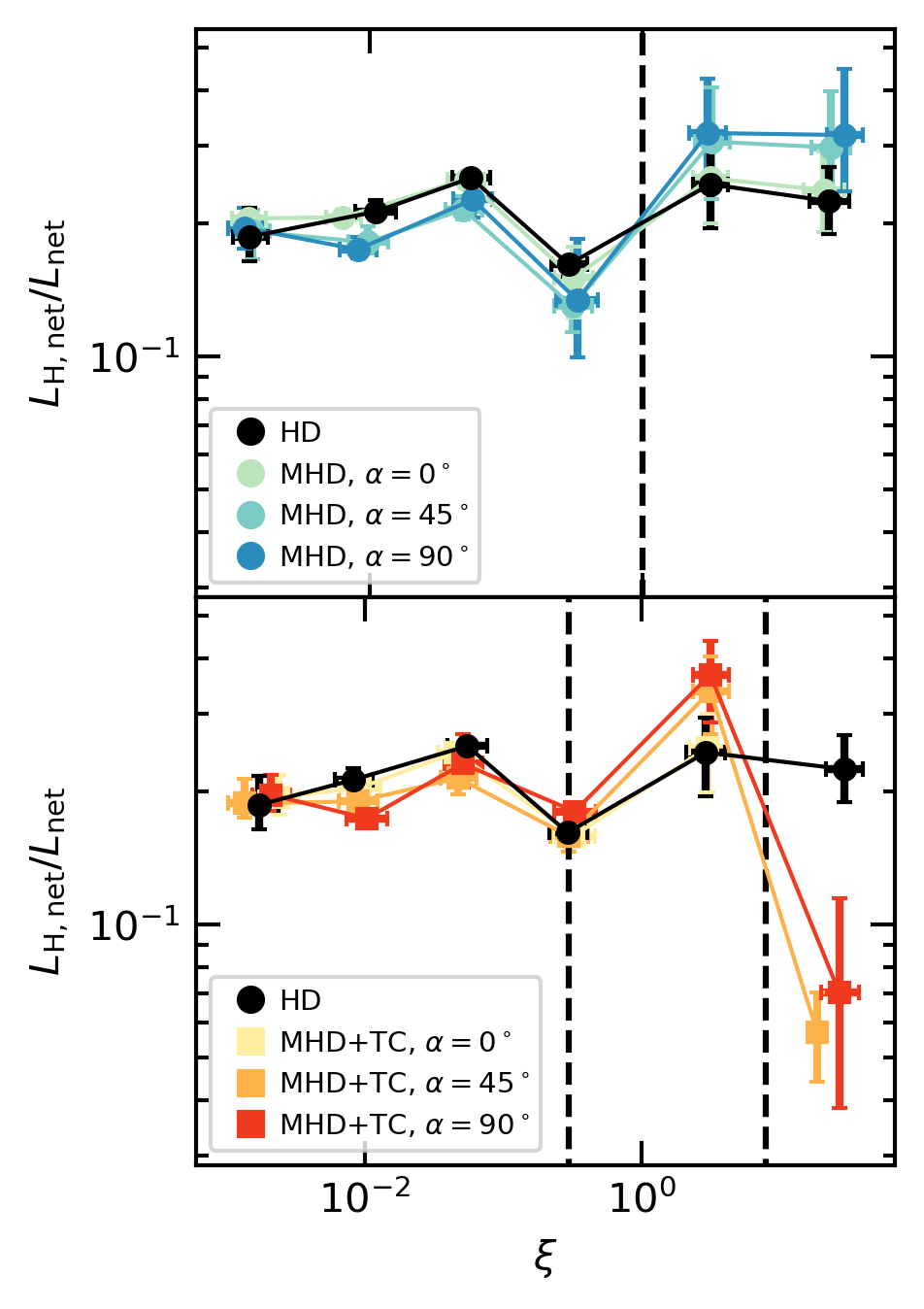}
    \caption{Ratio of the cooling emission from Hydrogen ($L_{\text{H}}$) over the total cooling emission for the mixing layer ($L_{\rm{net}}$) as a function of $\xi$ is shown. {\it Top panel}: Comparison between HD and MHD simulations. The vertical black dashed line represents the boundary between the {\it condensing stream} regime ($\xi<1$) and the {\it disrupting stream} regime. {\it Bottom panel}:  Comparison between HD and MHD+TC simulations. The black dashed lines represent the boundaries between the {\it condensing stream} regime ($\xi<0.3$), the {\it intermediate} regime ($0.3<\xi<8$), and the {\it diffusing stream} regime ($\xi>8$). 
    Each point represents an averaged value over the three Mach numbers $\mathcal{M}=0.5,1,2$, and the error bars indicate the range of minimum and maximum values. A small random offset is applied in the horizontal direction to improve visibility and distinguish each point.}
    \label{fig:rLH_Lnet}
\end{figure}

Fig.~\ref{fig:rLH_Lnet} compares the ratio $L_{\rm{H,net}} / L_{\rm net}$ between HD and MHD, and between HD and MHD+TC simulations as a function of $\xi=t_{\rm{cool}}/t_{\rm{sh}}$. For HD simulations, the ratio remains roughly constant across $\xi$ with a value $\sim 0.2$, showing little dependency on the stream velocity. The MHD simulations impact this ratio differently depending on the stream regime and for angles $\alpha \neq 0^\circ$. 
In the {\it condensing stream} regime ($\xi<1$), the hydrogen contribution is slightly lower, while in the {\it disrupting stream} regime ($\xi>1$), the hydrogen contribution increases to around $0.3$. It is worth noting that for increasing $\xi$, the stream velocities also impact the hydrogen contribution, resulting in $L_{\rm{H,net}} / L_{\rm net}$ up to $\sim 0.45$.
Thermal conduction does not significantly alter the results in the {\it condensing stream} regime. In the {\it intermediate} regime, MHD+TC appears to raise $L_{\rm{H,net}} / L_{\rm net} \lesssim 0.4$ as $\xi$ increases. Finally, in the {\it diffusing stream} regime, the hydrogen contribution in the cooling emissions drastically decreases to the range of $0.03\text{--}0.1$, depending on the stream velocity and the angle $\alpha$, as both parameters directly act on the conduction efficiency.

\section{Temperature thresholds}\label{app:Mass_thr}
We hereby checked that the temperature thresholds introduced in Sec.~\ref{sec4:stream_evo} are acceptable to compute the stream mass variables and the mixing layer variables.

\begin{figure}
	\includegraphics[width=1.0\columnwidth]{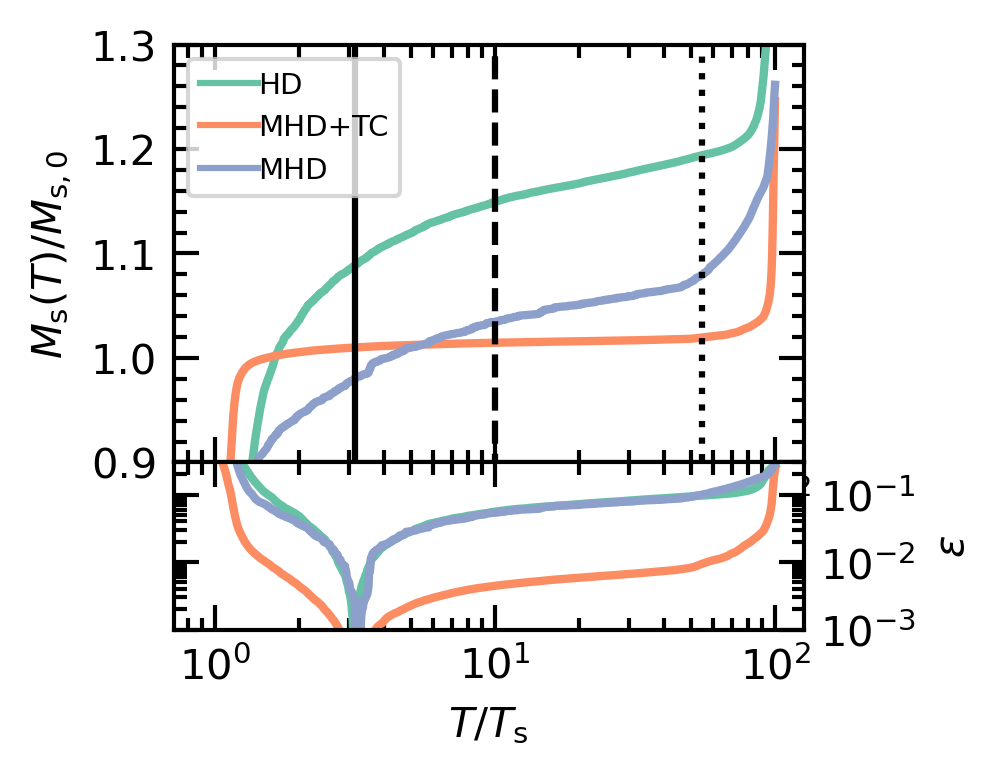} \\
	\includegraphics[width=1.0\columnwidth]{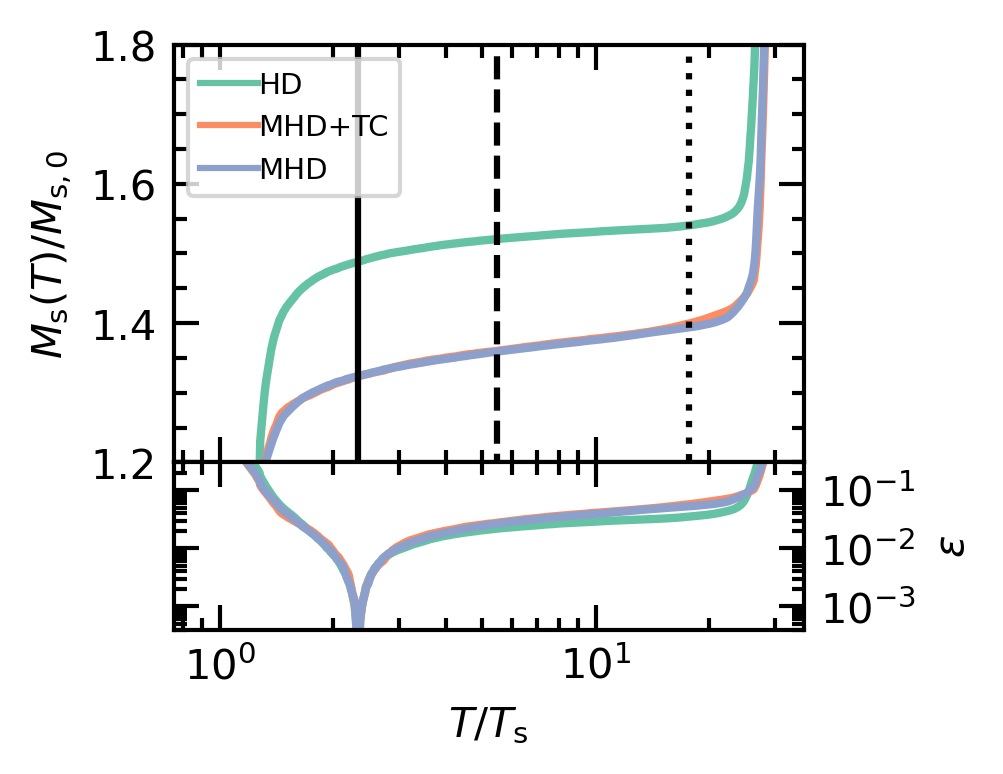}
    \caption{Mass threshold check. Mass in the all simulation domain encompassed inside under a given temperature in function of this temperature (upper subplot). The bottom curves show the relative error between our stream mass and the mass computed with the given temperature. (Top) HD, MHD, and MHD+TC simulations for $\delta=100$. (Down) HD, MHD, and MHD+TC simulations for $\delta=30$. In both panels, our chosen stream temperature threshold $T_{\rm{l}}$, $T_{\rm{mix}}$ and our CGM temperature threshold $T_{\rm{u}}$ are plotted with plain, dashed and dotted black lines. The relative error with respect to the mass $M_{\stream}\left(T\right)$ is shown beneath each panel.}
    \label{fig:A2_Mass_Thr}
\end{figure}
In Fig.~\ref{fig:A2_Mass_Thr}, we show the total mass of gas under a given temperature $T$ for HD, MHD, and MHD+TC simulations at $\mathcal{M}=1$ for density ratios $\delta = \rho_{\stream}/\rho_{\cgm}=100,30$. In each panels are plotted the thresholds $T_{\rm{l}}=(T_{\stream}T_{\rm{mix}})^{1/2}$, $T_{\rm{mix}}$, and $T_{\rm{u}}=(T_{\cgm}+T_{\rm{mix}})/2$. The relative error,
\begin{equation}
    \epsilon\left( T\right) =  \left\lVert \frac{ M_{\stream}\left( T\right) - M_{\stream}\left( T_{\rm{l}}\right) }{   M_{\stream}\left( T_{\rm{l}}\right) } \right\rVert,
\end{equation}
is also plotted beneath each panel.
For all curves, the mass encompassed inside the stream, the mixing layer and the CGM are clearly identifiable. The first rise of the mass slightly after $T_s$ results from the stream gas. The relatively flattened phase represents the gas in the mixing layer. The final rise around $T_{\cgm}=\delta T_{\stream}$ comes from the gas in the CGM.
The threshold mass $M_{\stream}\left(T_{\rm{l}}\right)$ seems acceptable as in all cases, variations of the threshold to temperatures $2T_{\stream}$ or $3T_{\stream}$ would change the mass by a few percent. Such an argument can also be made for the threshold $T_{\rm u}$. The only drawback is that such a threshold might introduce a systemic error discriminating mixing layer gas as stream gas. However, such error seems relatively small, and as it is systemic, it would not change significantly our results qualitatively.

Finally, Fig.~\ref{app:Mass_thr} can also give some useful insight about the physics difference discussed in Sec.~\ref{sec4:stream_evo}. The simulations with $\delta=30$ are in the {\it condensing stream} regimes, thus the similar results between HD, MHD and MHD+TC cases. However, for $\delta = 100$, the simulations are in the {\it intermediate} regime (MHD+TC) or the {\it condensing stream} regime (HD, and MHD). The impact of thermal conduction is directly visible as for the MHD+TC simulations, the mixing layer phase is completely flattened, meaning the absence of gas in this phase due to its diffusion which then causes the decrease of cooling emission in Fig.~\ref{fig:Lnet_comp}.

\section{Modelling of the Magnetic field growth}\label{app:Mag_Mod}
We hereby describe the modelling of the magnetic field amplification due to the shear velocity, which is the main mechanism responsible for the magnetic field growth shown in Sec.~\ref{sec4:mag}.
As our magnetic field is initially uniform, a simple way to consider the growth of the magnetic field is to focus solely on the direction of a magnetic field line and model it as a flux tube \citep[][]{Spruit2013} with an initial density $\rho_{\rm i}\left(x\right)$ and length $l_{\rm i}$.

The total length of the field line can be approximated as $l\left(t\right) \sim l_{\rm i} + v_{\stream,0}t$ with the elongation occurring only near the interface of the stream and the CGM such that $l_{\rm i}/R_{\stream} \sim 1$. We focus here on the portion of the field line, which is then elongated. As the stream moves forward, the line will bend at the interface between the stream and the CGM. We assume that this bending only happens for an initial portion of the line of length $R_{\stream}$ located between each side of the interface $[ 0.5R_{\stream},1.5R_{\stream}] $ \footnote{This is relatively consistent with our simulations, as in Fig.~\ref{fig:BfieldContour}, for a given stream cross-section, the amplification happens around the interface, and it does not occur near the stream centre.}.
As the field is frozen into the flow, the total mass along the field line and the magnetic flux should remain constant,
\begin{equation}\label{eq:tubeM}
    M\left(x\right) = \rho\left(x,t\right)\sigma\left(t\right)l\left(t\right) = M_0, \, \, \, \text{and}, \Phi = \sigma\left(t\right)B\left(t\right)=\Phi_0,
\end{equation}
where $\sigma(t)$ is the cross-section of the tube.

The total pressure should remain constant as $p_{\rm T}= p+B^2/2=p_0+B_0^2/2$. Considering a cold stream without perturbations, we can also assume the tube to have a steady temperature $T\left(x,t\right)= T_0\left(x\right)$.
Using the ideal gas law and introducing equation \ref{eq:tubeM} into the total pressure term, one may obtain the following solution for the magnetic field growth,
\begin{equation}\label{eq:Bmodel}
    \frac{B\left(t\right)}{B_0} = \frac{1}{2}\left[ \left(\beta \frac{l_{\rm{i}}}{l}\right)^2 + 4\left(\beta+1\right)\right]^{1/2} - \frac{1}{2} \beta \frac{l_{\rm{i}}}{l} ,
\end{equation}
where $\beta = p_0/\left(B_0^2/2\right)=10^5$ represents the ratio of the thermal over the magnetic pressure. It is important to note that in Equation \ref{eq:Bmodel}, the time dependency is contained inside $l$.

\begin{figure}
	\includegraphics[width=1.0\columnwidth]{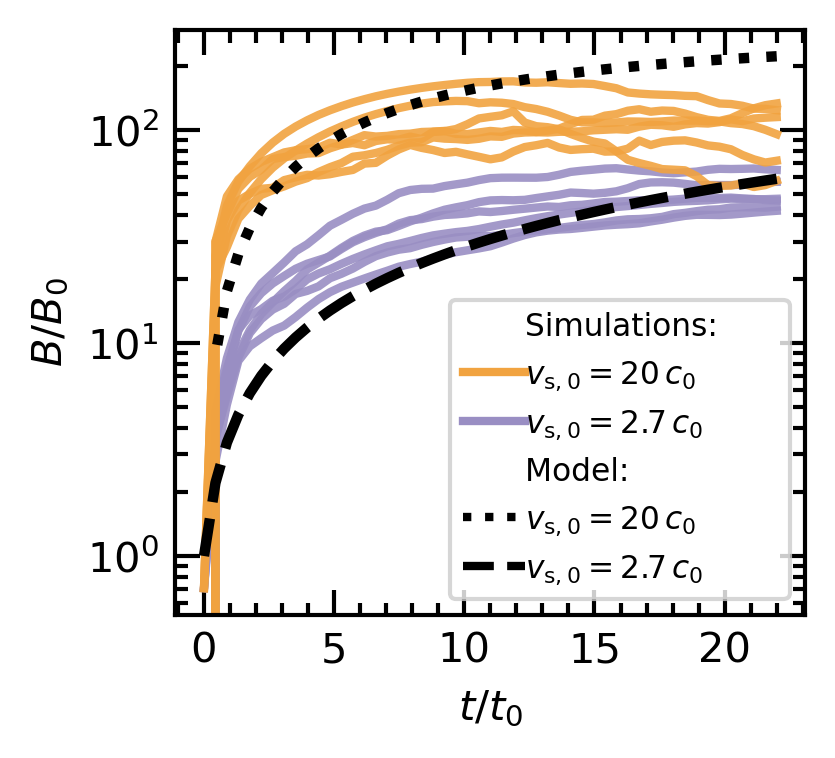}
    \caption{Comparison of the magnetic field evolution between the simulations and the model from Equation \ref{eq:Bmodel}. For clarity, we plot only MHD simulations, focusing on the maximum (orange) and minimum (blue-navy) expected growth cases, i.e., the cases with the highest and lowest stream velocities. All the lines for a same color are MHD simulations with the given $v_{\stream ,0}$ for $\alpha = 45^\circ, 90^\circ$ and $n_{\stream} = 10^{-3},10^{-2},10^{-1} \, \rm cm^{-3}$ (thus 6 lines per color).}
    \label{fig:B1_Bmodel}
\end{figure}

We compare the model with the results of the simulations in Fig.~\ref{fig:B1_Bmodel}. The magnetic field from the simulations is averaged inside the stream and the mixing layer as explained in Sec.~\ref{sec4:mag}. For clarity, the figure shows only the simulations where the magnetic field growth is expected to be the highest and lowest, i.e., the highest and lowest stream velocities, which correspond to simulations with $\mathcal{M}=2, \delta=100$ and $\mathcal{M}=0.5, \delta=30$, respectively.
The flux tube model is in relatively good agreement with the simulations. It captures the initial rapid growth up until the asymptotic behaviour after $t \sim 5t_0$. Notice the impact of the velocity which also shortens the time needed to reach the asymptotic trend.

The discrepancies between our simplified flux tube model and the actual magnetic field evolution in the simulations can be attributed to several factors.

Firstly, our model assumes that the magnetic field grows infinitely and uniformly in the stream. In reality, the field growth is constrained by the energy equipartition, where $B^2 /2 \sim \rho \Delta v^2 /2$ and is not uniform, with stronger growth as we approach the interface between the stream and the CGM. In the mixing layer, equipartition gives a constraint of $B \lesssim \left(300\text{--}500\right)\times \mathcal{M}$, which is either higher or similar to the maximum magnetic field observed in our simulations. Because our model reflects the mean magnetic field in the stream, it does not reach such maximum value and thus misses the point at which it should remain constant. 
Also, in realistic scenarios, the magnetic field can be amplified by turbulent eddies and compression from condensation. 
These additional mechanisms of magnetic field amplification are not accounted for in our simple model, leading to a faster asymptotic trend in the simulations compared to the model.
Furthermore, the simulations may exhibit non-monotonic behavior of the magnetic field, whereas our model assumes a constant magnetic field growth. Factors such as the deceleration of the stream, numerical magnetic diffusion, and the broadening of the mixing length can all influence the growth and behavior of the magnetic field, causing deviations from the idealized model predictions.


\bsp	
\label{lastpage}
\end{document}